\documentclass[3p,times,12pt]{elsarticle}

\journal{Nuclear Physics B}

\usepackage[colorlinks]{hyperref}
\usepackage{amsmath,amssymb}
\allowdisplaybreaks[2]

\usepackage{epstopdf}
\usepackage{amsmath}
\usepackage{amsfonts}
\usepackage{graphicx}
\usepackage{appendix}
\usepackage{dsfont}
\usepackage{amssymb}
\usepackage{bm}
\usepackage{color}
\usepackage{ulem}
\usepackage{soul}
\usepackage{natbib}
\usepackage{nccmath}
\usepackage{array}

\newcommand{\beq}{\begin{equation}}
\newcommand{\eneq}{\end{equation}}
\newcommand{\be}{\begin{equation}}
\newcommand{\ee}{\end{equation}}
\newcommand{\bea}{\begin{eqnarray}}
\newcommand{\eea}{\end{eqnarray}}

\usepackage{multirow}
\usepackage{wasysym}

\begin{document}

\begin{frontmatter}



\title{Real fermion modes, impurity entropy, and nontrivial fixed points in the phase diagram of junctions of interacting quantum wires 
and topological superconductors }


\author[unical,infncos]{Domenico Giuliano\corref{cor1}}
\ead{domenico.giuliano@fis.unical.it}
\author[ubc]{Ian Affleck}
\ead{iaffleck@phas.ubc.ca} 

\address[unical]{Dipartimento di Fisica, Universit\`a della Calabria,  
Arcavacata di Rende I-87036, Cosenza, Italy }
\address[infncos]{INFN, Gruppo collegato di Cosenza, 
Arcavacata di Rende I-87036, Cosenza, Italy }
\address[ubc]{ Department of Physics and Astronomy and Stewart Blusson Quantum Matter Institute,
University of British Columbia, Vancouver, B.C., Canada, V6T, 1Z1} 
\cortext[cor1]{Corresponding author}

\begin{abstract}
We discuss how to  extend the impurity entropy to systems with 
boundary interactions depending on zero-mode real fermion operators (Majorana modes as well as Klein factors). 
As specific applications of our method, we consider a junction between $N$ interacting quantum wires and a topological 
superconductor, as well as a Y-junction of three spinless interacting quantum wires.  In addition  we
find a  remarkable correspondence between  the $N=2$ topological superconductor junction 
and the  Y-junction.  On one hand, this  allows us to  
determine the range of the system parameters 
in which  a stable phase of the $N=2$ junction is realized as   a nontrivial, finite-coupling 
fixed point corresponding to the $M$-fixed point in the phase diagram of the Y-junction. 
On the other hand, it enables us to show the occurrence of a novel ``planar'' finite-coupling 
fixed point in the phase diagram of the Y-junction. Eventually, we  discuss how to set the system's parameters to realize the correspondence.  

\end{abstract}

\begin{keyword}


Quantum wires \sep Fermions in reduced dimensions \sep
Electron states and collective excitations in multilayers,
quantum wells, mesoscopic, and nanoscale systems  
\PACS 73.21.Hb \sep 71.10.Pm  \sep 73.21.-b 
\end{keyword}

\end{frontmatter}
 
\section{Introduction}
\label{intro}
 
Recently, considerable interest has arisen in junctions involving interacting quantum wires (QW's), both 
in the case of  spinless \cite{oca_2,oca,lal_1,reinhold_1,meden_1,bellazzini_1}, and of spinful 
wires \cite{claudio_1,shi,nava,amit}. This is mainly due to the fact that Landau's Fermi liquid 
paradigm typically breaks down in one-dimensional interacting electronic systems, whose low-energy, 
long-wavelength properties are rather described by means of  the Tomonaga-Luttinger liquid (TLL) framework 
 \cite{tll_1,tll_2}. Within the TLL-approach,   tunneling processes at a 
 junction are described in terms of  nonlinear vertex operators of the bosonic fields, 
 with nonuniversal scaling dimensions continuously depending  on the ``bulk'' interaction parameters \cite{hald_1,hald_2}. 
 This opens the way to a plethora of 
nonperturbative  features in the phase 
diagram of those systems, including the remarkable emergence of intermediate, finite-coupling 
fixed points (FCFP's), either describing phase transitions between different phases (repulsive fixed points),
or novel, nontrivial phases of the junction (attractive fixed points), thus generalizing to 
multi-wire junctions the Kane-Fisher FCFP  emerging at a junction between 
two spinful QW's  \cite{kafi}.  

In this context, the prediction  that localized Majorana modes (MM's) can appear at a junction between a 
normal QW  and a topological superconductor (TS)  \cite{kitaev} has 
opened additional brand-new scenarios, as the direct coupling between a quantum wire and a localized MM 
can potentially give rise to relevant boundary interactions, typically not allowed at junctions between 
normal wires \cite{alicea}. As a result, it has been possible to predict  the emergence of 
novel FCFP's in the phase diagram of  junctions between 
more-than-one interacting QW  and TS's \cite{giuaf_1,pikulin}. 
Moreover, due to ubiquity of the TLL-formalism, which successfully
describes (junctions of) quantum spin chains \cite{affleck_eggert,rossini,tsve_1,dgjjn}, 
Josephson junction networks \cite{giusoY,giusoju,gspairing,cirillo}, as well 
as topological, Kondo-like systems \cite{beritopo,erikson,betheb,alegger}, novel FCFP's have been 
predicted to emerge   in the   phase diagram of those systems, as well.  Besides their theoretical interest,
FCFP's   have been argued to 
correspond to ``decoherence-frustrated'' phases, in which   competing frustration effects 
can operate to reduce the unavoidable 
decoherence  in the boundary quantum degrees of freedom coupled to the ``bath'' of bulk modes 
\cite{novais_1,novais_2}, thus making the junction, regarded as a localized quantum impurity, 
 a  good candidate to work as a
frustration-protected quantum bit \cite{giusoY}. For this reason, it becomes of importance 
to search for FCFP's in the phase diagram of 
pertinently designed junctions of quantum wires.

An effective means to study the phase diagram of junctions of QW's is 
given by a cooperative combination of   perturbative renormalization group (RG) approach  \cite{kafi_0,kafi,aflud_1} and of  
the delayed evaluation of  boundary conditions (DEBC) technique \cite{oca,claudio_1}. 
In particular, as extensively discussed in   e.g. Ref. \cite{oca}, DEBC 
method  is based on constructing the boundary operators allowed by symmetries at a 
certain fixed point: the emergence of  (at least one) relevant operator  is, therefore, 
 evidence of the instability of that fixed point against some other one. In addition, 
as pointed out, for instance, in Ref. \cite{oca}, for a junction of normal QW's
and in Refs. \cite{giuaf_1,pikulin} for junctions between interacting QW's and TS's, 
  those methods are efficiently complemented by using the zero-temperature
impurity entropy (IE) to characterize the fixed points of the junction.
The IE was originally introduced as a mean to characterize and classify, 
at zero temperature,  quantum impurity systems that  are critical in the bulk  
at the fixed  points of their boundary phase diagram  \cite{aflud_1,aflud_2}.
Later on, it  has been shown to correspond to the impurity contribution to the groundstate 
entanglement entropy, which is particularly suitable for characterizing the phases of 
1+1-dimensional models via the Density Matrix Renormalization Group (DMRG)
technique \cite{pasqjohn}. The exponentiated IE $g$ 
yields the groundstate degeneracy of the system at a certain fixed point. In general, 
$g$ corresponds to the specific value of the ``$g$-function'', which 
always decreases along renormalization group trajectories (``$g$-theorem''); therefore,
if two boundary fixed points are characterized by groundstate degeneracies $g_1$ and 
$g_2$, with $g_1 > g_2$, provided there is the possibility of connecting the two of them with a renormalization 
group trajectory, this will always flow from $g_1$ to $g_2$.  \cite{aflud_1,aflud_2,friedan,friedan_2,friedan_3}.

Technically, to compute $g$ one first constructs the partition function over a 
strip of length $\ell$ for the model at a boundary critical point. Letting ${\cal A}$ label 
the   conformally invariant boundary conditions (CIBC's)  characterizing a specific boundary critical point, 
one  computes the partition function on the strips, ${\cal Z}_{AA}$, by assuming type ${\cal A}$-CIBC's at 
both boundaries. Specifically, one gets ${\cal Z}_{AA}=\sum_n\exp \left[ - \frac{ x_{AA}^n \beta u}{\ell} \right]$,
where the sum is taken over energy eigenstates of the whole system, 
$u$ is a velocity scale determined by the (critical) bulk of the system, $\beta$ is the 
inverse temperature, and  $x_{AA}^n$ are dimensionless numbers typical of the system. 
The $g$-function is derived by sending  
$\ell \to \infty$ at finite $\beta$. In this limit, one obtains 
${\cal Z}_{AA} \to g_A^2 e^{\frac{ \pi \ell c}{ 6 \beta u}}$,
with the dimensionless number $c$ being the conformal anomaly of the bulk 
critical theory. From the last result, $g_A$ can be readily extracted \cite{afosh}.

In this paper we employ a combined use of perturbative RG approach, DEBC-method and 
calculation of the $g$-function to study  junctions of interacting QW's and 
TS's and to spell out the correspondence between this sort of junctions and 
the Y-junction of three interacting quantum wires (Y3J) studied in Ref. \cite{oca} in the 
bulk ${\bf Z}_3$-symmetric version and later on discussed in Ref. \cite{claudio_2} in the general 
case in which the  ${\bf Z}_3$ symmetry is broken. In doing so, we 
have necessarily to take into account the emergence of real fermionic modes at
our junctions.   These are primarily provided 
by the localized MM's $\{\gamma_j \}$ emerging at the interface between a topological superconductor 
and a normal system \cite{kitaev,oreg,dassarma}. In addition,  real fermionic operators also appear
as Klein factors (KF's), $\{ \Gamma_j \}$, which have to be introduced when  
employing the bosonization approach to interacting one dimensional fermionic systems, to recover the correct (anti)commutation relations between fermionic fields for different 
wires, as well as between the fermionic field for each wire and the $\gamma_j$'s. 
In many cases, KF's play  no role, as the relevant multi-point 
correlation functions of the fermionic fields either contain an equal number of creation and annihilation operators
of fermions of the same kind (see, for instance, Ref. \cite{vondelft} and reference therein), or, at most, the net effect 
of KF's can be an extra minus sign which can be equally well accounted for by, for instance, 
 redefining the zero-mode operators of the bosonic fields \cite{leeex}. On the other hand, 
they are definitely essential to recover the correct phase diagram of e.g. a junction of three interacting 
QW's \cite{oca}, as well as to correctly account for the hybridization between MM's and normal 
electronic modes in a conductor \cite{pascal_1,pascal_2}, at junctions 
between normal wires and TS's  \cite{alicea,giuaf_1,giu_af_majo}, and in the remarkable 
``topological'' realization of the Kondo effect, in which MM's determine an effective impurity spin 
coupled to electronic modes by the normal contacts \cite{beri}. Thus, it is by now evident that they 
must be properly accounted for in applying the $g$-theorem to junctions of interacting quantum wires.
 
Besides generalizing the results of Refs. \cite{giuaf_1,oca} respectively  to a multiwire junction of interacting 
QW's and a TS's and to a non ${\bf Z}_3$-symmetric Y3J, we unveil the remarkable correspondence between the two models. In doing so, we prove how, extending the range of 
system's parameters with respect to the case discussed in Ref. \cite{giuaf_1}, it is possible for
a FCFP to correspond to the stable phase of the system.   Remarkably, 
this completely reverses the scenario found within the range of parameters discussed in  Ref. \cite{giuaf_1}, where we 
proved that it is necessary to fine-tune  the 
boundary couplings to the MM, to drive the RG flow towards the FCFP, which was   unstable against more ``trivial'' 
fixed points.  
Apart  being  interesting per se, this result appears also of relevance for engineering stable 
phases with frustrated decoherence, potentially amenable for applications to quantum computation.
An important point to stress about the correspondence is that it holds despite the 
obvious observation that the Y3J has three KF's, while the $N=2$ junction has 
just two KF's and one MM. In fact, as we discuss in the following, while one of 
the three fields of the Y3J decouples from the boundary interaction, its 
KF   gets ``left behind" and it plays the role
of the MM.  Therefore, the correspondence works perfectly well when considering 
a single junction in both systems. Remarkably, it also yields the right result when computing $g$
at the fixed points of the two model. In this case, as we discuss in detail below, one has to 
resort to a two-boundary version of the corresponding model Hamiltonian. Thus, one 
cannot ignore the intrinsic difference between MM's, which can be assumed to be basically local in real space, 
and KF's, which on the contrary are global, as one naturally associates them with the whole extent of 
a QW in real space. Despite this,  the correspondence works fine and allows for 
recovering a number of nontrivial results about the phase diagram of one model from what is known about
the phase diagram of the other.

The paper is organized as follows:

\begin{itemize}
 
\item In section \ref{pha_1} we introduce our procedure for computing the $g$-function
in boundary models with real fermionic modes emerging  in   the boundary interaction 
Hamiltonian. To illustrate our procedure, here we apply it to  a single
interacting spinless quantum wire connected to  two $p$-wave superconductors, by discussing in 
detail the subtleties in counting the degrees of freedom associated with the real fermionic modes and
how to deal with them;

\item In section \ref{phadia_1} we discuss the main features of the phase diagram and compute  the impurity 
entropy at the fixed points of a junction between two quantum 
wires and a topological superconductor and of the asymmetric Y junction of three spinless interacting quantum wires.
In both cases,  we mostly review known results \cite{giuaf_1,oca,claudio_2} which, nevertheless, are
important for the sake of the presentation of the following results;
 
\item In section \ref{pha_2} we discuss in detail  the correspondence between 
a junction with two quantum wires  and a topological superconductor and the asymmetric  Y3J.  
In particular, we show how the results derived in Ref. \cite{giuaf_1} for the former system can 
shed light on the phase diagram of the Y3J in the case of asymmetric bulk, as well as boundary interaction and, 
conversely, how the results of Refs. \cite{oca,claudio_2}  for the Y3J allow for extending the 
analysis of the phase diagram of the junction  with two quantum wires  and a topological superconductor 
to windows of values of the system's parameters which were not encompassed in the derivation 
of Ref. \cite{giuaf_1};

\item In section \ref{junction_N}, as a further application of our method for computing 
the $g$-function in boundary models with real fermion modes emerging at the boundary interaction, 
  we generalize the results of Ref. \cite{giuaf_1} by 
discussing the fixed points in the phase diagram, and the corresponding calculation of the 
$g$-function, in   a  junction between $N$ quantum wires  and a topological superconductor. 

\item In section \ref{concl} we provide our conclusions and discuss  possible further developments of our 
work.

\item In the various appendices, we provide mathematical details of our derivation.

\end{itemize}
 
To help following the various abbreviations, we list in table 
\ref{abbr} the meaning of  the ones we use most commonly throughout the paper.

\vspace{0.2cm}
 
\begin{table}
\centering
\begin{tabular}{| c | c |}
\hline 
TLL & Tomonaga-Luttinger liquid \\
 \hline
FCFP  & Finite-coupling fixed point \\
 \hline
MM & Majorana mode \\
 \hline
TS & Topological superconductor \\
 \hline
IE & Impurity entropy  \\
 \hline
CIBC & Conformally invariant boundary condition \\
 \hline
RG & Renormalization group \\
 \hline
KF & Klein factor \\
 \hline
QW & Quantum wire \\
 \hline
Y3J & Y-junction of three interacting quantum wires \\
 \hline 
\end{tabular}
\caption{Glossary of most commonly used abbreviations} 
   \label{abbr}
\end{table}
\noindent

 \section{Impurity entropy in a boundary model with real fermionic modes in the boundary interaction} 
\label{pha_1}

When bosonizing more than one species of fermion operators in one dimension, real fermionic Klein factors 
must be introduced, to properly account for the anticommutation relations between operators corresponding to 
different species of fermions. Typically, KF's  appear in
boundary Hamiltonians describing junctions of one-dimensional quantum wires  (which is appropriate, 
at points where different wires contact  each other) and, in many 
cases, they strongly affect the boundary dynamics of the junction  \cite{vondelft}. For instance, only by 
properly accounting for KF's in the boundary Hamiltonian, does one prove  the emergence of a FCFP in the Y3J discussed in 
Refs. \cite{oca_2,oca}, or in its spinful version \cite{claudio_1}. 
In addition to KF's, real fermion operators emerge as MM's at junctions between  
QW's and TS's \cite{kitaev}. The combined effect of KF's and MM's can 
eventually lead to the ``Majorana-Klein'' hybridization and, eventually, to 
a remarkable topological version of the Kondo effect \cite{beri,beritopo,betheb}, also discussed in 
its multi-channel realization \cite{altland,alegger,erikson}, as well as to novel phases corresponding to 
FCFP's in the phase diagram of junctions between QW's and TS's \cite{giuaf_1,pikulin}. 
Thus, despite their definition   as a mathematical means for properly doing bosonization, KF's affect the 
boundary dynamics of a junction exactly as ``physical'' MM's do and, accordingly, they must 
be properly accounted for, when computing the IE  of the junction. To demonstrate this point, in this section we compute the IE
in a paradigmatic system given by  a single spinless interacting wire connected to two TS's 
at its endpoints. This enables us to show how, in order to 
find results for the $g$-function 
consistent with the expected phase diagram of this systems, one has to
count the degrees of freedom associated with real fermions at the system boundaries. In doing so, 
we face an additional subtelty, which was originally put forward in Ref. \cite{wilczek},
which is  strictly related to  how to count real, zero-mode fermionic degrees of freedom. 

In general, two real fermionic modes, say $\gamma_a$ and 
$\gamma_b$, can be combined together into a complex (Dirac) fermionic mode $a = \frac{1}{2} ( \gamma_a + i \gamma_b )$, which 
leads to a single fermion energy level, which can be either empty, or full, eventually resulting 
in an additional degeneracy factor of 2 in the partition function. In a boundary theory, 
the procedure for computing the partition function (and, eventually extracting the $g$-function from
the result)  consists in making up a two-boundary version of the model Hamiltonian   by mirroring the
boundary interaction describing the junction  at the other boundary of a finite-size ($\ell$) version of the system
(see  Ref. \cite{rahmani} for details of the procedure).  While this procedure 
unavoidably leads to a doubling of the  MM's emerging at the interfaces (so they always contribute 
an even number of real fermionic modes) \cite{alicea,giuaf_1,giu_af_majo,pikulin},  when introducing KF's through bosonization
of the normal wires, the final total number of real fermionic modes can either be even, or odd. 
When it is odd, one has to face  an  ambiguity about   how
to count the left-over real fermion, which 
is strictly related to the need to account for fermion 
parity conservation in the presence of real fermion operators \cite{wilczek}. To overcome such a difficulty, 
 we introduce an additional decoupled  ``auxiliary'' 
fermionic wire which, in bosonization language, is characterized by its own Luttinger parameter $\bar{K}$ and
by its Klein factor $\bar{\Gamma}$. $\bar{\Gamma}$  enters the 
total counting of degrees of freedom related to real fermions, by pairing with the real mode that 
is left over after all the other ``physical'' ones have paired into Dirac complex modes. Of course, 
we expect this to affect the actual value of the $g$ function. Yet, as for any definition of entropy, what matters in 
the $g$-theorem is the entropy {\it difference} between two different fixed points or, which is the same, the ratio between the corresponding values of $g$. 
In fact, we expect our procedure to provide the correct result for the ratio and, to ground our 
speculation, in the following we provide a number of different examples of physical interest 
where we show that this is, in fact, the case. 

To illustrate our procedure, here we apply it to compute the $g$-function in a single
interacting spinless quantum wire connected to  two $p$-wave superconductors in
their topological phase \cite{kitaev} at its endpoints. At low energies, the superconducting leads can be traded for 
two MM's $\gamma_L , \gamma_R$, respectively residing at the left-hand side and at the 
right-hand side of the QW. In addition, the interacting QW is effectively described by 
resorting to the bosonization approach, which we review in   \ref{bomo}, in 
terms of the bosonic fields $\phi ( x ) , \theta ( x )$, whose dynamics is 
encoded in the Luttinger liquid Hamiltonian in Eq. (\ref{bomo.1}). In particular, 
when expressing the chiral fermion operators $\psi_R ( x ) , \psi_L ( x )$ in 
terms of $\phi ( x )$ and $\theta (x)$, one sets 

\beq
\psi_R ( x ) = \Gamma \: e^{ i \sqrt{\pi} [ \phi ( x ) + \theta ( x ) ] } \;\;\; , \;\; 
\psi_L ( x) = \Gamma \: e^{ i \sqrt{\pi} [ \phi ( x ) - \theta ( x ) ] } 
\:\:\:\: , 
\label{inte.1}
\eneq
\noindent
with the KF $\Gamma$ such that $\{ \Gamma , \gamma_L \} = \{ \Gamma, \gamma_R \} = 0$. 
Together with the MM's, the KF forms a set of three real fermionic zero-mode operators. 
This is the case in which, as we discuss above, to consistently count for the 
corresponding degrees of freedom, we introduce a second wire, described again by a 
 Luttinger liquid Hamiltonian such as the one in Eq. (\ref{bomo.1}), with 
bosonic fields $\bar{\phi} ( x ) , \bar{\theta} ( x )$, with 
 parameters $\bar{u} , \bar{K}$ and, more importantly, requiring the 
introduction of a second KF, $\bar{\Gamma}$. The second wire is 
fully decoupled from the rest of the system. Therefore, for any values of the boundary 
couplings, $\bar{\theta} ( x )$ is pinned at both $x = 0$ and $x=\ell$ and, accordingly, 
$\partial_x \bar{\phi} ( x = 0 ) = \partial_x \bar{\phi} ( x  = \ell ) = 0$ (type $N$ 
CIBC's). Therefore, $\bar{\phi} ( x ) , \bar{\theta} ( x ) $  are 
decomposed in normal modes according to Eqs. (\ref{bomo.2}), with velocity and Luttinger 
parameters $\bar{u} , \bar{K}$. The calculation of the factor that the
auxiliary QW contribute the total partition function, $\bar{\cal Z}$, is, therefore, a straightforward exercise in 
elementary algebra: the result is 

\beq
\bar{\cal Z} = \frac{1 }{\eta ( \bar{q} ) } \: \sum_m \bar{q}^\frac{m^2}{2 \bar{K}} 
\:\:\:\: , 
\label{aax.2}
\eneq
\noindent
with $\bar{q} = e^{ - \frac{\beta \pi \bar{u}  }{\ell} }$ (see   \ref{duapoi} for the definition 
of the Dedekind function $\eta ( q )$). At the disconnected fixed point of the phase diagram
the QW is fully decoupled from the TS's. Therefore, $\phi ( x ) , \theta ( x )$ take the expansion 
in normal modes  in Eqs. (\ref{bomo.2}), as well, and they will accordingly contribute the total 
partition function by a factor analogous to  $\bar{\cal Z}$ in Eq. (\ref{aax.2}). In addition, 
we have to account for the (4) real fermionic zero mode operators. As at the disconnected fixed point 
they are fully decoupled from each other, as well as from the dynamical degrees of freedom of the 
bulk, according to the above argument, we 
expect them to contribute to the partition function a factor equal to 2 elevated to the total number of 
real fermions divided by 2 (that is, 4). As a result, the total partition function at the disconnected fixed 
point is given by 

\beq 
\hat{\cal Z}_{\rm Disc} = \frac{4 }{\eta ( q ) \eta (\bar{q} )  } \: \sum_{m , m'} q^\frac{m^2}{2 K} \:   \bar{q}^\frac{(m')^2}{2 \bar{K}}
 \:\:\:\: ,
\label{aax.2bis}
\eneq
with  $ q = e^{ - \frac{\beta \pi  u  }{\ell} }$
Using the standard asymptotic expansions of the Dedekind function and of the elliptic functions in 
the $\ell \to \infty$ ($q \to 1$)-limit, one eventually finds that 

\beq
\hat{\cal Z}_{\rm Disc} \to_{\ell \to \infty} e^{ [  \pi \ell / ( 6 u ) ]   + [ \pi \ell / (6 \bar{u} )]   }\: [ 4 \sqrt{K \bar{K}} ] 
\:\:\:\: , 
\label{aax.ter}
\eneq
\noindent
which implies, for the $g$-function at the disconnected fixed point 

\beq
g_{\rm Disc} = 2 [ K \bar{K} ]^\frac{1}{4} 
\:\:\:\: . 
\label{aax.f}
\eneq
\noindent
Turning on the (two)-boundary coupling to the MM, taking into account the type $N$ CIBC's at 
the disconnected fixed point, the two-boundary Hamiltonian $H_b$ takes the form

\beq
H_b =  - 2 i t  \gamma_L \Gamma \: \cos [ \sqrt{\pi} \phi ( 0 ) ] - 
2 i t \gamma_R \Gamma \: \cos [ \sqrt{\pi} \phi ( \ell ) ] 
\:\:\:\: .
\label{aax.5}
\eneq
\noindent 
When $K> 1/2$, $H_b$  is a relevant operator, which  drives the system toward a fixed point 
in which  $\phi ( x )$ is pinned both at $x=0$ and at $x=\ell$, as $\phi ( 0 ) = \sqrt{\pi} \nu_0$, $\phi ( \ell ) = 
\sqrt{\pi} \nu_\ell$, with integer $\nu_0 , \nu_\ell$. Accordingly, $\phi ( x )$ and $\theta ( x )$ take the mode expansion in Eqs. (\ref{bomo.3}) and, 
in addition, taking into account the boundary conditions, one gets 
$H_b =   - 2 i t  \left\{  (-1)^{\nu_0} \gamma_L +   (-1)^{\nu_\ell} \gamma_R \right\}
\Gamma$. A mode expansion such as the one in Eqs. (\ref{bomo.3}) implies a corresponding contribution to the total partition function
such as the one in Eq. (\ref{aax.2}), except for switching $K$ with $K^{-1}$. In addition, minimizing 
$H_b$ locks together the real fermions $\Gamma$ and $\tilde{\gamma} = \frac{1}{\sqrt{2}} \gamma_L + (-1)^{\nu_0 - \nu_\ell} 
\frac{1}{\sqrt{2}} \gamma_R$ by the condition that physical states are annihilated by the 
complex fermionic operator $a = \frac{1}{2} ( \tilde{\gamma} + i \Gamma )$. This condition leaves unpaired 
the real fermion operator  $\eta = - \frac{1 }{\sqrt{2} } \gamma_L + (-1)^{\nu_0 - \nu_\ell}
 \frac{1 }{\sqrt{2} } \gamma_R$ which, together with $\bar{\Gamma}$, determines 
 an additional degeneracy factor of 2. Therefore, the partition function and the $g$-function at the corresponding 
 fixed point are  given by  
 
\beq
{\cal Z}_{A} = \frac{2}{\eta ( q ) \eta ( \bar{q} ) } \:   \sum_{m , m'} q^\frac{K m^2}{2 } \bar{q}^\frac{ (m')^2}{2 \bar{K} } \:  
\to_{\ell \to \infty} e^{ [  \pi \ell / ( 6 u ) ]   + [ \pi \ell / (6 \bar{u} )]   }\: \left[ 2 \sqrt{  \frac{ \bar{K}}{K}  } \right] 
\:\:\:\: . 
\label{aax.3}
\eneq
\noindent
From Eq.(\ref{aax.3}) we eventually extract the corresponding value of the $g$-function, 
$ g_{A} = \sqrt{2} \left[ \frac{ \bar{K}}{K} \right]^\frac{1}{4}$.
The derivation of Eq. (\ref{aax.3}) is a first example of implementation of the Majorana-Klein hybridization \cite{beri} in computing 
the $g$-function at the $A$ fixed point. Indeed, the real-fermion zero-mode operator we combined together with 
$\bar{\Gamma}$ to obtain the degeneracy factor of 2 originates from a linear combination of $\gamma_L $ 
and $\gamma_R$, with coefficients determined by the boundary conditions. In addition, we note that, introducing the auxiliary 
wire on one hand allowed us to perform the calculations in an unambiguous way, on the other hand gave us back 
$g_{\rm Disc}$ and $g_A$ up to an over-all arbitrary, multiplicative factor. To get rid of the factor, we 
normalize $g_A$ to $g_{\rm Disc}$ by considering the ratio  

\beq
\rho_A = \frac{g_A}{g_{\rm Disc}} = \frac{1}{\sqrt{2 K }}
\:\:\:\: , 
\label{aay}
\eneq
\noindent
which, on one hand shows that the ratio is independent of the (arbitrary) parameter $\bar{K}$, as  it must be, on the other hand 
that, consistently with the $g$-theorem, the renormalization group trajectories are expected to flow from the disconnected to the 
$A$ fixed point when $K > 1/2$. 

Later on in the paper, we generalize the results of this section to a junction between a generic number $N$ of QW's and a topological superconductor. 
In this case, the relevant Luttinger parameters are  $K_\rho = K / \sqrt{1 + \frac{(N-1) U K}{\pi u }}$ and 
$K_\sigma =  K / \sqrt{1 -  \frac{U K}{\pi u }}$, with $u, K$ being the Luttinger parameter of each QW and $U$ being the 
inter-wire interaction strength. (See   \ref{bomo} for a detailed derivation and discussion of the
Luttinger parameters for  the $N$-wire junction.)

\section{Review of the phase diagram of the $2$-wire junction with a topological superconductor 
for $\frac{1}{2} < K_\sigma < 1$ and of the Y-junction of interacting quantum wires}
\label{phadia_1}

To complement the calculation of the $g$-function and to pave the way to  the derivation of the 
correspondence between the $N$=2-wire junction with a TS and 
the Y3J, here we review known results 
about the phase diagram of the two systems.  In doing so, we also compute the $g$-function at the various fixed points 
of the (generically asymmetric) 
Y3J, and demonstrate the consistency between our results and what is
known about the phase diagram of that system \cite{claudio_2}. We now begin by briefly reviewing  
the phase diagram of the $N=2$-wire junction with a TS, which we extensively analyzed and discussed 
in Ref. \cite{giuaf_1} in the  regime $  K_\sigma < 1$.  

\subsection{Phase  diagram of the $2$-wire junction with a topological superconductor for
$   K_\sigma < 1$}
\label{phadiaN2}

We refer to section \ref{junction_N} for an extensive derivation, from the 
microscopic Hamiltonian, of the 
Luttinger liquid description of a generic junction between $N$-QW's and a 
TS and for the analysis of the corresponding phase diagram. Here, we just review the main features of the phase diagram in 
the case $N=2$  starting from the effective Luttinger liquid Hamiltonian, 
$H_{2,{\rm B}} = H_{2 , {\rm Bulk}} + H_{b , {\rm B} ,2}^{(1)}$, with the 
bulk Hamiltonian 

\beq
H_{2 , {\rm Bulk}} =\frac{u_\rho}{2}   \int_0^\ell \: d x \: \{ 
K_\rho ( \partial_x \phi_\rho ( x ))^2 + K_\rho^{-1} ( \partial_x \theta_\rho ( x ))^2 \} 
+ \frac{u_\sigma}{2}  \int_0^\ell \: d x \: \{ 
K_\sigma ( \partial_x \phi_\sigma  ( x ))^2 + K_\sigma^{-1} ( \partial_x \theta_\sigma ( x ))^2 \}
\;\;\;\; , 
\label{pd.1}
\eneq
\noindent
with the plasmon velocities and the Luttinger parameters defined in terms 
of the microscopic system parameters as per Eqs. (\ref{1w.19}) for $N=2$,
and $\phi_\rho  ( x ) = \frac{1}{\sqrt{2}} [ \phi_1 ( x ) + \phi_2 ( x ) ] , \phi_\sigma ( x ) = \frac{1}{\sqrt{2}} [ \phi_1 ( x ) - 
\phi_2 ( x )]$, $\phi_1 ( x ) , \phi_2 ( x )$ being the fields in the two wires, and 
analogous formulas for $\theta_\rho  ( x ) , \theta_\sigma ( x )$ in terms of $\theta_1 ( x ) , \theta_2 ( x ) $. 
By construction, $H_{2 , {\rm Bulk}}$ in Eq. (\ref{pd.1}) corresponds to the symmetric 
version of the model Hamiltonian  of  Ref. \cite{giuaf_1}, in which the Luttinger parameters
of the two QW's are equal to each other, $K_1 = K_2=K,u_1= u_2=u$, 
and $\phi_\rho ( x ) , \theta_\rho ( x )$ ($\phi_\sigma ( x ) , \theta_\sigma ( x )$) simply correspond 
to the symmetric (asymmetric) linear combinations of $\phi_1 ( x ) , \phi_2 ( x )$ and $\theta_1 ( x ) , \theta_2 ( x )$. 
  Accordingly, the chiral fermionic fields in the QW's are 
bosonized as (see     \ref{bomo} for details)

\begin{eqnarray}
 \psi_{R , a } ( x ) &=& \Gamma_a e^{i \sqrt{\pi} [ \phi_a ( x ) + \theta_a ( x ) ] } \nonumber \\
 \psi_{ L, a } ( x ) &=& \Gamma_a e^{ i \sqrt{\pi} [ \phi_a ( x ) - \theta_a ( x ) ] } 
 \:\:\:\: ,
 \label{pd.1a}
\end{eqnarray}
\noindent
with $a = 1,2$ and $\Gamma_1 , \Gamma_2$ fermionic KF's. 
Having stated this, the phase diagram for $K_\sigma < 1$ is 
readily recovered from the analysis of Ref. \cite{giuaf_1}. 

The simplest fixed point corresponds to setting all the boundary couplings $t_a = 0$. 
This is the disconnected fixed point, which, in a sample defined over a segment of 
length $\ell$, corresponds to pinning $\theta_1 ( x ) , \theta_2 ( x )$ at both boundaries
$x = 0 , \ell$. The corresponding value of the $g$-function can be computed   as done in Ref. \cite{giuaf_1}, 
yielding, as a special case 
of the general formula we derive in  section 
\ref{junction_N}, the result 

\beq
g_{\rm Disc} = 2 [ K_\rho K_\sigma ]^\frac{1}{4} 
\:\:\:\: , 
\label{pd.2}
\eneq
\noindent
with the factor 2 stemming from the fourfold degeneracy due to the two MM's in the two-boundary 
version of the model and to the KF's associated to the two QW's. (Note that, differently from 
what was done in Ref. \cite{giuaf_1}, here we do not count the additional degeneracy 
associated with the MM's at the outer boundaries of the superconducting lead, as they contribute 
to the $g$-function just an overall factor of $2$, which drops from the physically meaningful 
ratios between $g$ computed at two different fixed points). At the disconnected fixed point, 
the boundary Hamiltonian describing the coupling of the QW's to the MM is presented in 
bosonic coordinates as

\beq
H_ {b, {\rm B} , 2 }^{(1)}  =     - 2 i  t_1  \gamma_L \Gamma_1 \: \cos \left[ \sqrt{\frac{\pi}{2}} ( \phi_\rho ( 0 ) + \phi_\sigma ( 0 ) ) \right]
-    2 i  t_2  \gamma_L \Gamma_2 \: \cos \left[ \sqrt{\frac{\pi}{2}} ( \phi_\rho ( 0 ) - \phi_\sigma ( 0 ) ) \right]
\;\;\;\; . 
\label{1w.202}
\eneq
\noindent
The scaling dimension of $H_ {b, {\rm B} , 2 }^{(1)}$, $d_{b }$, can be 
readily derived using the transformation in Eqs. (\ref{1w.8bis}). The result is 
$d_{b }  = \frac{1}{4 K_\rho} + \frac{ 1}{4 K_\sigma} $. 
For $1/2 < K < 1$ and for $U>0$, we find $\frac{1}{2} < d_b < 1$, which implies that
 $H_ {b, {\rm B} , 2 }^{(1)} $ always corresponds to a relevant boundary interaction. 
In addition,   other boundary interactions, though not present in the original 
(``bare'') Hamiltonian, can be generated from the RG. The first one  corresponds to 
   boundary normal intra-wire backscattering,  described by operators of the form $\psi_{R , a}^\dagger ( 0 ) \psi_{L , a } ( 0 ) + {\rm h.c.}$,
which are not effective, due to Dirichlet boundary conditions on $\theta_a ( x )$ at $x=0$. Then, one has  inter-wire normal backscattering, 
corresponding to  operators of the form $V_{{\rm Normal} , (1,2)} = \psi_{R , 1}^\dagger ( 0 ) \psi_{L , 2 } ( 0 ) + {\rm h.c.} \propto \Gamma_1 \Gamma_2 \cos [ \sqrt{2 \pi} \phi_\sigma ( 0 ) ]$, 
with corresponding scaling dimension $d_{{\rm Normal},(1,2)} = K_\sigma^{-1}$. Finally, one has inter-channel pairing,
corresponding to operators of the form  $V_{{\rm Pair} , (1,2)} = \psi_{R , 1 } ( 0 ) \psi_{L , 2 }  ( 0) + {\rm h.c.}
\propto \Gamma_1 \Gamma_2 \cos [ \sqrt{2 \pi} \phi_\rho ( 0 ) ]$, with scaling dimension $d_{{\rm Pair},(1,2)} = K_\rho^{-1}$. Thus, 
as long as $1/2 < K_\rho , K_\sigma < 1$, no relevant operators are allowed at the disconnected fixed point but the 
boundary coupling to the MM, $H_ {b, {\rm B} , 2 }^{(1)} $. 

The relevance of the operators in $H_ {b, {\rm B} , 2 }^{(1)}$ implies that, as soon  as (at least one of) the $t_a$'s are
$\neq 0$, the corresponding operator(s) trigger an RG flow away from the disconnected fixed point. In Ref. \cite{giuaf_1}
it is shown that, for $K_\sigma < 1$, the junction either flow towards a fixed point with type $A (N)$ boundary 
conditions in channel-1(2) (the $A \otimes N$ fixed point), or towards the complementary, $N \otimes A$ fixed point. In both cases, 
a straightforward implementation of our method yields, for the $g$ function 

\beq
g_{A \otimes N} = g_{N \otimes A} = \frac{ \sqrt{2K}}{ ( K_\rho K_\sigma)^\frac{1}{4} }  \Rightarrow \rho_{1,1} = \frac{g_{A \otimes N}}{g_{\rm Disc}} = \frac{1}{\sqrt{2K}}
\left[ 1 - \left( \frac{K U}{\pi u} \right)^2 \right]^\frac{1}{4} 
\;\:\:\:.
\label{pd.3}
\eneq
\noindent
The result in Eq. (\ref{pd.3}) corresponds to the symmetric limit of the junction discussed in Ref. \cite{giuaf_1}.
Note that it also fixes an error in that reference, though without affecting the final 
result. The important point is the over-all factor $\sqrt{2}$, which 
ensures that $\rho_{1,1} < 1$ as long as $1/2 < K$. It actually comes from the correct counting of the 
degrees of freedom associated with zero-mode real fermion operators. In particular, 
for the  RG flow towards the $A \otimes N$ ($N \otimes A$) fixed point to occur, the bare couplings must be such that 
$t_1 > t_2$ ($t_1 < t_2$). In this case, the RG makes the running coupling corresponding to the larger bare coupling constant flow 
all the way to $\infty$. Let us assume this is $t_1$. Accordingly, to recover the $A \otimes N$ fixed point, one considers the  two-boundary version of 
$H_ {b, {\rm B} , 2 }^{(1)}$, in which   $t_1 \to \infty$.   In fact, 
this implies ``locking'' two of the four real fermions
(2 KF's plus 2 MM's) into a linear combination annihilating the physical states, leaving the 
other two decoupled from the boundary interaction, with a total degeneracy factor of 2 in the 
total partition function and, therefore, a factor $\sqrt{2}$ in $g_{A \otimes N}$ \cite{giuaf_1}. 
The set of allowed boundary operators at the $A \otimes N$ fixed point includes the same 
operators we listed at the disconnected fixed point, though realized differently, and with 
different scaling dimensions, due to the change in the boundary conditions for the bosonic
fields \cite{giuaf_1}. In particular, normal intra-channel 1 backscattering corresponds to 
an operator $V_{{\rm Intra} , 1} \propto \cos [ 2 \sqrt{\pi} \theta_1 ( 0 ) ]$, with 
scaling dimension $d_{{\rm Intra} , 1} = d_b^{-1}$. At the same time,   inter-channel normal 
backscattering and inter-channel pairing are described by boundary operators that are linear 
combinations of $V_{a, (1,2)} $ and of $V_{b , (1,2)}$, respectively given by 

\begin{eqnarray}
 V_{a , (1,2)} &=& \Gamma_1 \Gamma_2 e^{ i \sqrt{\pi} [ \theta_1 ( 0 ) - \phi_2 ( 0 ) ]} \nonumber \\
 V_{b, (1,2)} &=& \Gamma_1 \Gamma_2  e^{ i \sqrt{\pi} [ \theta_1 ( 0 ) + \phi_2 ( 0 ) ]} 
 \;\;\;\; , 
 \label{pds.1}
\end{eqnarray}
\noindent
together with their Hermitean conjugates. The operators in Eqs. (\ref{pds.1}) have the 
same scaling dimension, $d_{a , (1,2)} = d_{b , (1,2)}$, which is consistent with  
the result in Eq. (D.26) of Ref. \cite{giuaf_1}. As a 
result, one obtains $d_{a , (1,2)} = d_{b , (1,2)}=  \frac{1 + K_\rho K_\sigma}{K_\rho + K_\sigma}$. For $N=2$, one obtains 
$K_\rho = K / \sqrt{1 + \frac{UK}{\pi u}}$, which is always $<1$ for $1/2 < K < 1$ and $U>0$. Thus, one obtains 
that $ \frac{1 + K_\rho K_\sigma}{K_\rho + K_\sigma} >1$, as long as $K_\sigma < 1$. Therefore, we conclude that both inter-channel normal 
backscattering and inter-channel pairing are described by irrelevant operators. 
Finally, an additional boundary operator arises from the residual boundary coupling of channel 2 
to the MM. This corresponds to the term $\propto t_2$ in Eq. (\ref{1w.202}). 
Despite the fact that it appears to correspond to a relevant operator, due to the 
hybridization between $\gamma_L$ and $\Gamma_1$ in the state that sets 
in at the $A \otimes N$ fixed point, it becomes effective only to order $t_2^2$, 
corresponding to an operator  $ V_{2 , {\rm Res}} \propto \cos [ 2 \sqrt{\pi} \phi_2 ( 0 ) ]$, 
with scaling dimension $d_{2 , {\rm Res}} = \frac{ 4  }{K_\rho + K_\sigma} >1$. This eventually 
proves that, for $K_\sigma < 1$, the stable phase of the $N=2$ junction either corresponds to 
the $A \otimes N$, or to the $N \otimes A$ fixed point. 

An additional possibility is provided by the $A \otimes A$ fixed point of Ref. \cite{giuaf_1},
with  type $A$ boundary conditions  in both channels. The $g$-function at 
the $A \otimes A$ fixed point can be readily derived either from the analysis 
of Ref. \cite{giuaf_1} (up to an over-all $\sqrt{2}$, as discussed above), or 
from the general result of section \ref{junction_N}, 
taken for $N=2$ and for $N_a=2,N_n=0$. As a result, one obtains 

\beq
g_{A \otimes A} = \frac{2}{ [ K_\rho K_\sigma]^\frac{1}{4}} \Rightarrow \Biggl\{ 
\begin{array}{l} \rho_{A \otimes A} = \frac{g_{A \otimes A}}{g_{\rm Disc}} = \frac{1}{ \sqrt{ K_\rho K_\sigma} } 
\\ 
\hat{\rho}_{A \otimes A} =  \frac{g_{A \otimes A}}{g_{A \otimes N}} = \sqrt{\frac{2}{K}}
\end{array}
\:\:\:\: .
\label{pd.4}
\eneq
\noindent
From Eqs. (\ref{pd.4}) one readily checks that, as long as $K_\sigma <1$, one obtains $\rho_{A \otimes A} < 1$, as 
well as $\hat{\rho}_{A \otimes A} < 1$, which implies that the $A \otimes A$ fixed point is unstable to both 
the $N \otimes N$, as well as to the $A \otimes N$ fixed point. This is consistent with the 
DEBC results about the set of allowed boundary operators at the $ A \otimes A$ fixed point. 
Indeed, implementing type $A$ boundary conditions at $x=0$ for both $\phi_1 ( x )$ and $\phi_2 ( x )$, we
see that the boundary operators  describing inter-channel normal backscattering, as
well as inter-channel pairing, are in general expressed as linear combinations of 
the operators $V_{a , (1,2)} , V_{b , (1,2)}$, respectively given by 

\begin{eqnarray}
V_{a , (1,2)}  &=& \Gamma_1 \Gamma_2 e^{ i \sqrt{2 \pi} \theta_\sigma ( 0 ) }  \nonumber \\
V_{ b , (1,2)}  &=& \Gamma_1 \Gamma_2 e^{ i \sqrt{2 \pi} \theta_\rho ( 0 ) } 
 \:\:\:\: ,
\label{2a.7}
\end{eqnarray}
\noindent
together with their Hermitean conjugates. Their scaling dimensions are 
accordingly given by  $d_{a , (1,2)}  = K_\sigma$, $d_{b , (1,2)} = K_\rho$. Thus, 
we see that, for $K_\sigma < 1$,  they both correspond to relevant boundary interactions.
Other  boundary interaction terms are determined by the 
operators $\tilde{V}_{{\rm Res} , 1} , \tilde{V}_{{\rm Res} , 2}$ describing 
the residual coupling to the MM which, as discussed in detail in  
\ref{stabl}, in this case can be effective to first-order in the boundary 
interaction strengths, different from what happens at the $A \otimes N$ fixed 
point. In particular, on applying the bosonization procedure to the 
operators derived in   \ref{stabl}, one obtains 

\begin{eqnarray}
 \tilde{V}_{{\rm Res} , 1} &\propto& e^{ i \sqrt{\frac{\pi}{2} }  [ \theta_\rho ( 0 ) + \theta_\sigma ( 0 ) ] } \nonumber \\
 \tilde{V}_{{\rm Res} , 2} &\propto& e^{ i\sqrt{\frac{\pi}{2} } [ \theta_\rho ( 0 ) - \theta_\sigma ( 0 ) ] }
 \:\:\:\: . 
 \label{addit.1}
\end{eqnarray}
\noindent
The corresponding scaling dimensions are readily derived to be equal to each other 
and given by $d_{{\rm Res} , 1 } = d_{{\rm Res} , 2} = \frac{K_\rho + K_\sigma}{4}$. 
Given the definition of $K_\rho$ and $K_\sigma$ in section \ref{pha_1}, we see that they are both relevant,
 as long as $K_\sigma < 1$.
Incidentally, we note that the other allowed boundary operators, corresponding to 
intra-channel boundary backscattering processes, are realized as 
$V_{{\rm Intra} , 1 (2)} \propto \cos [ 2 \sqrt{\pi} \theta_{1 (2)} ( 0 ) ] 
= \cos [ \sqrt{2 \pi} ( \theta_\rho ( 0 ) \pm \theta_\sigma ( 0 ) ) ]$. Accordingly, 
they have the same scaling dimension, $d_{{\rm Intra} , 1} = d_{{\rm Intra} , 2} = 
K_\rho + K_\sigma$ and, therefore, they are both irrelevant, for $1/2 < K$ and $U > 0$. 

The conclusion that, for $K_\sigma < 1$, there are two equivalent stable fixed points 
in the phase diagram of the $N=2$ junction (the $A \otimes N$ and the $N \otimes A$ fixed points 
discussed above) implies that there must be a 
phase transition between the two of them. In Ref. \cite{giuaf_1}, the 
phase transition has been identified at a FCFP in the phase diagram of the junction, which is 
attractive along the line in parameter space corresponding, in the symmetric case, to  
$t_1 = t_2$, and otherwise repulsive. To show this, an effective means is to 
resort to the perturbative RG approach within the $\epsilon$-expansion method. 
Basically, one assumes that the junction parameters are such that $d_b = 1 - \epsilon$, with 
$0 < \epsilon \ll 1$, and accordingly derives the RG equations to the first nonlinear
order in the boundary couplings, so as to recover nontrivial zeroes for the $\beta$-functions 
corresponding to the FCFP. For the details of the systematic derivation of 
the corresponding RG equations we refer to Ref. \cite{giuaf_1} in the specific 
case $N=2$, as well as to   \ref{renge} for the generalization of the procedure 
to a generic $N$, while here we just quote the final result. Specifically, as 
extensively discussed in   \ref{renge}, one introduces the dimensionless
running couplings $\bar{t}_a = t_a \tau_0^\epsilon$, with the cutoff $\tau_0 \propto D_0^{-1}$,
$D_0$ being a high-energy (band) cut-off for the system. Therefore, letting the
scale run from $D_0$ down to the scale parameter $D < D_0$, one obtains that the 
corresponding RG trajectories of the running couplings are determined by the 
differential equations

\begin{eqnarray}
\frac{ d \bar{t}_{1 } }{d l} &=& \epsilon \bar{t}_{1  } - {\cal F} \left( \frac{1}{2 K_\rho} - \frac{1}{2 K_\sigma}  \right) \bar{t}_{1  }   \bar{t}_{ 2  }^2 
\nonumber \\
\frac{ d \bar{t}_{2} }{d l} &=& \epsilon \bar{t}_{2  } - {\cal F} \left( \frac{1}{2 K_\rho} - \frac{1}{2 K_\sigma}  \right) \bar{t}_{2  }   \bar{t}_{ 1  }^2 
\:\:\:\: ,
\label{1w.262}
\end{eqnarray}
\noindent
with $l =  \ln ( D_0 / D )$, $D$ being the running energy scale, 
and the function ${\cal F} ( \nu)$ defined in Eq. (\ref{renge.8}). In general, for small initial values of the $\bar t_i$'s, Eq. (\ref{1w.262}) implies a 
growth of the $\bar{t}_a$ along the RG  trajectories. Along the symmetric line $t_1 = t_2$ in 
parameter space, this takes the 
system to the  FCFP  discussed in Ref. \cite{giuaf_1}, which corresponds to the nontrivial zeroes of the right-hand 
sides of Eqs. (\ref{1w.262}) at $\bar{t}_1 = \bar{t}_2 = t_* = \sqrt{\epsilon /  {\cal F} \left( \frac{1}{2 K_\rho} - \frac{1}{2 K_\sigma}  \right) }$. 
Alternatively, if the initial condition lies off the symmetric line, the RG trajectories flow towards either the 
$A \otimes N$, or the $N \otimes A$, fixed point, according to whether, at $D = D_0$, one 
has $t_1 > t_2 $, or $t_1 < t_2$. While an exact description of the FCFP is still missing, within the $\epsilon$-expansion method it is possible to estimate the 
corresponding value of the $g$ function to leading order in the $\epsilon$, obtaining \cite{giuaf_1}

\beq
g_{FCFP} = g_{N \otimes N} \: \left\{ 1 - \frac{2 \pi \epsilon^2}{ {\cal F} \left( \frac{1}{2 K_\rho} - \frac{1}{2 K_\sigma}  \right) } \right\}
\:\:\:\: , 
\label{next.1}
\eneq
\noindent
which implies $\frac{g_{FCFP}}{g_{N \otimes N} } < 1$, consistently with the RG flow from the disconnected fixed point to the 
FCFP, for $\epsilon > 0$.

\subsection{Phase diagram of the Y-junction of three spinless interacting normal wires}
\label{j3}

The Y3J has been introduced and extensively   discussed 
in  Ref. \cite{oca} in the fully ${\bf Z}_3$-symmetric case (bulk and boundary interaction).
Later on, in Ref. \cite{claudio_2}, the effects of relaxing the bulk ${\bf Z}_3$-symmetry have 
been considered. Here, we consider the most general situation in which 
the ${\bf Z}_3$ symmetry between the QW's can be broken by the boundary interaction, 
or by the bulk Hamiltonian \cite{claudio_2}, or both. Accordingly, we use as  bulk Hamiltonian of the (asymmetric) Y3J, $H_{\rm Bulk}$, 
given by 

\beq
H_{\rm Bulk} = \sum_{a = 1,2} \: \frac{u }{2} \: \int_0^\ell \: d x \: [ 
K ( \partial_x \phi_a ( x ))^2 + K^{-1} ( \partial_x \theta_a ( x ))^2 ] +  \frac{u }{2} \: \int_0^\ell \: d x \: [ 
K_3 ( \partial_x \phi_3 ( x ))^2 + K_3^{-1} ( \partial_x \theta_3 ( x ))^2 ]
\:\:\:\: ,
\label{3w.4}
\eneq
\noindent
with $ K, K_3 $ Luttinger parameters of the QW's and the velocity $u$ set equal in all three 
channels to avoid unnecessary formal complications. 

In the absence of a boundary interaction, the $g$-function for the Y3J  can be readily computed following the recipe 
of section \ref{pha_1} for $N=3$. In particular, since $N$ is odd, we add  an   auxiliary disconnected wire, with 
parameters $\bar{u}$ and $\bar{K}$ and  Klein factor $\bar{\Gamma}$, so to recover a total even number of KF's. 
Taking into account that, for a generic $K_3$,  the $g$-function  
at the disconnected fixed point, $g_{\rm Disc}  ( K,K_3 ) $,  is given by 
 
\beq
g_{\rm Disc}  ( K,K_3 ) = 2 [ K_3 K^2 \bar{K} ]^\frac{1}{4}  
\:\:\:\:,
\label{capuccy.5}
\eneq
\noindent
with the factor $2$ due to the four real fermions that determine a total degeneracy of 
4. Keeping all the  $\theta_a ( 0 )$ pinned and turning on a (non ${\bf Z}_3$-symmetric) 
boundary interaction, one may readily present the corresponding boundary Hamiltonian $H_b$ by 
implementing the transformation matrix ${\bf M}_N$ in Eq. (\ref{1w.8bis}) with $N=3$ to resort 
to the center of mass- and to the relative-field basis,   that is, by setting 

\beq
\left[ \begin{array}{c}
\Phi ( x ) \\ \varphi_1 ( x ) \\ \varphi_2 ( x )         
       \end{array} \right] = \left[ \begin{array}{ccc}
\frac{1}{\sqrt{3}} & \frac{1}{\sqrt{3}} & \frac{1}{\sqrt{3}} \\
\frac{1}{\sqrt{2}} & - \frac{1}{\sqrt{2}} & 0 \\
\frac{1}{\sqrt{6}} & \frac{1}{\sqrt{6}} & - \frac{2}{\sqrt{6}}
                                    \end{array} \right]
\: \left[ \begin{array}{c}
\phi_1 ( x ) \\ \phi_2 ( x ) \\ \phi_3 ( x )            
          \end{array} \right] \;\;\;  , \;\; 
 \left[ \begin{array}{c}
\Theta ( x ) \\ \vartheta_1 ( x ) \\ \vartheta_2 ( x )         
       \end{array} \right] = \left[ \begin{array}{ccc}
\frac{1}{\sqrt{3}} & \frac{1}{\sqrt{3}} & \frac{1}{\sqrt{3}} \\
\frac{1}{\sqrt{2}} & - \frac{1}{\sqrt{2}} & 0 \\
\frac{1}{\sqrt{6}} & \frac{1}{\sqrt{6}} & - \frac{2}{\sqrt{6}}
                                    \end{array} \right]
\: \left[ \begin{array}{c}
\theta_1 ( x ) \\ \theta_2 ( x ) \\ \theta_3 ( x )            
          \end{array} \right]         
\:\:\:\: . 
\label{next.1a}
\eneq
\noindent 
  As a result, one obtains 

\begin{eqnarray} 
H_b  &=&   
\{ t_{2,1} \Gamma_2 \Gamma_1 e^{ - i \left[ \sqrt{2 \pi} \varphi_1 ( 0 )   \right] }  
+ t_{3,2} \Gamma_3 \Gamma_2 e^{ - i  \left[ \sqrt{2 \pi} \left( - \frac{1}{2} \varphi_1 ( 0 ) + \frac{\sqrt{3}}{2} \varphi_2 ( 0 )  \right) 
\right] } \nonumber \\
&+& t_{1,3} \Gamma_1 \Gamma_3 e^{ - i \left[ \sqrt{2 \pi} \left( - \frac{1}{2} \varphi_1 ( 0 ) - \frac{\sqrt{3}}{2} \varphi_2 ( 0 )  \right) 
\right]  }
\} + {\rm h.c.} \equiv \sum_{a = 1}^3 t_{a + 1,a} V_{a , a+1} + {\rm h.c.}
\:\:\:\: , 
\label{3w.5bis}
\end{eqnarray}
\noindent
with $t_{a + 1,a}, a=1,2,3$ being the boundary interaction strengths (assuming the convention $a+1 \equiv 1$ for $a=3$) 
and the $\{ \Gamma_a \}$'s being the three KF's required to bosonize the fermionic fields
of the three wires, which shows that $H_b$ only depends on the relative fields $\varphi_1 ( x ) , \varphi_2 ( x )$.
Therefore, in constructing other boundary fixed points, we only act on the boundary conditions on 
$\varphi_1 ( x ) , \varphi_2 ( x )$, which are type $N$ at the disconnected fixed point but, in general, can 
change at other fixed points \cite{oca,claudio_2}. 
Incidentally, we note that the right-hand side of Eq. (\ref{3w.5bis}) is the leading  boundary operator 
allowed at the disconnected fixed point. It is a linear combination of the operators $V_{a , a+1}$ (plus their
Hermitean conjugates), with scaling dimensions respectively given by $d_{V_{1,2}} = \frac{1}{K}$, 
$d_{V_{2,3}}= d_{V_{3,1}} = \frac{1}{2K} + \frac{1}{2 K_3}$. So, a necessary condition for the 
disconnected fixed point not to be stable is that either one of the scaling dimensions above (or both) 
become $<1$. 

A first, alternative, fixed point is recovered by assuming type $A$ boundary conditions for both 
$\varphi_1 ( x )$ and $\varphi_2 ( x )$. This corresponds to the $D_P$ fixed point of  
Refs. \cite{oca,claudio_2}.  To  stabilize $D_P$, we 
introduce  a two-boundary  pairing potential $V_P$, given by 

\begin{eqnarray}
V_P &=& - \Delta \sum_{ a = 1}^3 \: \{ \psi_{R , a } ( 0 ) \psi_{L , a } ( 0 ) \psi_{L , a + 1}^\dagger ( 0 ) 
\psi_{R , a +1}^\dagger ( 0 ) \} -
 \Delta \sum_{ a = 1}^3 \: \{ \psi_{R , a } ( \ell ) \psi_{L , a } ( \ell ) \psi_{L , a + 1}^\dagger ( \ell ) 
\psi_{R , a +1}^\dagger ( \ell ) \} + {\rm h.c.} \nonumber \\
&=&  - \Delta \sum_{a = 1}^3  e^{ 2 i \sqrt{\pi} [ \phi_{a} ( 0 ) - \phi_{a+1} ( 0 ) ]  }  - 
\Delta  \sum_{a = 1}^3  e^{ 2 i \sqrt{\pi} [ \phi_{a } ( \ell ) - \phi_{ a+1} ( \ell ) ] }   + {\rm h.c.}
\:\:\:\: , 
\label{vp.1}
\end{eqnarray}
\noindent
and eventually send  $\Delta \to \infty$.   Sending 
$\Delta \to \infty$,  one pins $\varphi_1 ( x )$ and $\varphi_2 ( x )$ at both boundaries.
Taking this into account, one determines the corresponding spectrum of the zero-mode operators and, 
repeating the calculation of the $g$-function 
at the $D_P$ fixed point, one obtains 

\beq
g_{D_P} ( K,K_3 ) = \frac{ 2 \sqrt{2 K + K_3} \bar{K}^\frac{1}{4}}{( K_3 K^2)^\frac{1}{4}}
\:\:\:\: , 
\label{capuccy.11}
\eneq
\noindent
which yields the  ratio  

\beq
 \rho_{D_P} (  K,K_3 )  =  \frac{g_{D_P} ( K,K_3) }{g_{\rm Disc} ( K,K_3 ) } = 
  \frac{1}{K} \sqrt{\frac{2K+ K_3}{K_3}}  
  \:\:\:\: . 
  \label{capuccy.12}
\eneq
\noindent 
An important comment about Eq. (\ref{capuccy.12}) is that, despite the fact that, for $K=K_3$,  neither 
$g_{\rm Disc} ( K,K_3 )$, nor $g_{D_P} (K,K_3 )$, are equal to the values derived in 
Ref. \cite{oca}, the ratio between the two of them is the same as 
one would get by using the results obtained there. This is due to the fact that, 
in our derivation, we count the degrees of freedom associated with the real KF's, including 
the auxiliary one and, in addition, do not restrict our derivation to the sector involving 
the relative fields only. Nevertheless, the ratio between the two of them is consistent 
with Ref. \cite{oca}. Clearly, this further enforces out intuition that, despite 
the arbitrary aspects of our procedure, the ratios between values of the $g$-function at 
different fixed points do always give back the right, physical result. 

The leading dimension boundary  interaction at the disconnected fixed point is $H_b$ in Eq. (\ref{3w.5bis}).
It is a linear combination of operators with scaling dimensions $d_{V_{1,2}} = \frac{1}{K}$,
$d_{V_{2,3}}=d_{V_{3,1}} = \frac{1}{2K} \left( \frac{K_3+K}{K_3} \right)$. At variance, at the $D_P$ fixed point, the
leading dimension boundary operators
are given by \cite{oca,claudio_2}

\begin{eqnarray}
 T_{1,2} &=& \Gamma_1 \Gamma_2 \: e^{ -i \sqrt{ \pi} [ \theta_1 ( 0 ) + \theta_2 ( 0 ) ] } \nonumber \\
 T_{2,3} &=& \Gamma_2 \Gamma_3 \: e^{  -i \sqrt{ \pi} [ \theta_2 ( 0 ) + \theta_3 ( 0 ) ]  } \nonumber \\
  T_{3,1} &=& \Gamma_3 \Gamma_1 \: e^{   -i \sqrt{ \pi} [ \theta_3 ( 0 ) + \theta_1 ( 0 ) ] } 
 \;\;\;\; , 
 \label{dp3.5}
\end{eqnarray}
\noindent
together with their Hermitean conjugates.    To derive the corresponding scaling dimensions, one has to resort to 
the center of mass- and relative field basis by using Eqs.(\ref{next.1a}) and to take into account that $\Theta ( 0 )$ is 
always pinned, as  $\Phi ( 0 )$ never appears in the boundary interaction \cite{oca}. As a result, one obtains  \cite{claudio_2}
$d_{T_{1,2}} = \frac{K K_3 }{2 K + K_3}$ and $d_{T_{2,3}} = d_{T_{3,1}} = \frac{K ( K_3 + K)}{2 ( K_3 + 2 K)}$.  
From the scaling dimensions of the  boundary operators we see that
there is a $K_3/K$-dependent  window of values of $K$ in which neither the disconnected, nor 
the $D_P$ fixed point, is stable. Specifically, this happens for $1 < K < 3$ for $K=K_3$ \cite{oca}
and, more generically, for $K > K_{\rm min} = {\rm max} \left\{ 1 , \frac{2 K_3}{K + K_3} \right\}$ 
and $K< K_{\rm max} = \left( \frac{2 K +K_3}{K_3} \right)$ for a generic $K_3$ \cite{claudio_2}. The absence of time-reversal breaking in 
$H_b$ in Eq. (\ref{3w.5bis}) rules out the possibility of stabilizing the ``chiral'' $\chi_{\pm}$ fixed 
points: thus, one concludes that, for $K_{\rm min} < K < K_{\rm max}$, the stable phase of the 
system either corresponds to one of the asymmetric $A_a$-points emerging in  the Y3J when 
the ${\bf Z}_3$-symmetry between the channels is broken, or to a generically asymmetric version 
of the $M$-FCFP found in the ${\bf Z}_3$-symmetric Y3J in the time-reversal symmetric case 
\cite{oca}. 

The symmetries of the bulk Hamiltonian in Eq. (\ref{3w.4}) naturally lead to two different types of
asymmetric fixed points: the $A_3$ fixed point corresponds to QW-3 
disconnected from the junction,  while the two-wire junction between QW's -1 and -2 is ``healed'' (which is 
a natural consequence of having $K>1$, once QW-3 is disconnected from the junction \cite{kafi}),
and the (equivalent, up to swapping QW-1 and QW-2 with each other)
$A_1$ and $A_2$ fixed points, in which respectively QW-1 and QW-2 are disconnected from 
the junction.  Mathematically, disconnecting QW-$a$  corresponds to imposing type $N$ (type $A$) boundary conditions on 
$\phi_a ( x )$ ($\theta_a ( x )$), as well as type $A$ (type $N$) boundary conditions on 
$\phi_{a+1} ( x ) - \phi_{a+2} ( x )$ ($\theta_{a+1} ( x ) - \theta_{a+2} ( x ) $). Accordingly, the calculation of 
the corresponding value of the $g$-function  can be readily carried out, 
providing the result

\begin{eqnarray}
g_{ A_3}  ( K,K_3 ) &=& 2 [ K_3 \bar{K} ]^\frac{1}{4}  \nonumber \\
g_{ A_1} ( K,K_3)  &=& g_{A_2}  ( K,K_3 ) =  2 [ K_3  \bar{K} ]^\frac{1}{4} \: \left[ \frac{1}{2  } + \frac{K}{2 K_3 } \right]^\frac{1}{2}
\:\:\:\: . 
\label{capuccy.7}
\end{eqnarray}
\noindent
yielding the ratios 

\begin{eqnarray}
 \rho_{A_3} ( K,K_3 ) &=& \frac{g_{A_3} ( K,K_3) }{g_{\rm Disc} (K,K_3) } = K^{ - \frac{1}{2} } \nonumber \\
 \rho_{A_1} (K,K_3 ) &=&  \rho_{A_2} ( K,K_3 ) = \frac{g_{A_1} ( K,K_3) }{g_{D_P} (K,K_3) } = K^{ - \frac{1}{2} } \: \sqrt{\frac{K + K_3}{2 K_3}}
 \:\:\:\: . 
 \label{capuccy.8}
\end{eqnarray}
\noindent
It is useful to also compute the ratios with $g_{D_P} ( K,K_3 )$. The result is

\begin{eqnarray}
\tilde{\rho}_{A_3} ( K,K_3 ) &=& \frac{g_{A_3} ( K,K_3 ) }{g_{D_P} (K,K_3) } = K^{  \frac{1}{2} } \: \sqrt{\frac{K_3}{2 K + K_3}} \nonumber \\
\tilde{\rho}_{ A_1} ( K,K_3 )  &=& \tilde{\rho}_{A_2}  ( K,K_3 )  = \frac{g_{A_1} (K,K_3 ) }{g_{D_P} (K,K_3) } =  K^{  \frac{1}{2} } \:
\sqrt{\frac{2 K  + K_3}{ 2(K+K_3) }}
\:\:\:\: . 
\label{capuccy.7bis}
\end{eqnarray}
\noindent
An effective mean to infer the  stability of the $A_a$ fixed points against the disconnected and the $D_P$ fixed point 
consists in using Eqs. (\ref{capuccy.8},\ref{capuccy.7bis}) in combination of the DEBC analysis of the corresponding allowed boundary 
operators. To construct the leading boundary perturbation at the $A_a$ fixed point, one considers the operators 
$T_{a,a+1} , \tilde{T}_{a , a+1}$, respectively given by \cite{claudio_2}

\begin{eqnarray}
 T_{a,a+1} &=& \Gamma_a \Gamma_{a+1} \: e^{ - i \sqrt{ \pi}  [ \phi_a ( 0 ) - \phi_{a+1} ( 0 ) ] - i \sqrt{\pi} [ \theta_a ( 0 ) + \theta_{a+1} ( 0 ) ] }
 \nonumber \\
 \tilde{T}_{a,a+1} &=& \Gamma_a \Gamma_{a+1} \: e^{ - i \sqrt{ \pi}  [ \phi_a ( 0 ) - \phi_{a+1} ( 0 ) ] + i \sqrt{\pi} [ \theta_a ( 0 ) + \theta_{a+1} ( 0 ) ] }
  \:\:\:\: , 
  \label{capuccy.x1}
\end{eqnarray}
\noindent
together with their Hermitean conjugates. $ T_{a,a+1}$ and $ \tilde{T}_{a,a+1}$ 
respectively correspond to the boundary operators bilinear in the $\{ \psi_{R , a} , \psi_{L , a} \}$'s given by 
$\psi_{R , a }^\dagger ( 0 ) \psi_{ L , a+1} ( 0 )$ and $\psi_{L , a}^\dagger ( 0 ) \psi_{R , a + 1} ( 0 )$. 
Once the appropriate CIBC's are implemented, they only depend on the linear combinations of the $\phi_a (0)$'s and 
of the $\theta_a ( 0 )$'s that are not pinned at the corresponding fixed point.   In particular, the CIBC's corresponding to the $A_a$ fixed point are
recovered by pinning the arguments of both $T_{a+1 , a+2}$ and $\tilde{T}_{a+1,a+2}$. The $T_{a,a+1} , \tilde{T}_{a , a+1}$-operators 
are the  only   operators that may become relevant at the $A_a$ fixed point, 
with scaling dimension  $d_{A_3} = \frac{2 K + K_3  + K_3 K^2}{4 K_3 K}$, and $d_{A_1} = d_{A_2} =  
\frac{2 K + K_3 + K_3 K^2}{2 ( K + K_3 ) K}$ \cite{claudio_2}.  As a result, from Eqs. (\ref{capuccy.8},\ref{capuccy.7bis}) one 
finds that, in order for the $A_3$ fixed point to be stable with respect to both the 
disconnected and the $D_P$ fixed point, the condition $1 < K < 1 + \frac{2 K}{K_3}$ has to be 
satisfied. In addition, there must be no relevant boundary operators allowed at  
$A_3$ in order for it to correspond to the actual stable fixed point of the Y3J.
This leads to the additional condition $d_{A_3} > 1$, that is, $K_3 < 2 K /(4 K - 1 - K^2)$. 
In particular, for $1< K < 3$, the last condition implies $K_3<K$, which yields $ g_{A_3} ( K,K_3 ) / g_{A_1} ( K,K_3) = \rho_{A_3} ( K,K_3 ) / \rho_{A_1} ( K,K_3 ) = 
\sqrt{\frac{2K_3}{K + K_3  }} < 1$, thus showing that $A_3$ is stable against both $A_1$ and $A_2$, as well. 
This is ultimately consistent with the 
results plotted in Fig.5 of Ref. \cite{claudio_2}, as well as with the observation that, 
a small enough $K_3/K$ eventually makes the interaction in wire-3 to be effectively repulsive, thus 
triggering the disconnection of this wire from the junction, in agreement with the known 
results about junctions of Luttinger liquids \cite{kafi_0,kafi,oca}. Conversely, in 
order for either $A_1$, or $A_2$, to be stable against the disconnected, as well as the 
$D_P$ fixed point, the condition $\frac{K + K_3}{2 K_3} < K < \frac{2 (K+K_3)}{2 K + K_3 }$ has to be met. 
In addition, the condition $d_{A_1} > 1$ implies $ K_3 > \frac{2 K}{K-1}$. In particular, 
the above conditions yield $ g_{A_3} ( K,K_3 ) / g_{A_1} ( K,K_3 ) = \sqrt{\frac{2K_3}{K+K_3}}$, 
which shows that having  $K_3<K$ ($K_3>K$)  is a necessary  condition to make $A_3$ ($A_1$) stable against $A_1$ ($A_3$).
 
Finally, we note that there are regions in parameter space in which, though 
one has that  one of the conditions $\rho_{A_3} ( K,K_3 ) < 1$, or $\rho_{A_1} ( K,K_3 ) < 1$ is met, 
none of the above fixed points  is stable. This happens for $\frac{2K}{4K-1-K^2} < K_3 < K$, as well as 
for $K < K_3 < \frac{2 K}{K-1}$. In this case, based on the well-grounded results of Ref. \cite{oca} about the ${\bf Z}_3$-symmetric 
Y3J, we expect that the stable phase of the system corresponds to a 
(possibly non-${\bf Z}_3$-symmetric) FCFP, which generalize the  $M$-FCFP of  Ref. \cite{oca}.
In the ${\bf Z}_3$-symmetric case $K = K_3$, the emergence  of the $M$-FCFP can be 
inferred from the perturbative RG equations in Eqs. (\ref{puccy.x1}) of  \ref{failure},
given by

\begin{eqnarray}
 \frac{d \bar{t}_{2,1} }{d l  } &=& \epsilon \{ \bar{t}_{2,1} - \bar{t}_{2,1} [ b ( \bar{t}_{2,1} )^2 + c ( ( \bar{t}_{3,2} )^2 + (  \bar{t}_{1,3} )^2 ) ] \} \equiv 
 \beta_1 [ \bar{t}_{2,1} , \bar{t}_{3,2} , \bar{t}_{1,3}]  \nonumber \\
  \frac{d \bar{t}_{3,2} }{d l  } &=& \epsilon \{ \bar{t}_{3,2} - \bar{t}_{3,2} [ b ( \bar{t}_{3,2} )^2 + c ( ( \bar{t}_{1,3} )^2 + ( \bar{t}_{2,1} )^2 ) ] \} \equiv 
  \beta_2   [ \bar{t}_{2,1} , \bar{t}_{3,2} , \bar{t}_{1,3}]  \nonumber \\
  \frac{d \bar{t}_{1,3} }{d l } &=& \epsilon \{ \bar{t}_{1,3} - \bar{t}_{1,3} [ b ( \bar{t}_{1,3} )^2 + c ( ( \bar{t}_{2,1} )^2 + ( \bar{t}_{3,2} )^2 ) ] \} \equiv 
  \beta_3  [ \bar{t}_{2,1} , \bar{t}_{3,2} , \bar{t}_{1,3}]  \
 \:\:\:\: , 
 \label{puccy.x1bis}
\end{eqnarray}
\noindent
with $0 < \epsilon (=1-K^{-1}) \ll 1$, and the parameters $b$ and $c$  estimated in   \ref{failure} to be 
$b \approx 26.32, c \approx 16.45$. An important point about the $\beta$-functions in Eqs. (\ref{puccy.x1bis}) is 
that they are over-all   $\propto \epsilon$. As a result, the $M$-FCFP is found to 
lie at   $\bar{t}_{2,1} = \bar{t}_{3,2} = \bar{t}_{1,3} = t_* = 1 /\sqrt{  b + 2 c }$, independent of $\epsilon$. 
On one hand, this result points in the right direction. Indeed,  analytical \cite{oca}, as well as numerical \cite{rahmani}, 
results for the conductance tensor at  the $M$-FCFP ultimately show that it has  to be finite, as $\epsilon \to 0$. 
Had we found an $M$-FCFP at $t_*$ going to zero as  $\epsilon \to 0$, we
would unavoidably get  a conductance tensor going to zero as $\epsilon \to 0 $, as well,  which would be  incorrect
\cite{oca,rahmani}. On the other hand, since there is no ``small parameter'', such as $\epsilon$, that 
can be used to control the coupling strengths at the FCFP's, one cannot really expect Eqs. (\ref{puccy.x1bis}) to be 
reliable to make quantitative predictions on e.g. the conductance tensor at the FCFP, or on the $g$-function (at variance 
with what happens for the junction between $N$ QW's and a TS). Yet, besides the emergence of the $M$-FCFP itself, other 
remarkable conclusions can be derived from Eqs. (\ref{puccy.x1bis}), such as that 
the RG-trajectories  always point towards the  ${\bf Z}_3$-symmetric $M$-FCFP, that is,   
any asymmetry in the boundary couplings is an irrelevant perturbation of the RG 
flow trajectories. In fact, this is a remarkable feature that, in the ${\bf Z}_3$-symmetric case, the Y3J  shares with 
the topological Kondo effect, in which the magnetic impurity is realized in terms of localized
MM's \cite{beritopo}. In the general case $K_3 \neq K$, we rather refer to the corresponding generalization of 
Eqs. (\ref{puccy.x1bis}) that we provide in Eqs. (\ref{puccy.y1}) of   \ref{failure}. In particular, 
looking for nontrivial zeroes of the $\hat{\beta}_a$-functions  of Eqs. (\ref{puccy.y1}), we see 
that the predicted values of the running couplings corresponding to the $M$-FCFP are either characterized by an ``easy 
plane'' asymmetry for $K_3>K$ (which implies $\bar{t}_{2,1 , * } < \bar{t}_{3,2 , * } = \bar{t}_{1,3 , *}$), or by an 
``easy axis'' asymmetry in the complementary case, $K_3<K$ (which implies $\bar{t}_{2,1 , * } > \bar{t}_{3,2 , * } = \bar{t}_{1,3 , *}$).
In both cases, the flow towards the $M$-FCFP always requires the relevance of all the $V_{a , a+1}$-operators
entering $H_b$, as we discuss in detail in section \ref{pha_2}, when spelling out  the correspondence between 
the $N=2$ junction and the Y3J.

\section{Correspondence between an $N=2$ junction with a  topological superconductor and  
a Y-junction of three spinless quantum wires}
\label{pha_2}

In this section we discuss in detail the various aspects of the correspondence between 
a junction with 2 QW's and a topological superconductor and the (generically asymmetric) Y3J.  
For the purpose of this work, the correspondence is of the utmost importance 
for several reasons. First of all, it works as a sort of ``model duality'', allowing for 
recovering results about the phase diagram of one of the two systems from the known (and controlled)
features of the phase diagram of the other, in the various regions of 
the system parameters. Moreover, the correspondence is useful in computing the $g$-function of 
one model from known results on the other one. About this point, it is worth stressing that, as 
in our work we attribute physical meaning only to the ratio between the $g$-function at different 
fixed points of the phase diagram, contributions from modes not entering  the correspondence 
factorize and cancel, when computing the ratios, which enforces the reliability of the correspondence 
to computing the IE. Finally, as the correspondence requires defining MM's in the $N=2$ junction in 
terms of KF's in the Y3J, and vice versa, it provides also  strong evidence for the fact that 
both real fermionic modes have to be taken into account, and considered on the same footing, 
when computing the $g$-function, which is one of the main points we make here. 

 For  clarity, in the following we split the presentation of the correspondence in two 
sub-sections. In sub-section \ref{lul_0}, we explicitly construct the mapping from the 
 $N=2$ junction to the Y3J. This allows us to  use  known results about the phase diagram
of the $N=2$ junction \cite{giuaf_1} to unveil specific features in the phase diagram of the 
Y3J, such as   emergence of ``planar'' FCFP's (that is, with one of the 
boundary coupling strengths set to 0). In sub-section \ref{lul}, we derive the mapping from 
the Y3J to the $N=2$ junction. Reversing the direction of the correspondence allows us to employ 
  the known results about the phase diagram of 
the non-${\bf Z}_3$-symmetric Y3J to derive the phase diagram of the $N=2$ junction for 
$K_\sigma > 1$, a range of values of the system's parameters which was not discussed in 
Ref. \cite{giuaf_1}.

\subsection{From the $N=2$ junction with a topological superconductor to the Y-junction} 
\label{lul_0}

We now   consider the asymmetric Y3J with $K$ and $K_3$ set so
that  $V_{1,2}$, defined in Eq. (\ref{3w.5bis}), becomes  irrelevant, while $V_{2,3} , V_{3,1}$ both stay  
relevant, that is, so that $d_{V_{1,2}} = \frac{1}{K} > 1$, while $d_{V_{2,3}} = d_{V_{3,1}} = \frac{1}{2K} + 
\frac{1 }{2 K_3} < 1$ (which amounts to choosing $r$ so that  $r^{-1} < 2K-1$). For the asymmetric 
Y3J, the boundary coupling flow is determined by  the perturbative RG  Eqs. (\ref{puccy.y1}). Since $V_{1,2}$
is irrelevant, 
$\bar{t}_{2,1}$ expected to renormalize to 0 for $D_0 / D \to \infty$ and, as a result, one may recover the phase diagram 
of the Y3J in this regime by restricting the analysis to the plane $\bar{t}_{2,1} = 0$ in parameter space. Setting 
$\bar{t}_{2,1} = 0$ in  the  second- and in the third ones  of Eqs. (\ref{puccy.y1}),  we obtain the system 
of two coupled RG equations given by 

\begin{eqnarray}
    \frac{d \bar{t}_{3,2}}{d \ln ( D_0 / D ) } &=& \left( 1 - \frac{1}{2K} - \frac{1}{2 r K} \right) \bar{t}_{3,2} -{\cal B} \left[ \frac{1}{2K} + \frac{1}{2 r K} \right] (\bar{t}_{3,2})^3 
    -  {\cal C} \left[ \frac{1}{K} , \frac{1}{r K} \right]   \bar{t}_{3,2}( \bar{t}_{1,3})^2    \nonumber \\
    \frac{d \bar{t}_{1,3}}{d \ln ( D_0 / D ) } &=& \left( 1 - \frac{1}{2K} - \frac{1}{2 r K} \right) \bar{t}_{1,3} -{\cal B} \left[ \frac{1}{2K} + \frac{1}{2 r K } \right] (\bar{t}_{1,3})^{3} 
    -  {\cal C} \left[ \frac{1}{K} , \frac{1}{r K} \right]   \bar{t}_{1,3} (\bar{t}_{3,2})^{2}   
    \: \:\:\: .
    \label{puccy.y1bis}
\end{eqnarray}
\noindent
Remarkably, Eqs. (\ref{puccy.y1bis}) can now be consistently dealt with within the $\epsilon$-expansion method, 
by setting $\frac{1}{2K} + \frac{1}{2 K r} = 1 - \epsilon$, with $0 < \epsilon \ll 1$. Expanding to linear order in $\epsilon$  and neglecting subleading contributions (in $\epsilon$) to nonlinear terms in the 
$\hat{\beta}$-functions, according to  Eqs. (\ref{puccy.y1bis})  and to the definition of the function ${\cal B}$ in Eq. (\ref{pp.4quater}),  which 
implies that terms $\propto {\cal B}$ at the right-hand side of Eqs. (\ref{puccy.y1bis})  are all  $\propto \epsilon$, 
we trade Eqs. (\ref{puccy.y1bis}) for the system 

\begin{eqnarray}
\frac{d \bar{t}_{3,2}}{d \ln ( D_0 / D ) } &=&  \epsilon \bar{t}_{3,2} - {\cal F} \left( 2 - \frac{1}{K} \right) \bar{t}_{3,2} ( \bar{t}_{1,3} )^2  \nonumber \\
\frac{d \bar{t}_{1,3}}{d \ln (  D_0 / D) } &=&  \epsilon \bar{t}_{1,3} - {\cal F} \left( 2 - \frac{1}{K} \right) \bar{t}_{1,3} ( \bar{t}_{3,2} )^2 
\:\:\:\: 
\label{puccy.y2bis}
\end{eqnarray}
where ${\cal F}$ is defined in Eq. (\ref{renge.8}). 
\noindent
Apparently, Eqs. (\ref{puccy.y2bis}) correspond to the perturbative RG equations of an $N=2$ junction with $\frac{1}{4 K_\rho} + \frac{1}{4 K_\sigma} = 1 - \epsilon$ and 
$\frac{1}{2 K_\rho} - \frac{1}{2 K_\sigma} = 2 - \frac{1}{K}$. The correspondence is clearly not accidental. Indeed, on 
performing the canonical transformations 

\begin{eqnarray}
 \bar{\phi}_{1,2}  ( x ) &=& \sqrt{K} \phi_{1,2} ( x ) \;\; , \; \bar{\theta}_{1,2} ( x ) = \theta_{1,2} ( x ) / \sqrt{K} \nonumber \\
  \bar{\phi}_{3}  ( x ) &=& \sqrt{K_3} \phi_{3} ( x ) \;\; , \; \bar{\theta}_{3} ( x ) = \theta_{3} ( x ) / \sqrt{K_3}
  \:\:\:\: , 
  \label{ppy.y3}
\end{eqnarray}
\noindent
followed by the rotation 

\beq
\left[ \begin{array}{c}
\hat{\phi}_\sigma ( x) \\ \hat{\phi}_\rho  ( x ) \\ \hat{\phi}_\chi ( x )         
       \end{array} \right] = 
\left[ \begin{array}{ccc}
- \frac{1}{\sqrt{2}} & \frac{1}{\sqrt{2}} & 0 \\
\frac{1 / \sqrt{K}}{\sqrt{2/K + 4 /K_3}} & \frac{1 / \sqrt{K}}{\sqrt{2/K + 4 /K_3}} & \frac{-2 / \sqrt{K_3}}{\sqrt{2/K + 4 /K_3}} \\
\frac{1 / \sqrt{K_3}}{\sqrt{1/K + 2 /K_3}} & \frac{1 / \sqrt{K_3}}{\sqrt{1/K + 2 /K_3}} & \frac{1 / \sqrt{K}}{\sqrt{1/K + 2 /K_3}} 
       \end{array} \right] 
\: \left[ \begin{array}{c}
\bar{\phi}_1 ( x ) \\ \bar{\phi}_2  ( x ) \\ \bar{\phi}_3 ( x )            
          \end{array} \right] 
\:\:\:\: , 
\label{ppy.y4}
\eneq
\noindent
and analogous rotation from $\{ \bar{\theta}_1 ( x ) , \bar{\theta}_2 ( x ) , \bar{\theta}_3 ( x )\}$ to 
$\{ \hat{\theta}_\sigma ( x ) , \hat{\theta}_\rho ( x ) , \hat{\theta}_\chi ( x ) \}$, one obtains, for the bulk Hamiltonian 

\beq
H_{\rm Bulk} = \frac{u}{2} \: \sum_{a = r , c , \chi} \: \int_0^\ell \: d x \: \{ ( \partial_x \hat{\phi}_a ( x ))^2 +
( \partial_x \hat{\theta}_a ( x ))^2 \} 
\:\:\:\: . 
\label{ppy.y5}
\eneq
\noindent
Once $\bar{t}_{2,1}$ is set to 0  in the boundary Hamiltonian (which corresponds to dropping the term 
$\propto V_{1,2}$ in Eq.(\ref{3w.5bis})), $H_b$  becomes

\begin{eqnarray}
H_b &=& t_{3,2} \Gamma_2 \Gamma_3 \: \exp \left\{ - i \sqrt{\frac{\pi}{2}} \left[ \frac{1}{\sqrt{K}} \hat{\phi}_\sigma ( 0 ) + \sqrt{\frac{1}{K} + \frac{2}{K_3}} \hat{\phi}_\rho ( 0 ) \right] \right\} 
\nonumber \\
&+&t_{1,3} \Gamma_3 \Gamma_1 \: \exp \left\{ - i  \sqrt{\frac{\pi}{2}}  \left[ \frac{1}{\sqrt{K}} \hat{\phi}_\sigma ( 0 ) - \sqrt{\frac{1}{K} + \frac{2}{K_3}} \hat{\phi}_\rho ( 0 ) \right] \right\} 
+ {\rm h.c.} 
\:\:\:\: . 
\label{ppy.y6}
\end{eqnarray}
\noindent
Apparently, $\hat{\phi}_\chi ( x ) , \hat{\theta}_\chi ( x )$ fully decouple from $H_b$ in Eq. (\ref{ppy.y6}). Moreover, 
shifting $\hat{\phi}_\rho ( x )$ by a constant, so that 
$ \sqrt{\frac{\pi}{2}} \sqrt{\frac{1}{K} + \frac{2}{K_3}} \hat{\phi}_\rho ( 0 ) \to 
 \sqrt{\frac{\pi}{2}} \sqrt{\frac{1}{K} + \frac{2}{K_3}} \hat{\phi}_\rho ( 0 ) + \frac{\pi}{2}$, one obtains

\begin{eqnarray}
H_b &\to& - 2 i t_{3,2} \Gamma_2 \Gamma_3 \: \cos \left\{   \sqrt{\frac{\pi}{2}}  \left[ \frac{1}{\sqrt{K}} \hat{\phi}_\sigma ( 0 ) + \sqrt{\frac{1}{K} + \frac{2}{K_3}} \hat{\phi}_\rho ( 0 ) \right] \right\} 
\nonumber \\
&+&2 i t_{1,3} \Gamma_3 \Gamma_1 \: \cos \left\{   \sqrt{\frac{\pi}{2}}  \left[ \frac{1}{\sqrt{K}} \hat{\phi}_\sigma ( 0 ) - \sqrt{\frac{1}{K} + \frac{2}{K_3}} \hat{\phi}_\rho ( 0 ) \right] \right\} 
\:\:\:\: . 
\label{ppy.y7}
\end{eqnarray}
\noindent
Finally, performing the reverse canonical rescaling given by 

\begin{eqnarray} 
 \phi_\rho( x ) &=&  \sqrt{\frac{1}{K} + \frac{2}{K_3}} \hat{\phi}_\rho  ( x ) \;\; , \; \phi_\sigma ( x ) =  \frac{1}{\sqrt{K}} \hat{\phi}_\sigma ( x ) \nonumber \\
  \theta_\rho ( x ) &=& \hat{\theta}_\rho ( x ) / \left[   \sqrt{\frac{1}{K} + \frac{2}{K_3}} \right]  \;\; , \; \theta_\sigma  ( x ) =    \sqrt{K}  \hat{\theta}_\sigma ( x )
  \:\:\:\: , 
  \label{ppy.y8}
\end{eqnarray}
\noindent
and setting $\Gamma_3 \to \gamma_L$, 
we recover the Hamiltonian for the $N=2$ junction with a topological superconductor, with 
  Luttinger parameters  given by 

\begin{eqnarray}
K_\rho &=& \frac{ K K_3}{   ( K_3 + 2 K) } \nonumber \\
K_\sigma &=& K (< 1)
\:\:\:\: . 
\label{ppy.y9}
\end{eqnarray}
\noindent
Besides the mapping procedure involving $H_b$, to further ground the correspondence 
we now extend it to all the allowed boundary operators in the Y3J and in the $N=2$ junction, 
at each fixed point in the boundary phase diagram of the two systems that we  
discuss in section \ref{phadia_1}. 

Starting with the disconnected fixed point, due to the condition $K<1$, the only 
relevant allowed boundary operators are $V_{2,3}$ and  $V_{3,1}$ entering  $H_b$ in Eq.(\ref{ppy.y6}). 
Consistently with their scaling dimensions, these 
are identified with the operators at the second and third line of the table 
in appendix A.a of Ref. \cite{claudio_2}.  Another operator, 
which is irrelevant due to our choice of the system parameter, is the boundary operator 
$V_{1,2}$ of the asymmetric Y3J, with scaling dimension $1/K$. Referring to the    table 
in appendix A.a of Ref. \cite{claudio_2}, it apparently corresponds to any of the operators 
listed at the first line, taken at the disconnected fixed point.
According to the analysis of  section \ref{phadiaN2},  its counterpart in the $N=2$ junction 
is the normal boundary backscattering operator $V_{{\rm Normal},(1,2)}$, 
with scaling dimension $1 / K_\sigma = 1/K$. Additional boundary operators can potentially appear in the Y3J,
which are   quartic in the fermionic fields of the Y3J such as, for instance (in the notation 
of Ref. \cite{claudio_2}) $T = T_{31}^{LR} T_{32}^{RL}$. Quartic operators do not appear 
in the   table  in appendix A.a of Ref. \cite{claudio_2}, which only contains
quadratic operators: to make them relevant a strong, bulk inter-channel attractive interaction 
is required, which we exclude here, as we only focus on repulsive, inter-channel bulk interactions. 
Yet, to complete the correspondence with the $N=2$ junction, we see that $T$ is the second boundary 
operator that must be identified with  the boundary pairing operator of the $N=2$ junction,  $V_{{\rm Pair}, (1,2)}$,
with scaling dimension $1 / K_\rho$.  

Moving to the $A \otimes N$ and to the $N \otimes A$ fixed points of the $N=2$ junction 
\cite{giuaf_1}, based on the analysis of section \ref{phadia_1}, we naturally identify them with 
respectively the $A_1$- and the $A_2$-asymmetric fixed point of the Y3J. To further 
corroborate our identification, we now show that it is realized as a one-to-one correspondence 
between boundary operators in the two systems. Here, we only discuss the 
correspondence between  the $A \otimes N$ and the $A_1$ fixed point. The complementary 
one can be readily recoverd by symmetry. At the $A \otimes N$ fixed point of the 
$N=2$ junction, the first pair of allowed boundary operators corresponds to boundary 
inter-channel backscattering/pairing between channels 1 and 2.
The corresponding operators are realized as a linear combination of  $V_{a , (1,2)} , V_{b , (1,2)}$, respectively 
given by   

\begin{eqnarray}
V_{a , (1,2)}&= &   \Gamma_1 \Gamma_2 e^{ i \sqrt{\pi} [ \phi_2 ( 0 ) - \theta_1 ( 0 ) ] }   = 
  \Gamma_1 \Gamma_2 e^{ i \sqrt{\frac{\pi}{2} } [ \phi_\rho ( 0 ) - \phi_\sigma ( 0 ) - \theta_\rho ( 0 ) - \theta_\sigma ( 0 ) ] }  \nonumber \\
  V_{b , (1,2)} &=&   \Gamma_1 \Gamma_2 e^{ i \sqrt{\pi} [ \phi_2 ( 0 ) +\theta_1 ( 0 ) ] }    = 
  \Gamma_1 \Gamma_2 e^{ i \sqrt{\frac{\pi}{2} } [ \phi_\rho ( 0 ) - \phi_\sigma ( 0 ) + \theta_\rho ( 0 ) + \theta_\sigma ( 0 ) ] }  
 \:\:\:\: , 
 \label{ppy.x1}
\end{eqnarray}
\noindent
plus their Hermitean conjugates.  (Note that, due to the boundary conditions at the $A \otimes N$ fixed point,
$V_{a , (1,2)}$ and $V_{b , (1,2)}$ do no more correspond respectively to normal boundary scattering and to boundary 
pairing, as they instead do at the disconnected fixed point -- see the discussion after Eq.(\ref{1w.202}) in sub-section
\ref{phadiaN2}. Instead, as we state above, normal boundary scattering and   boundary 
pairing operators are realized as linear combinations of the two of them.)  
Inverting the transformations above to get back to the fields of the Y3J, it is not difficult to check 
that the operators in Eqs. (\ref{ppy.x1}) respectively correspond to the $T_{21}^{RL}$- and to the $T_{21}^{RR}$-operators at the 
$A_1$ fixed point of the asymmetric Y3J, plus their Hermitean conjugates
(see appendix A.d of Ref. \cite{claudio_2}). This is further confirmed by  the observation that, using 
the results of section \ref{phadia_1} for the $N=2$ junction, the scaling dimension of the operators in
Eqs. (\ref{ppy.x1}) are the same and are given by  $d_{a, (1,2)} = d_{b, (1,2)} = \frac{1+K_\sigma K_\rho}{K_\sigma + K_\rho}= 
\frac{2 K+ K_3+ K_3 K^2}{2 (K + K_3 ) K}$, which is the correct result for the    $T_{21}^{RL}$- and for  the $T_{21}^{RR}$-operators at the 
$A_1$ fixed point of the Y3J \cite{claudio_2}.  

A second class of boundary operators at the $A \otimes N$ fixed point corresponds to  intra-channel 1 normal backscattering 
processes, that is, to the operator 

\beq
V_{{\rm Intra},1} \propto \cos [   \sqrt{2 \pi} ( \theta_\rho ( 0 ) + \theta_\sigma ( 0 ) )]
\:\:\:\: . 
\label{ppy.x2}
\eneq
\noindent
Considering that $\hat{\phi}_\chi ( x ) , \hat{\theta}_\chi ( x)$ are fully decoupled 
from $H_b$, it is natural to assume that, throughout the whole phase diagram of the 
system, $\hat{\theta}_\chi ( x )$ is pinned at $x=0$. As a result, going again backwards along the 
sequence of transformations discussed above, we express $V_{{\rm Intra},1}$ in terms of 
the fields of the Y3J at $x=0$ as 

\beq
V_{{\rm Intra} , 1} \propto \cos [ 2 \sqrt{\pi} \theta_2 ( 0 ) ] 
\:\:\:\: . 
\label{ppy.x3}
\eneq
\noindent
The right hand side of Eq. (\ref{ppy.x3}) corresponds to the $T_{22}^{RL}$-operator at 
the second line of the table at appendix A.d of Ref. \cite{claudio_2} (plus its Hermitean conjugate), 
as witnessed by the perfect agreement between the scaling dimension of that operator and 
the result of section \ref{phadia_1} for the dimension of $V_{{\rm Intra} , 1}$,  $d_{{\rm Intra} , 1} = \frac{2 K K_3}{K + K_3}$. 

Finally,  in the $N=2$ junction one has the residual coupling of channel-2 to the MM. When properly accounting for the 
``Schr\"odinger cat'' nature of the state formed out of the hybridization between the MM and the KF $\Gamma_1$, one obtains,
as corresponding boundary operator, $V_{2 , {\rm Res}}$  given by \cite{giuaf_1}

\beq
V_{2 , {\rm Res}} \propto \cos [ \sqrt{2 \pi} ( \phi_\rho ( 0 ) - \phi_\sigma ( 0 ) )] 
\:\:\:\: . 
\label{ppy.x4}
\eneq
\noindent
Again, just as for the other operators listed above, one can readily check that
the right-hand side of Eq. (\ref{ppy.x4}) corresponds to the operator $T_{21}^{RL} [ T_{12}^{RL} ]^\dagger $
at  the third line of the table in appendix A.d of Ref. \cite{claudio_2} (plus its Hermitean conjugate),
consistent with the result for the scaling dimension of $V_{2 , {\rm Res}}$ we derived in 
section \ref{phadia_1},   $d_{2 , {\rm Res}} = \frac{4}{K_\rho+K_\sigma} = \frac{2 ( 2 K + K_3  )}{K ( K + K_3 )}$.

The $A \otimes A$ fixed point of the $N=2$ junction corresponds to pinning both $\phi_\rho ( x ) $ and 
$\phi_\sigma ( x )$ at $x=0$. Accordingly, one naturally identifies it with the $D_P$ fixed point of
the Y3J. To double-check the identification between the two fixed point, we note that the 
leading boundary operators at the $A \otimes A$ fixed point of the $N=2$ junction corresponds to the 
$V_{b , (1,2)}$ operator at the second line of Eq. (\ref{2a.7}), as well as $\tilde{V}_{{\rm Res} , 1}$
and $\tilde{V}_{{\rm Res} , 2}$ in Eqs. (\ref{addit.1}). Following the correspondence between
the parameters of the $N=2$ junction and the ones of the Y3J, one finds that the corresponding 
scaling dimensions are given by 

\begin{eqnarray}
d_{b , (1,2)} &=& K_\rho = K \left[ \frac{ K_3 }{ 2 K  + K_3 } \right]  \nonumber \\
d_{{\rm Res} , 1} &=& d_{{\rm Res} , 2 } = \frac{K_\rho + K_\sigma}{4} = K \: \left[ \frac{K + K_3}{2 ( 2 K + K_3 )} \right]  
 \:\:\:\: . 
 \label{addit.2}
\end{eqnarray}
\noindent
A comparison of the results in Eqs. (\ref{addit.2}) with the table at appendix A.b of 
Ref. \cite{claudio_2} again supports the identification of $A \otimes A$ with 
the $D_P$ fixed point in the Y3J.  (Note that, according to Eqs. (\ref{addit.2}),
 $A \otimes A$ becomes a stable fixed point as soon as the conditions 
$d_{b , (1,2)} > 1$ and $ d_{{\rm Res} , 1} > 1$ are both satisfied. This may clearly happen only
for $K_\sigma > 1$, without contradicting the conclusions of Ref. \cite{giuaf_1} and of section \ref{phadiaN2}, 
which were reached under the assumption that $K_\sigma < 1$).  

In view of the perfect correspondence of the $N=2$ junction and the asymmetric Y3J with $K<1$, $2K > 1 + K / K_3 $, 
one naturally concludes that  an analog of the FCFP found in the 
$N=2$ junction along the symmetric line $\bar{t}_2 = \bar{t}_3$ exists in the phase diagram of the 
Y3J, as well. In particular, we infer that, for $K<1$, $2K > 1 + K / K_3$ the stable fixed point of the Y3J is either 
the $A_1$ or the $A_2$ asymmetric fixed point of Ref. \cite{claudio_2}, depending on whether,
at the reference scale,  $\bar{t}_{3,2} > \bar{t}_{1,3}$, or $\bar{t}_{3,2} < \bar{t}_{1,3}$, or the FCFP of 
the $N=2$ junction located, according to the analysis of Ref. \cite{giuaf_1}, at $\bar{t}_{3,2 , * } = \bar{t}_{1,3 , * } = \epsilon / {\cal F}  
\left( 2 - K^{-1} \right)$.
 
Before concluding this sub-section, two remarks are in order. First of all, we would like to stress that the condition $K < 1$, which we 
have assumed at the start of the discussion here, has the mere effect of making $V_{1,2}$ irrelevant, thus allowing for dropping 
terms $\propto \bar{t}_1$ from the following discussion. In fact, this condition can be relaxed and one can extend all the conclusions 
we derive here to the case $K> 1$, as well, but only provided $\bar{t}_1 $ is fine-tuned to 0 from the very beginning, and remains =0 along 
the RG trajectories. Secondly, we would like to emphasize that, in order to make the mapping effective, we identified one of the 
three KF's, specifically $\Gamma_3$, with the MM $\gamma_L$ emerging at the $N=2$ junction with a topological superconductor.
This points out, once more, that, to make a comprehensive discussion of the physics of real fermions at junctions of one-dimensional
interacting electronic systems and/or topological superconductors, KF's and MM's have to be considered altogether as 
actual degrees of freedom, despite the apparent conventional definition of the former ones as a mere mathematical 
means to properly follow the bosonization procedure.

\subsection{From the  Y-junction to the $N=2$ junction with a topological superconductor}
\label{lul}

In the previous section we   used the known results about the $N=2$ junction 
\cite{giuaf_1} to infer  the emergence of a planar FCFP in 
the phase diagram of the asymmetric Y3J for $K<1$, $\frac{1}{2K} + \frac{1}{2K_3} < 1$. 
Nevertheless, the identification of $K$ with the Luttinger parameter 
$K_\sigma$ of the $N=2$ junction makes it impossible to directly extend the correspondence to the case   $K>1$.
In fact, the analysis of Ref. \cite{giuaf_1} is 
limited to the regime $K_\sigma<1$. For $K_\sigma > 1$, two key things happen. First, the identification of the argument $\nu$ of the 
${\cal F}$-function in the perturbative RG  Eqs. (\ref{puccy.y2bis}) with $2 - K^{-1}$ implies that 
$\nu > 1$ for $K>1$. For $\nu>1$, ${\cal F} ( \nu ) < 0$, with the corresponding disappearance of the 
FCFP along the diagonal in the $\bar{t}_{3,2} - \bar{t}_{1,3}$-plane. Second, DEBC method shows the emergence of a
relevant operator at the $A \otimes N$, as well as at the $N \otimes A$, fixed point, given by  the inter-channel 
normal backscattering operator, $V_{{\rm Normal} , (1,2) }$, which implies that, unless one fine-tunes to 0 the 
coupling strength in front of $V_{{\rm Normal} , (1,2) }$, neither  $A \otimes N$, nor $N \otimes A$, are 
stable fixed points anymore. To figure out what the phase diagram of the $N=2$ junction looks like for 
$K_\sigma > 1$, in this section we reformulate the correspondence  with the Y3J, but this time to retrieve informations 
about the $N=2$ junction from what is known about the phase diagram of the asymmetric Y3J. 

To begin with, we refer to the disconnected fixed point. There, as stated above, for $K_\sigma > 1$, the leading boundary operators for the $N=2$ junction are the 
couplings of the two wires to the MM, $V_{b,1(2)}$, and the inter-channel normal backscattering operator, 
$V_{{\rm Normal} , (1,2) }$, given by

\begin{eqnarray}
V_{b,1} &=&  2 i t_1 \: \gamma_L \Gamma_1 \cos [ \sqrt{\pi} \phi_1 ( 0 ) ] \nonumber \\
V_{b,2} &=&  2 i t_2 \: \gamma_L \Gamma_1 \cos [ \sqrt{\pi} \phi_2 ( 0 ) ] \nonumber \\
 V_{{\rm Normal} , (1,2) } &=& v_{1,2} \Gamma_1 \Gamma_2 e^{ i \sqrt{\pi} [ \phi_1 ( 0 ) - \phi_2 ( 0 ) ] } + {\rm h.c.} 
 \:\:\:\: , 
 \label{compa.3}
\end{eqnarray}
\noindent
of scaling dimensions respectively given by $d_{b,1} = d_{b,2} = \frac{1}{4 K_\rho} + \frac{1}{4 K_\sigma}$, $d_{{\rm Normal} , (1,2 )} = \frac{1}{K_\sigma}$. 
At a given $U$, the condition $K_\sigma > 1$ is recovered by setting    $\frac{1}{2} < K < K_* ( U )$, with 

\beq
K_* ( U ) = - \frac{U}{2 \pi u } + \sqrt{1 + \left( \frac{U}{2 \pi u } \right)^2} \leq 1
\:\:\:\: . 
\label{lui.1}
\eneq
\noindent
Thus, we conclude that, for  $\frac{1}{2} < K < K_* ( U )$, the disconnected fixed point 
is unstable, with three allowed independent relevant boundary operators. 

Moving to the $A \otimes N$ fixed point (and/or to the complementary $N \otimes A$ fixed point),
we see that, referring to the operators  $V_{a , (1,2)} ,   V_{b , (1,2)}$, their scaling 
dimension can be rewritten as

\beq
d_{a , (1,2)} = d_{b  , (1,2) }= 
\frac{1 + K_\sigma K_\rho}{K_\sigma + K_\rho} = 1 + \frac{(1-K_\sigma )(1-K_\rho)}{K_\sigma+K_\rho} 
\;\;\;\; , 
\label{lei.1}
\eneq
\noindent
which is clearly $<1$ for $K_\sigma > 1, K_\rho < 1$. Therefore, we conclude that 
both inter-channel normal boundary backscattering, as well as boundary 
pairing, provide relevant perturbations at the $A \otimes N$ fixed point, 
as soon as $K_\sigma > 1$. 

Finally, we readily see that the $A \otimes A$ fixed point is not stable either, 
as a relevant boundary perturbation is provided by the operator $V_{b , (1,2)}$ of Eq. (\ref{2a.7}), 
with scaling dimension $K_\rho < 1$.  To get some insight on the phase diagram of 
the $N=2$ junction for $K_\sigma > 1$, we  now employ  the correspondence with the 
Y3J. In order to do so, we start by assuming, for the time being, that 
$V_{b,1} , V_{b , 2}$ and $V_{{\rm Normal} , (1,2)}$ all 
have the same scaling dimension, that is, $d_b =  d_{{\rm Normal} , (1,2)}$. At 
$N=2$, this condition requires 

\beq
\frac{1}{4 K_\rho} = \frac{3}{4 K_\sigma} \Rightarrow \frac{K U}{ \pi u } = \frac{4}{5}
\:\:\:\: . 
\label{lui.3}
\eneq
\noindent
Let $\hat{K} ( U )$  be the value of $K$ that satisfies Eq. (\ref{lui.3}) at given  $U$.  
We obtain  $\frac{\hat{K} ( U )  U}{ \pi u } = \frac{4}{5}$ and, in addition, due to 
the assumption of a repulsive intra-wire interaction, which implies 
$\frac{1}{2} < \hat{K}_2 ( U ) < 1$, we also get $\frac{1}{\sqrt{5}} < d_b < \frac{2}{\sqrt{5}}$. 
At  $K = \hat{K} ( U )$, one obtains   
$K_\rho = \hat{K} ( U ) \left[ \frac{\sqrt{5}}{3} \right]$ and 
$K_\sigma = \hat{K} ( U ) \: \sqrt{5}$, that is, $K_\rho = K_\sigma / 3$, and  
$u_\rho = u_\sigma \frac{K_\sigma}{K_\rho} = 3 u_\sigma$. Accordingly, $H_{2 , {\rm Bulk}}$  now takes the 
form 

\beq
H_{2 , {\rm Bulk}} = \frac{u_\sigma}{2} \: \int_0^\ell \: d x \: \left\{ K_\sigma [ ( \partial_x \phi_\rho ( x ))^2 + 
( \partial_x \phi_\sigma ( x ))^2 ] + K_\sigma^{-1} \left[9 ( \partial_x \theta_\rho ( x ))^2 +  ( \partial_x \theta_\sigma ( x ))^2 
\right] \right\}
\:\:\:\: . 
\label{lui.4}
\eneq
\noindent
In deriving the boundary phase diagram one has to work on the semi-infinite system. Thus, 
the upper integration bound ($\ell$) in the integrals at the right-hand side of Eq. (\ref{lui.4})
must be sent to $\infty$. Taking this into account, we now perform the canonical transformation 

\begin{eqnarray}
 \Phi ( x ) &=&  \phi_\rho ( x )  / \sqrt{3}  \;\;\; , \;\; 
\Theta ( x )   = \sqrt{3} \theta_\rho ( x ) \nonumber \\
 \varphi ( x ) &=& \phi_\sigma ( x ) \;\;\; , \;\;
 \vartheta ( x ) =  \theta_\sigma ( x ) 
 \:\:\:\: , 
 \label{lui.5}
\end{eqnarray}
\noindent
followed by a rescaling by $\sqrt{3}$ of the $x$ coordinate in the integrals involving 
the center-of-mass fields. As a result, Eq. (\ref{lui.4}) becomes

\beq
H_{2 , {\rm Bulk}} = \frac{u_\sigma}{2} \: \int_0^\ell \: d x \: \left\{ K_\sigma [ ( \partial_x \Phi ( x ))^2 + 
( \partial_x \varphi ( x ))^2 ] + K_\sigma^{-1}  [ ( \partial_x \Theta ( x ))^2 + 
( \partial_x \vartheta ( x ))^2 ]  \right\}
\:\:\:\: , 
\label{lui.6}
\eneq
\noindent
while, upon also redefining the field $\Phi ( x )$ according to  
$\sqrt{3}\Phi ( x ) \to \sqrt{3} \Phi ( x ) +\sqrt{\frac{\pi}{2}}$, 
the most general boundary interaction at the disconnected fixed point, $\hat{H}_b$, can be 
written as a linear combination of $V_{b,1} , V_{b , 2}$ and $V_{{\rm Normal} , (1,2)}$ as 
 
\begin{eqnarray}
\hat{H}_b &\to&  - 2 i t_1 \gamma_L \Gamma_1 \sin \left[ \sqrt{\frac{\pi}{2} }  \left( - \varphi ( 0 ) -\sqrt{3}\Phi ( 0 )  \right) \right]
- 2 i t_2   \Gamma_2 \gamma_L \sin \left[ \sqrt{\frac{\pi}{2} } \left( - \varphi ( 0 ) +\sqrt{3}\Phi ( 0 ) \right) \right] \nonumber \\
&-& 2 i v_{1,2} \Gamma_1 \Gamma_2 \sin [ \sqrt{2 \pi} \varphi ( 0 ) ] 
\:\:\:\: . 
\label{lui.8}
\end{eqnarray}
\noindent
The right-hand side of Eq. (\ref{lui.8}) corresponds to the boundary Hamiltonian of a    
Y3J with relative fields $\varphi ( x ) , \Phi ( x )$, boundary 
couplings $t_1 , t_2 , v_{1,2}$ and Luttinger parameters $K = K_3 = K_\sigma$. In section \ref{j3}, we argue that,  
 as soon as all three of the boundary couplings are $\neq 0$, for $1 < K_\sigma < 3$ (as 
is the case here), the system flows towards the ${\bf Z}_3$-symmetric $M$-FCFP of the symmetric Y3J \cite{oca}. 
Thus, we conclude that, for $K_\sigma > 1$, as soon as $v_{1,2} \neq 0$, the stable fixed point of the 
$N=2$ junction is realized outside of the $t_1 - t_2$-plane. It sets in at a finite value of the three 
boundary couplings $\bar{t}_1 = \bar{t}_2 = v_{1,2} = t_*$, with $t_*$  corresponding to 
the $M$-FCFP of the ${\bf Z}_3$-symmetric Y3J with Luttinger parameter $K = K_\sigma$. The $M$-FCFP 
is the endpoint of RG flow lines that, were $v_{1,2} = 0$, would instead end up at the $A \otimes A$ fixed 
point of the $N=2$ junction.

To complete our derivation, we now discuss the phase diagram of the $N=2$ junction when  
$K_\rho = \lambda  K_\sigma / 3$, with $\lambda \neq 1$, and $K_\sigma > 1$. In this case,
going backwards along the mapping we derived in section \ref{lul_0}, we see that 
the $N=2$ junction maps onto the  (bulk) asymmetric Y3J we review in section \ref{j3},
with $K = K_\sigma$ and $K_3 = \left( \frac{2 \lambda}{3 - \lambda}  \right) K$. 
Accordingly, one obtains for the scaling dimensions 
$d_{b , 1} = d_{b , 2 } = \frac{1}{4 K_\sigma} \left( 1 + \frac{3}{\lambda} \right)$. 
This implies $d_{b , 1} >$ $(<) d_{{\rm Normal}, (1,2)}$ depending on whether 
$\lambda < 1$ ($\lambda > 1$). Apparently, the difference  between $d_{b,1 (2)} $ and $d_{{\rm Normal} , (1,2)}$ could, in 
principle, trigger RG trajectories either towards an $A_a$-asymmetric fixed point, or towards an
asymmetric version of the $M$-FCFP.  As a function of $\lambda$ and of $K_\sigma$, one obtains 
for the scaling dimensions of the relevant boundary operators at the $A_3$ and at the $A_1 , A_2$ fixed 
points of the corresponding Y3J the expressions 

\begin{eqnarray}
 d_{A_3} ( \lambda , K_\sigma ) &=& \frac{3 + \lambda K_\sigma^2}{4 \lambda K_\sigma} \nonumber \\
 d_{A_1} ( \lambda , K_\sigma ) &=& d_{A_2} ( \lambda , K_\sigma ) = \frac{3 + \lambda K_\sigma^2}{( 3 + \lambda ) K_\sigma}
 \:\:\:\: . 
 \label{addit.y1}
\end{eqnarray}
\noindent
Given the assumption $K_\sigma > 1$, based on the discussion of sections \ref{phadiaN2},\ref{j3}, in the following, referring to 
the fixed points of the Y3J, we assume that
the stable fixed point has to be identified with $A_3$ if $d_{A_3} ( \lambda , K_\sigma )  > 1 , d_{A_1} ( \lambda , K_\sigma ) < 1$, with 
either one of $A_1$, or $A_2$ (or with a planar FCFP) if $d_{A_3} ( \lambda , K_\sigma )  < 1 , d_{A_1} ( \lambda , K_\sigma ) > 1$,
with the $D_P$ fixed point if  $d_{A_3} ( \lambda , K_\sigma ) > 1 , d_{A_1} ( \lambda , K_\sigma ) > 1$ and, finally,
with the $M$-FCFP if   $d_{A_3} ( \lambda , K_\sigma ) < 1 , d_{A_1} ( \lambda , K_\sigma ) < 1$. Eventually, from 
the correspondence rules of section \ref{lul_0}, we make the appropriate identifications with the 
fixed points of the $N=2$ junction. A straightforward algebraic derivation leads us to conclude that,
depending on whether $\lambda > 1$, or $\lambda < 1$, there is the possibility of stabilizing 
either the $A_3$, or the $A_1 , A_2$ fixed points. In particular, without considering 
additional constraints on the various parameters, one would obtain

\begin{itemize}

 \item $\lambda > 1$: 
 
 In this case, for $K_\sigma < \frac{3}{4} + \frac{9}{4 \lambda^2}$ the stable fixed point corresponds to 
 the $M$-FCFP of the Y3J. For $\frac{3}{4} + \frac{9}{4 \lambda^2 } < K_\sigma < 6 - \frac{3}{\lambda}$ 
 the stable fixed point would correspond to either the $A_1$ or the $A_2$ fixed point and, 
 for $K_\sigma > 6 - \frac{3}{\lambda}$, to the $D_P$ fixed point. However, these last two possibilities 
 are ruled out by the observation that, by definition, one has $K_\sigma = \frac{3}{\lambda} K_\rho < \frac{3}{\lambda}$.
Therefore, one obtains that $\frac{3}{4} + \frac{9}{4 \lambda^2 } < K_\sigma  \Rightarrow \frac{1}{3} < \lambda < 1$, against 
the initial assumption. As a result, for $\lambda > 1$ only the $M$-FCFP of the Y3J corresponds to a stable 
fixed point of the $N=2$ junction.

\item $\lambda < 1$:

 In this case, the $M$-FCFP of the Y3J corresponds to the stable phase of the system for $K_\sigma < 6 - \frac{3}{\lambda}$. 
For $6 - \frac{3}{\lambda} < K_\sigma <  
 \frac{3}{4} + \frac{9}{4 \lambda^2}$ the stable fixed point corresponds to 
 the $A_3$ fixed point. 
For $K_\sigma >  \frac{3}{4} + \frac{9}{4 \lambda^2}$, $D_P$ would become the stable fixed point. 
Again, we rule out this last possibility, due to the observation that, by 
definition, $K_\sigma^{-2} + K_\rho^{-2} = 2 (K_j )^{-2}$ where, to avoid confusion, we here 
use $K_j$ to mean the Luttinger parameter of each QW in the $N=2$ junction 
in the absence of bulk, inter-wire interaction. At this stage, 
we are assuming, just as in Ref. \cite{giuaf_1}, $\frac{1}{2} < K_j < 1$ (later on we discuss an extension of 
our analysis to $K_j > 1$ in the absence of inter-wire interaction). As a 
result, we find the condition $\frac{9}{4 \lambda^2} = \frac{K_\sigma^2}{2 K_j^2} - \frac{1}{4}
< 2 K_\sigma^2 - \frac{1}{4}$. Taking this into account, we see that $K_\sigma >  
 \frac{3}{4} + \frac{9}{4 \lambda^2} \Rightarrow 2 \left( K_\sigma - \frac{1}{2} \right)^2 < 0$,
clearly impossible. Therefore, the only allowed phase transition happens for  $6 - \frac{3}{\lambda} = K_\sigma$,
where the stable fixed point of the system changes from the $M$-FCFP  of the Y3J (for $K_\sigma < 6 - \frac{3}{\lambda}$) to the $A_3$ fixed point
(for $K_\sigma > 6 - \frac{3}{\lambda}$). 
\end{itemize} 
In terms of the parameters of the $N=2$ junction, such a fixed point corresponding to setting $\bar{v}_{1,2} \to \infty$, and 
$\bar{t}_1 = \bar{t}_2 = 0$. Accordingly, we see that it corresponds to 
the perfect healing of the junction between wires-1 and -2, with the MM decoupled from the 
two wires.  To double-check its stability, we 
resort to DEBC-approach, by imposing type $N$ boundary conditions on $\Phi ( x ) , \vartheta ( x )$ at $x=0$ and,
accordingly, type $A$ boundary conditions on $\varphi ( x ) , \Theta ( x )$. In this case, the leading boundary 
perturbation is indeed realized as a linear combination of the $V_{b , 1}$ and $V_{b , 2 }$-operators in 
Eqs. (\ref{compa.3}), that is, by the hybridization between the normal wires and the MM, which now take the 
form 

\begin{eqnarray}
 V_{b , 1} &\to& i t_1 \gamma_L \Gamma_1 e^{ i \frac{\sqrt{\pi}}{2} [ \sqrt{3} \Phi ( 0 ) \pm \vartheta ( 0 ) ] } + {\rm h.c.} \nonumber \\
 V_{b , 2} &\to& i t_2 \gamma_L \Gamma_2 e^{ i \frac{\sqrt{\pi}}{2} [ \sqrt{3} \Phi ( 0 ) \pm \vartheta ( 0 ) ] } + {\rm h.c.} 
 \;\;\;\; , 
 \label{final.1}
\end{eqnarray}
\noindent
both with scaling dimension $d_b = \frac{3  +\lambda  K_\sigma^2}{4 \lambda K_\sigma} > 1$. Therefore, we conclude that, 
as soon as $K_\sigma > 6 - \frac{3}{\lambda}$,   the system 
is attracted towards the $A_3$-like fixed point, in which the MM is ``pushed out'' of the quantum wires, which hybridize with each other to 
an effectively uniform wire, out of which lies the decoupled MM. 
 
\begin{figure} 
\includegraphics*[width=.6\linewidth]{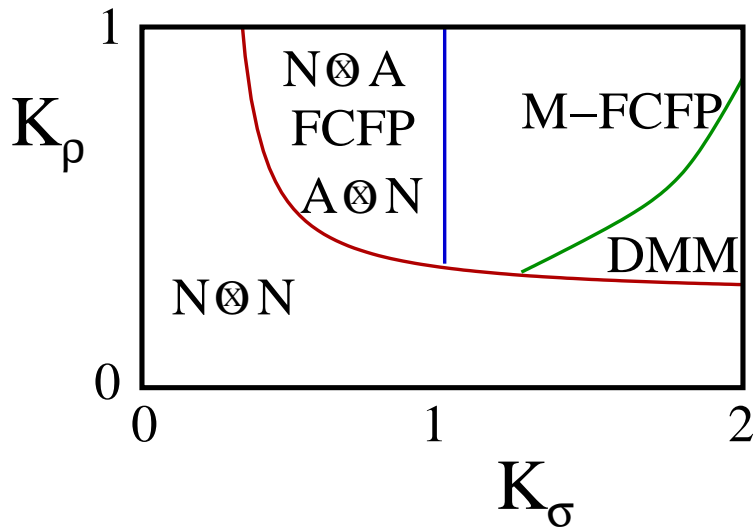}
\caption{Sketch of the phase diagram of the $N=2$ junction for $K_\rho < 1$. The red curve corresponds to 
$\frac{1}{4 K_\rho } + \frac{1}{4 K_\sigma} = 1$: it separates the phase corresponding to the disconnected fixed point
( $\frac{1}{4 K_\rho } + \frac{1}{4 K_\sigma} < 1$) from the other, ``nontrivial" phases. For $K_\sigma < 1$, the 
phase of the junction corresponds to either the $A \otimes N$, or to the $N \otimes A$, fixed point, according 
to whether the initial boundary coupling $t_1$ is larger, or smaller, than $t_2$. In the symmetric case
$t_1 = t_2$, the system's phase corresponds to the FCFP of Ref. \cite{giuaf_1}. When $K_\sigma$ becomes 
$>1$, a phase that maps onto a (generically non-${\bf Z}_3$ symmetric) deformation of the M-FCFP of the 
Y3J opens till, after crossing the green line, corresponding to the curve $K_\sigma \left( 1 + K_\rho^{-1} \right) = 6$, 
an additional phase opens (DMM), corresponding  to the MM fully decoupled from the junction with the other two wires.
 }  \label{pn_2}
\end{figure}

In conclusion, we have shown that the condition $K_\sigma > 1$ is enough to reverse the phase diagram of the junction 
between two quantum wires and a topological superconductor, with respect to 
the result derived in   Ref. \cite{giuaf_1} for $K_\sigma < 1$. Specifically, at $K_\sigma > 1$, the 
FCFP corresponds to the true stable phase of the system and is eventually identified with 
the  (in general non-${\bf Z}_3$-symmetric) $M$-FCFP of the Y3J. Further increasing $K_\sigma$ with respect to 
$K_\rho$ may eventually trigger an additional phase transition towards a phase corresponding to 
the perfect healing of the junction between wires-1 and -2, with the MM decoupled from the 
two wires. To evidence the new phases we find in the $N=2$ junction by means of the correspondence 
with the Y3J, in Fig. \ref{pn_2} we draw  the phase diagram of the $N=2$ junction for $K_\rho < 1$ by 
including, in addition to what we found in Ref. \cite{giuaf_1} for $K_\sigma < 1$,  the   phases 
emerging when the parameter $K_\sigma$ is $>1$.

 \noindent

  Before concluding this sub-section, it is worth remarking how the correspondence between the $N=2$ junction and 
the Y3J also allows for inferring the phase diagram of the former system in the case of zero inter-wire bulk interaction
($U=0$), and attractive intra-wire interaction ($K>1$). To do so, we first of all note that $U=0$ implies $K_\rho = K_\sigma = K$. 
From Eqs.(\ref{ppy.y9}), one sees that this condition is recovered in the $\frac{K_3}{K} \to \infty$ limit, which yields 
$K_\rho = K_\sigma = K$. Accordingly, one sees that, at the disconnected fixed point, the leading boundary operator 
is again provided by a generic linear combination of $V_{2,3}$ and of $V_{3,1}$ in Eq. (\ref{3w.5bis}) (plus their 
Hermitean conjugates), all with scaling dimensions $d_{V_{2,3}} = d_{V_{3,1}} = \frac{1}{2K}$. Thus, we 
recover the expected result that the disconnected fixed point is unstable for $K > 1/2$ \cite{giuaf_1}. 
At the $A \otimes N$ fixed point, the most general allowed boundary interaction contains $V_{a , (1,2)}$ and 
$V_{b , (1,2)}$ in Eqs. (\ref{ppy.x1}), both with scaling dimension $d_{a , (1,2)} = d_{b , (1,2)} = \frac{1 + K^2}{2 K}$,
of the intra-channel 1 normal backscattering operator in Eq. (\ref{ppy.x3}), with scaling dimension $d_{{\rm Intra} , 1} = 2 K$, 
and the operator $V_{2 , {\rm Res}}$  in Eq. (\ref{ppy.x4}) describing the residual coupling to channel 2, with scaling dimension 
$d_{2 , {\rm Res}} = \frac{2}{K}$. Among all those operators, the only one that can become relevant for $K>1$ is $V_{2 , {\rm Res}}$, 
whose scaling dimension becomes $<1$ as $K > 2$. Finally, at the $A \otimes A$ fixed point, the leading boundary 
interaction is provided by a linear combination of  $\tilde{V}_{{\rm Res} , 1}$
and $\tilde{V}_{{\rm Res} , 2}$ in Eqs. (\ref{addit.1}), with corresponding scaling dimension $d_{\rm Res} = \frac{K}{2}$. Putting 
the above results all together, one therefore infers that, for $U=0, K > 1$, the stable fixed point of the $N=2$ junction 
either corresponds to $A \otimes N$, or to $N \otimes A$, for $K<2$, depending on the initial values of the boundary 
coupling strengths. For $K > 2$, $ A \otimes A$ becomes the stable fixed point of the junction. Moreover, consistently
with the discussion of sub-section \ref{phadiaN2}, as well as with the results of Ref. \cite{giuaf_1}, one 
expects $A \otimes N$ and $N \otimes A$ to be separated by some intermediate phase(s). Whether this corresponds to 
just a FCFP, as it happens for $K<1$ \cite{giuaf_1}, or to more than one FCFP's, or even to a continuous line 
of fixed points, cannot be firmly stated with our method in this range of values of system's parameters and, very likely, 
to discriminate among the various possible options will require resorting to a nonperturbative,  numerical approach to   the 
problem.   

To summarize the   correspondence  between phases of the $N=2$ junction and of the Y3J, in table \ref{phases_corresp}
we provide a synoptic view of corresponding fixed points in the two models, using a different color (red, instead than 
blue) to highlight phases that in either model are predicted by means of the correspondence with the other. 

\vspace{0.2cm}
 
\begin{table}
\centering
\begin{tabular}{| c | c |}
\hline 
{\bf $N=2$ junction}  &  {\bf Y3J}   \\
 \hline {\color{blue} 
Disconnected fixed point ($N \otimes N$) }  &{\color{blue}  Disconnected fixed point   }\\
 \hline {\color{blue} 
$A \otimes N$ fixed point  } &{\color{blue}  $A_1$ fixed point }  \\
 \hline {\color{blue} 
$N \otimes A$ fixed point }  &{\color{blue}  $A_2$ fixed point }  \\
 \hline {\color{blue} 
FCFP at  $\bar{t}_1 = \bar{t}_2 = t_* = \left[ \epsilon /  {\cal F} \left( \frac{1}{2 K_\rho} - \frac{1}{2 K_\sigma}  \right) 
\right]^\frac{1}{2}$ } &  {\color{red} ``Planar'' FCFP at 
$\bar{t}_{1,2} = 0$} \\
 \hline
{\color{blue} $A \otimes A$ fixed point } & {\color{blue} $D_P$ fixed point }   \\
\hline
{\color{red} ``Off-planar'' FCFP } & {\color{blue} (Non ${\bf Z}_3$-symmetric) M-FCFP}   \\
\hline 
{\color{red} Disconnected MM fixed point } & {\color{blue} $A_3$ fixed point}   \\
 \hline 
\end{tabular}
\caption{Table of correspondence  between phases (fixed points) of the $N=2$ junction and of the Y3J. 
The fixed points inferred in each one of the systems by means of the correspondence with the other one are 
highlighted in red. The fixed points already known in both systems are highlighted in blue (see Ref.\cite{giuaf_1}
for the $N=2$ junction, Refs.\cite{oca,claudio_1} for the Y3J).} 
   \label{phases_corresp}
\end{table}
\noindent

To complement the results of the previous sub-sections, we now briefly discuss how the correspondence between the $N=2$ junction 
and the Y3J has to be implemented in computing the $g$-function at corresponding fixed points of the 
two models.

\subsection{Calculation of the $g$-function at corresponding fixed points}
\label{calcg_corr}

The simplest fixed point in both the $N=2$ junction and in the Y3J is the disconnected one, in
which all the boundary interaction strengths are set to 0. As from Eqs. (\ref{pd.2},\ref{capuccy.5}), at the disconnected fixed point, one obtains 
$g_{\rm Disc} = 2 [ K_\rho K_\sigma ]^\frac{1}{4}$ in the $N=2$ junction, and $g_{\rm Disc} = 2 [ K_3 K^2  \bar{K} ]^\frac{1}{4}$ in 
the Y3J. On comparing the two results, the first observation is that, to recover the over-all factor of 2 in the 
case of the Y3J, one has to include in the calculation the auxiliary KF $\bar{\Gamma}$, as well. Besides that, the 
two results are apparently not related to each other via the correspondence between the Luttinger parameters in 
the two models in Eqs. (\ref{ppy.y9}). This is due to the fact that the Y3J $g$-function receives contributions from overall degrees of 
freedom not entering the correspondence with the $N=2$ junction, that is, the auxiliary field and the center of mass 
field $\phi_\chi ( x )$. On recomputing $g_{\rm Disc}$ in the Y3J by dropping   those contributions, one eventually 
obtains the asymmetric version of the result of Ref. \cite{oca}, that is

\beq
g_{\rm Disc} = 2 \left[ \frac{K^2 K_3}{2 K + K_3} \right]^\frac{1}{4}
\:\:\:\: , 
\label{cag.1}
\eneq
\noindent
that is, exactly the result one obtains when inserting Eqs. (\ref{ppy.y9}) into the formula 
for $g_{\rm Disc}$ in the $N=2$ junction. Having stated the correspondence between the $g$-function at 
the disconnected fixed points, in the following we consider the $g$-function at alternative fixed 
points always normalized to $g_{\rm Disc}$, in both models.

The $A \otimes A$ fixed point in the $N=2$ junction corresponds to the $D_P$ fixed point of the Y3J. Here, 
despite the counting of the real fermionic degrees of freedom working differently in the two models, the results 
for the $g$-function are again consistent with each other. While in   \ref{stabl} we discuss in detail the derivation of 
the corresponding degeneracy factor in the $N=2$ junction, it is worth recalling  how one recovers it in  the Y3J. 
Setting for simplicity $t_{2,1} = 0$, when mirroring $H_b$ in Eq. (\ref{ppy.y7}), one obtains
its two-boundary version, $H^{(2)}_b$, given by

\begin{eqnarray}
H_b^{(2)}  &=&  - 2 i t_{3,2} \Gamma_2 \Gamma_3 \: \cos \left\{   \sqrt{\frac{\pi}{2}}  \left[ \phi_\sigma ( 0 ) + \phi_\rho ( 0 ) \right] \right\} 
- 2 i t_{1,3}  \Gamma_1 \Gamma_3\: \cos \left\{   \sqrt{\frac{\pi}{2}}  \left[ \phi_\sigma ( 0 ) - \phi_\rho ( 0 ) \right] \right\} 
\nonumber \\
 &-&   2 i t_{3,2} \Gamma_2 \Gamma_3 \: \cos \left\{   \sqrt{\frac{\pi}{2}}  \left[ \phi_\sigma ( \ell  ) + \phi_\rho ( \ell  ) \right] \right\} 
- 2 i t_{1,3}  \Gamma_1 \Gamma_3\: \cos \left\{   \sqrt{\frac{\pi}{2}}  \left[ \phi_\sigma ( \ell ) - \phi_\rho ( \ell  ) \right] \right\} 
\:\:\:\: . 
\label{ppy.y7addi}
\end{eqnarray}
\noindent
This is the two-boundary Hamiltonian for the $N=2$ junction, except that now one has $\gamma_L = \gamma_R = \Gamma_3$. 
When discussing the degeneracy factor due to the zero-mode real fermion operators, in   \ref{stabl} we separately consider this case, 
concluding that, when both $\phi_\rho ( 0 )$ and $\phi_\sigma ( 0 )$ are properly pinned, one recovers a total number 
of three real-fermion zero modes which, put together with $\bar{\Gamma}$, provide the degeneracy factor of 4 (2) to
the total partition function (to the $g$-function). Once the correct degeneracy factors have been taken into account, in the $N=2$ junction 
one obtains that $\rho_{A \otimes A} = g_{A \otimes A} / g_{N \otimes N} = [ K_\rho K_\sigma ]^{-\frac{1}{4}}$, which,
using  Eqs. (\ref{ppy.y9}), one readily shows to be the   same as  the result of Eq. (\ref{capuccy.12}) for the Y3J. 

When computing the $g$-function at the $ A \otimes N$ fixed point of the $N=2$ junction, we have to pertinently modify the 
result in Eq. (\ref{pd.3}), due to the identity $u_\sigma = u_\rho = u$, which is a direct consequence of the mapping from the Y3J. In this 
case, implementing the approach of Ref. \cite{giuaf_1}, ore readily derives, using 
 Eqs. (\ref{capuccy.8}), the identity  

\beq
g_{A \otimes N } =    \left[  \frac{1}{4 K_\rho}  + \frac{1}{4 K_\sigma}\right]^\frac{1}{2} g_{\rm Disc} = 
   \left[  \frac{1}{2 K }  + \frac{1}{2 K_3}\right]^\frac{1}{2} g_{\rm Disc} 
\:\:\:\: ,
\label{adgi.1}
\eneq
\noindent
which shows that, once normalized to the $g$-function at the disconnected fixed point, 
$\frac{ g_{A \otimes N} }{ g_{\rm Disc}}$ in the $N=2$ junction is equal to 
$\frac{g_{A_1}}{g_{\rm Disc}}$ in the Y3J, as expected from the correspondence 
between the two models (note the apparent difference between the right-hand side of Eq. (\ref{adgi.1}) for 
$g_{A \otimes N}$ and the result in Eq. (\ref{pd.3}). This is due to the condition 
$u_\rho = u_\sigma = u$ which naturally arises from the mapping and, in this case, takes the place 
of the formulas one generally derives from Eqs. (\ref{1w.19}) of   \ref{bomo}.)
 
  Before concluding this section, a comment is in order about the
correspondence between real fermion operators in the $N=2$ junction and in the Y3J. 
On one hand,  we  see that it is rather straightforward in the single-boundary version of the models, 
as, in that case, one simply uses the observation that the center of mass field of the 
Y3J decouples from the junction dynamics and, therefore, the ``left-over'' KF can be formally
mapped onto the MM in the $N=2$ junction. On the other hand, the correspondence is not 
anymore straight when resorting to the  two-boundary version of the model Hamiltonian to 
compute $g$.   In this case, the different nature of the   MM's, which are local in real space, 
and of the  KF's, which  are global along the full extent of a QW, results, for instance, in 
that, while in the $N=2$ junction one has a TSS at each end of the system and, therefore, two 
MM's, in the Y3J one still has only the KF associated to the center of mass field. 
Remarkably, as we discuss above, this mismatch can be fixed by adding the auxiliary KF $\bar{\Gamma}$ to 
the count of the degrees of freedom of the Y3J, though without the possibility of rigorously extending 
the Hamiltonian mapping to the two-boundary systems, as well. Yet, the very fact that the 
results are the same in the two models, apparently further  supports the extension of the actual correspondence 
between the $N=2$ junction and the Y3J also to the $g$-function at corresponding fixed points,
which can be eventually regarded as a double check  of the results about 
the mapping of the phase diagram of one system to the other.

\section{Phase diagram and impurity entropy of a junction between $N$ quantum wires and a topological superconductor}
\label{junction_N}

\begin{figure}
 \center
\includegraphics*[width=1. \linewidth]{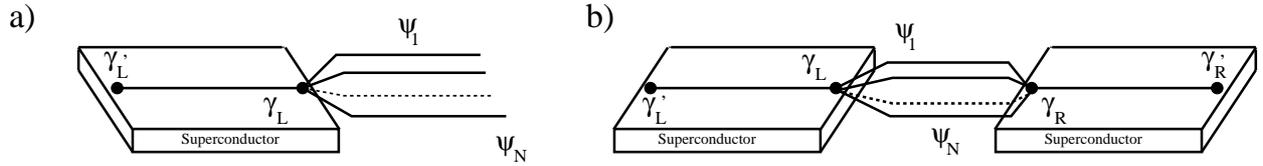}
\caption{
{\bf a)}: Sketch of the junction between $N$ interacting quantum wires and one topological superconductor. 
The one-dimensional topological superconductor is realized by depositing a semiconducting nanowire 
on top of a bulk, $s$-wave topological supercondutor and applying a magnetic field
(see Refs. \cite{oreg,dassarma} for 
details).  $\psi_a$ denotes the field of electrons in wire-$a$, $\gamma_L$ is the Majorana mode emerging 
at the interface between the quantum wires and the topological superconductor, $\gamma_L^{'}$ is the 
second Majorana mode expected to emerge at the opposide side of the topological superconductor \cite{kitaev};
{\bf b)}: Two-boundary version of the device sketched in {\bf a)}: $\gamma_R , \gamma_R^{'}$ now denote the 
Majorana modes at the edpoints of the topological superconductor connected to the right endpoint of the normal 
wires.} 
\label{device_1}
\end{figure}

As a further example of application of our method for computing the $g$-function, in this section we 
discuss the fixed points in the phase diagram, and the corresponding calculation of the 
IE, in   a  junction between $N$ QW's and a topological superconductor. For the sake of 
simplicity, in the following we make the symmetric assumption that the Luttinger parameters
$u , K$ are the same for each QW. In this respect, 
this is a symmetric multiwire generalization of the junction   discussed, for $K_\sigma < 1$,  
$N=2$ and (partially) for $N=3$,    in Ref. \cite{giuaf_1}.  
(An error occurred there in the final numerical estimate of the $g$-function, 
which we amend here; it did not affect the final conclusions.) 
Referring to the TLL-model Hamiltonian for the junction in Eq. (\ref{lutw.1})
of   \ref{bomo}, in the following, we assume that both intra-wire 
and inter-wire bulk interactions are repulsive, which implies $K<1, U>0$. In addition, we 
assume that $1 /2 < K$, which is a necessary condition to assure the relevance of 
the boundary coupling to the MM \cite{alicea,giuaf_1}.
 In Fig.\ref{device_1} we provide a sketch of the junction between $N$ interacting QW's and a TS
 in the single-boundary version (which we use to discuss the phase 
diagram) and in the two-boundary version (which we use to compute the $g$-function). 
We now provide a discussion of the phase diagram, which basically generalizes 
the analysis of Ref. \cite{giuaf_1} to a generic $N$.
 
 \noindent

\subsection{Phase  diagram of the $N$-wire junction with a topological superconductor}
\label{phadiaN}

The simplest fixed point in the phase diagram corresponds to having  all the wires 
disconnected from the TS. This implies 
type $N$ boundary conditions at $x=0$ for all the channels. When turning on nonzero 
couplings to the TS, $\{ t_a \}$, taking into account the boundary conditions, 
one may write the bosonized boundary Hamiltonian at 
the disconnected fixed point, $H_ {b, {\rm B} , N }^{(1)} $, in the form 

\beq
H_ {b, {\rm B} , N }^{(1)}  =     2 i \sum_{a = 1}^N t_a  \gamma_L \Gamma_a \: \cos [ \sqrt{\pi} \phi_a ( 0 ) ] 
\;\;\;\; . 
\label{1w.20}
\eneq
\noindent
The scaling dimension of $H_ {b, {\rm B} , N }^{(1)}$, $d_{b }$, can be 
readily derived using the transformation in Eqs. (\ref{1w.8bis}). The result is 
$d_{b }  = \frac{1}{2 N K_\rho} + \frac{N-1}{2NK_\sigma} $. 
For $1/2 < K < 1$ and for $U>0$, we find $\frac{1}{2} < d_b < 1$ $\forall N$, which implies that
$H_ {b, {\rm B} , N }^{(1)} $ always corresponds to a relevant boundary interaction for the range of
parameters considered. 
Thus, having nonzero $t_a$'s paves the way to the opening of new phases, corresponding to 
additional fixed points in the phase diagram of the junction. To discuss them and especially 
their stability under RG flow, one has to first identify the corresponding CIBC's, and then to 
employ them to construct, within the spirit of DEBC approach, all the allowed boundary operators 
at a given fixed point and eventually to check whether any of them corresponds to a relevant 
perturbation.   This can be readily done within the imaginary-time framework, in which we describe the wires using the  
Euclidean action $S_{\rm Eff}$ only depending on the fields at $x=0$. To derive $S_{\rm Eff}$, one has to integrate over the bulk fields 
everywhere in real space, except at $x=0$. Doing so, due to the duality between the $\phi$- and the $\theta$-fields on 
the semi-infinite line,  $S_{\rm Eff}$  can be either expressed in terms of the  fields  
$\phi_a ( \tau ) = \phi_a ( x = 0 , \tau)$, or of the fields  $\theta_a ( \tau ) = \theta_a ( x = 0 , \tau )$ \cite{oca}, as 

\begin{eqnarray}
S_{\rm Eff} [  \{ \phi_a \} ] &=& \frac{1}{2 \pi} \: \int \: d \Omega \: | \Omega | \: \vec{\phi}^\dagger ( \Omega ) 
 \: [ {\bf M}_N ]^T 
{\bf K}_N {\bf M}_N  \: \vec{\phi} ( \Omega ) \nonumber \\
S_{\rm Eff} [  \{ \theta_a \} ] &=& \frac{1}{2 \pi} \: \int \: d \Omega \: | \Omega |  \vec{\theta}^\dagger ( \Omega )  \: [ {\bf M}_N ]^T
{\bf K}_N^{-1} {\bf M}_N \: \vec{\theta} ( \Omega ) 
\:\:\:\: , 
\label{scadi.7}
\end{eqnarray}
\noindent
with $\vec{\phi} ( \Omega ) =  \int \: d \tau \: e^{  i \Omega \tau } [ \phi_1 ( \tau ) , \phi_2 ( \tau ) ,  \ldots ,  \phi_{N} ( \tau ) ]^T$,
$\vec{\theta} ( \Omega ) =  \int \: d \tau \: e^{  i \Omega \tau } [ \theta_1 ( \tau ) , \theta_2 ( \tau ) ,  \ldots ,  \theta_{N} ( \tau ) ]^T$, 
the matrix ${\bf M}_N$ defined in Eq.(\ref{1w.8ter}) of   \ref{bomo}, 
and the matrix ${\bf K}_N$ given by 

\beq
{\bf K}_N = \left[ \begin{array}{cccc}
K_\rho & 0 & \ldots & 0 \\ 
0 & K_\sigma & \ldots & 0 \\
\vdots & \vdots & \vdots & \vdots \\
0 & 0 & \ldots & K_\sigma 
                   \end{array} \right]
\:\:\:\: . 
\label{scadi.6}
\eneq
\noindent  
 Eqs. (\ref{scadi.7}) are true in general. The specific choice of either one of the actions in Eqs. (\ref{scadi.7}) depends on the boundary 
 conditions on  the  various fields.  

At the disconnected fixed point, all the $\theta$-fields obey Dirichlet boundary 
conditions at $x=0$. Therefore,  we must use $S_{\rm Eff} [  \{ \phi_a \} ]$ 
in Eq. (\ref{scadi.7}), with all the $\phi$'s different from 0. A straightforward calculation 
allows us to derive the scaling dimension of $H_{b ,{\rm B} , N}^{(1)}$, which is simply given by 
$ \frac{1}{2} \{ [ [ {\bf M}_N ]^T  
{\bf K}_N {\bf M}_N ]^{-1} \}_{1,1} = d_{b }$, where we have used the identity 
$ [ [ {\bf M}_N ]^T {\bf K}_N {\bf M}_N ]^{-1} =    [ {\bf M}_N ]^T 
{\bf K}_N^{-1} {\bf M}_N$.  In addition, while normal intra-wire backscattering 
plays no role, due to Dirichlet boundary conditions on $\theta_a ( 0 ) $, inter-wire backscattering and 
inter-wire pairing between channels $a$ and $b$ respectively correspond to the operators $V_{{\rm Normal} , (a,b)}$ and 
$V_{{\rm Pair} , (a , b) }$, given by

\begin{eqnarray}
 V_{{\rm Normal} , (a , b ) } &=& \Gamma_a \Gamma_b e^{ - i \sqrt{\pi} [ \phi_a ( 0 ) - \phi_b ( 0 ) ] } \nonumber \\
 V_{{\rm Pair} , (a,b)} &=& \Gamma_a \Gamma_b e^{ - i \sqrt{\pi} [ \phi_a ( 0 ) + \phi_b ( 0 ) ] }
 \;\;\;\; , 
 \label{scadu.1}
\end{eqnarray}
\noindent
plus their Hermitean conjugates. Their scaling dimensions are respectively given by 
(assuming $a \neq b$)

\begin{eqnarray}
 d_{{\rm Normal} , (a,b)} &=&  \{ [ [{\bf M}_N]^T   {\bf K}_N^{-1}   {\bf M}_N ]_{a,a} - [ [{\bf M}_N]^T   {\bf K}_N^{-1}   {\bf M}_N ]_{a,b} \} 
 = K_\sigma^{-1} \nonumber \\
  d_{{\rm Pair} , (a,b)} &=&  \{ [ [{\bf M}_N]^T   {\bf K}_N^{-1}   {\bf M}_N ]_{a,a} +  [ [{\bf M}_N]^T   {\bf K}_N^{-1}   {\bf M}_N ]_{a,b} \} 
 = K_\sigma^{-1} + \frac{2}{N} \left( K_\rho^{-1} - K_\sigma^{-1} \right)
 \:\:\:\: . 
 \label{scadu.2}
\end{eqnarray}
\noindent
 Given the assumption that 
$K_\sigma < 1$ (which we relaxed, when discussing the correspondence with the Y3J in section \ref{pha_2}), 
we conclude that both  $d_{{\rm Normal}, (a,b)}$ and $d_{{\rm Pair}, (a , b ) }$ are $>1$ and, accordingly, 
 $H_ {b, {\rm B} , N }^{(1)} $ provides the only relevant perturbation at the disconnected fixed point. 
 
To move away from the disconnected fixed point, in analogy to our derivation of Ref. \cite{giuaf_1}, we employ the perturbative 
RG approach within $\epsilon$-expansion method, which we 
briefly review in in   \ref{renge}. The corresponding RG equations for the running couplings are given by 

\beq
\frac{ d \bar{t}_{a } }{d l} = \epsilon \bar{t}_{a  } - {\cal F} ( \nu ) \bar{t}_{a  }  \: [ \sum_{b \neq a} \bar{t}_{ b  }^2 ]
\:\:\:\: ,
\label{1w.26}
\eneq
\noindent
with $l = - \ln ( D / D_0 )$, $D$ being the running energy scale, $D_0$ the high-energy (band) cutoff $\sim \tau_0^{-1}$,
and the function ${\cal F} ( \nu)$ defined in Eq. (\ref{renge.8}), with 
$\nu =  \frac{1}{ N K_\rho} - \frac{1}{N K_\sigma}$. In general, Eq. (\ref{1w.26}) implies a 
growth of the $\bar{t}_a$ along the RG  trajectories. This may either take the 
system to some FCFP, which generalizes the one discussed in 
Ref. \cite{giuaf_1} for $N=2,3$, or to  pinning $N_a$ $\phi_a ( 0 ) $'s, leaving the 
corresponding $\theta_a ( 0 )$ unpinned. Because of the symmetry between the channels, 
in the following we assume always that such a fixed point corresponds to pinning  
the first $N_a$ $\phi_a ( x )$'s  at $x=0$, leaving the remaining  $N_n$  unpinned
(so that, for instance, the disconnected fixed point corresponds to $N_a = 0 , N_n = N$). 

Let us consider the $N_a = 1 , N_n = N-1$ fixed point.  From Eq. (\ref{1w.20}), we see that this  is recovered  by sending $\bar{t}_{1 } \to \infty$
and minimizing the corresponding contribution to $H_ {b, {\rm B} , N }^{(1)}$. Accordingly, 
besides pinning $\phi_1 ( 0 )$, this also requires ``locking'' the system into a state either annihilated by the 
Dirac fermion $a_1 = \frac{1}{2} [ \gamma_L + i \Gamma_1 ]$, or by $a_1^\dagger$, depending on the value at 
which $\phi_1 ( 0 )$ is pinned. Taking this into account, we may list the various allowed boundary operators 
at that fixed point. First of all,  the intra-channel normal backscattering
operator in channel-1 is realized as   $V_{{\rm Intra},1} \sim V_a \cos [ 2 \sqrt{\pi} \theta_1 ( 0 ) ]$. To derive  
the corresponding scaling dimension, $d_{{\rm Intra} , 1}$, in  Eq. (\ref{scadi.7}) we set to 0 all the $\theta_a$'s but 
$\theta_1$. As a result, we eventually find 

\beq
d_{{\rm Intra} , 1} = 2 \{ [  [ {\bf M}_N ]^T 
{\bf K}_N^{-1} {\bf M}_N  ]_{1, 1} \}^{-1}  = \frac{2}{ \left[ \frac{1}{N K_\rho} + \frac{N-1}{N K_\sigma} \right] } = [ d_{b, {\rm Disc}} ]^{-1}
\:\:\:\: , 
\label{scadi.9}
\eneq
\noindent
which implies  
$d_{{\rm Intra} , 1} = [ d_{b, {\rm Disc}} ]^{-1}$. Switching to boundary inter-channel 
normal backscattering/pairing operators, we have to separately consider whether those 
processes involve channel-1, or not. In the latter case, assuming that  both $a$ and $b$ $\neq 1$, the corresponding boundary operators are 
again  linear combinations of the ones in Eqs. (\ref{scadu.1}), with the corresponding 
scaling dimensions in Eqs. (\ref{scadu.2}) proving their irrelevance. At variance, when 
e.g. $a=1$, the analog of the operators in Eqs. (\ref{scadu.1}) are given by 
linear combinations of the operators 

\begin{eqnarray}
 V_{a , (1 , b ) } &=& \Gamma_1 \Gamma_b e^{ - i \sqrt{\pi} [ \theta_1 ( 0 ) - \phi_b ( 0 ) ] } \nonumber \\
 V_{b , (1,b)} &=& \Gamma_1 \Gamma_b e^{ - i \sqrt{\pi} [ \theta_1 ( 0 ) + \phi_b ( 0 ) ] }
 \;\;\;\; , 
 \label{scadu.3}
\end{eqnarray}
\noindent
plus their Hermitean conjugates.
To compute the corresponding scaling dimensions, we have  to account for the Dirichlet boundary conditions on $\phi_1 ( 0 )$.
To do so, we get rid 
of the corresponding field in the Euclidean action, by emplying the ``reduced'' action $S_{{\rm Eff} , (1)} [ \{ \phi_b \} ]$, given by

\beq
S_{{\rm Eff} , (1)} [ \{ \phi_b \} ]= \frac{1}{2 \pi} \: \int \: d \Omega \: | \Omega | \tilde{\phi}^\dagger ( \Omega ) \:   \tilde{{\cal K}}_N  \:
\tilde{\phi} ( \Omega) 
\:\:\:\: , 
\label{scadi.11}
\eneq
\noindent
with 
$\tilde{\phi} ( \Omega ) =  \int \: d \tau \: e^{  i \Omega \tau } [ \phi_2 ( \tau ) ,    \ldots ,  \phi_{N} ( \tau ) ]^T$ and 
$ \tilde{{\cal K}}_N  = \frac{K_\rho + (N-1)K_\sigma}{N} {\bf I}_{N-1} + \frac{K_\rho - K_\sigma}{N} \tilde{I}_{N-1} $, and 
${\bf I}_N$ being the $N$-dimensional identity matrix and $\tilde{I}_N$ being the $N$-dimensional square matrix 
with all the entries equal to 1 but the ones at the diagonal, which are equal to 0. Accordingly, we 
find 

\beq
d_{a , (1,b)} = d_{b , (1,b)} = \frac{1}{2} \{ [  [ {\bf M}_N ]^T 
{\bf K}_N^{-1} {\bf M}_N  ]_{1, 1} \}^{-1} +  \{ [ \tilde{{\cal K}}_N ]^{-1} \}_{b,b}
\:\:\:\: . 
\label{scadu.4}
\eneq
\noindent
 By mathematical recursion, one may show that 

\beq
\tilde{{\cal K}_N}^{-1} = \left[ \frac{ (N-2)K_\rho + 2 K_\sigma }{ (N-1)K_\rho K_\sigma + K_\sigma^2} \right] {\bf I}_{N-1} - 
\left[  \frac{  K_\rho -    K_\sigma }{ (N-1)K_\rho K_\sigma + K_\sigma^2} \right]  \tilde{I}_{N-1}  
 \:\:\:\: , 
 \label{scadi.13}
\eneq
\noindent
which eventually leads to the final result  

\beq
d_{{\rm Normal}, (1,b)} = d_{{\rm Pair}, (1,b)} = \frac{1}{4} \left\{ \frac{1}{ d_{b , {\rm Disc}} } + 2 \frac{ (N-2)K_\rho + 2 K_\sigma }{ (N-1)K_\rho K_\sigma + K_\sigma^2}
\right\} 
 \:\:\:\: . 
 \label{scadi.14}
\eneq
\noindent
For $1/2 < K_\rho , K_\sigma < 1$ and for $U>0$ one obtains  $d_{{\rm Inter}, (1,b)} > 1$, thus showing the irrelevance of the corresponding operators.
Finally, we consider the   residual coupling to the MM. In 
this case, as discussed at length in Refs. \cite{alicea,giuaf_1}, though the residual coupling between, say, channel-2 and the 
Majorana mode seems to provide a relevant perturbation, in fact, it does not, due to the condition that the 
physical states must either be annihilated by $a_1$, or by $a_1^\dagger$ defined above. This makes an operator such as 
$2 i \bar{t}_2 \gamma \Gamma_2 \cos [ \sqrt{\pi} \phi_2 ( 0 ) ]$  become effective only to second order 
in $\bar{t}_2$, where it effectively behaves like an operator $ V_{2 , {\rm Res}} \propto \cos [2 \sqrt{\pi} \phi_2 ( 0 )]$, with 
scaling dimension $d_{{\rm Res} , 2 } = 4 \{ [ \tilde{{\cal K}}_N ]^{-1} \}_{2,2} = 
2 \frac{ (N-2)K_\rho + 2 K_\sigma }{ (N-1)K_\rho K_\sigma + K_\sigma^2} > 1$  for  $1/2 < K < 1$ and for $U>0$. Accordingly, 
this is an irrelevant operator, which leads us to conclude that, as long as $K_\sigma < 1$, the stable phase of the 
$N$-wire junction with a topological superconductor always corresponds to a $N_a = 1 , N_n = N-1$ fixed point. 

While our above analysis can in principle be readily extended to any $N_a \geq 2$, in the following we limit ourselves to 
the case $N_a = 2$ to show how, in this case, at least two relevant boundary operators emerge at the corresponding 
fixed point. Eventually, this leads to the conclusion that the corresponding fixed point is not stable, 
consistently with the result of Ref. \cite{giuaf_1} for $N=2$. The instability of fixed points with 
an $N_a \geq 3$ can eventually be inferred by means of similar arguments. Assuming $N_a = 2$, the key operators 
correspond to normal boundary backscattering/pairing involving channels-1 and -2. Within DEBC approach, they 
are readily recovered as a linear combination of the operators  $V_{a , (1 , 2 ) }$
and $V_{b , (1 , 2 ) }$, given by

\begin{eqnarray}
 V_{a , (1 , 2 ) } &=& \Gamma_1 \Gamma_2 e^{ - i \sqrt{\pi} [ \theta_1 ( 0 ) - \theta_2 ( 0 ) ] } \nonumber \\
 V_{b , (1,2)} &=& \Gamma_1 \Gamma_b e^{ - i \sqrt{\pi} [ \theta_1 ( 0 ) + \theta_2 ( 0 ) ] }
 \;\;\;\; , 
 \label{scadu.5}
\end{eqnarray}
\noindent
plus their Hermitean conjugates. The corresponding scaling dimensions are accordingly 
given by 

\begin{eqnarray}
 d_{a , (1,2)} &=&  \{ [  [ {\bf M}_N ]^T 
{\bf K}_N^{-1} {\bf M}_N  ]_{1, 1} \}^{-1} -  \{ [  [ {\bf M}_N ]^T 
{\bf K}_N^{-1} {\bf M}_N  ]_{1, 2} \}^{-1} = \frac{N}{2 K_\rho^{-1} + (N -2 ) K_\sigma^{-1} } \leq K_\rho \nonumber \\
d_{b , (1,2)} &=&  \{ [  [ {\bf M}_N ]^T 
{\bf K}_N^{-1} {\bf M}_N  ]_{1, 1} \}^{-1} +  \{ [  [ {\bf M}_N ]^T 
{\bf K}_N^{-1} {\bf M}_N  ]_{1, 2} \}^{-1} = K_\sigma 
\;\:\:\: . 
\label{scadu.6}
\end{eqnarray}
\noindent
Both $d_{a , (1,2)}$ and $d_{b , (1,2)}$ are $<1$, implying that boundary operators encoding 
normal inter-channel backscattering and pairing both correspond to relevant boundary 
interactions. Accordingly, we conclude that the $N_a=2$ fixed point is unstable and, 
by means of an obvious extension of the argument, that any fixed point with $N_a \geq 3$ is 
unstable, as well. In conclusion, we see that, also for $N > 2$, the only stable fixed points
in the phase diagram of the $N$-wire junction are the $N$-ones with $N_a = 1$. As those are 
all equivalent to each other, there must be intermediate FCFP's separating the corresponding phases.
 
\begin{figure}
 \center
\includegraphics*[width=0.4 \linewidth]{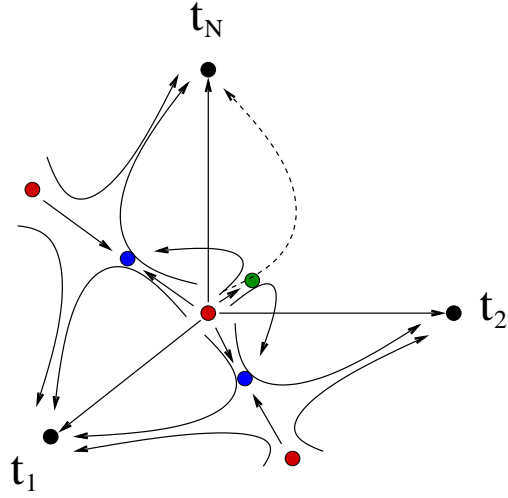}
\caption{  Sketch of the typical renormalization group flow diagram of a junction of $N$ interacting quantum wires with 
a topological superconductor drawn for $U>0$ and $  K_\sigma <1$ (see text).  In this range of 
parameters, we see that  the disconnected fixed point, as well as the fixed points with $N_a > 1$,   are unstable against turning on a nonzero boundary interaction. 
In the presence of a symmetry between two, or more than two, boundary couplings, the renormalization 
group trajectories flow towards FCFP's (drawn in green and blue). Eventually, when one boundary coupling
takes over all the others, the system flows towards one of the $N_a=1$ fixed points.} 
\label{flodia_1}
\end{figure}

FCFP's have been argued to potentially host ``decoherence-frustrated'' phases
with reduced  
decoherence effects in the boundary quantum degrees of freedom 
\cite{novais_1,novais_2,giusoY}. In our case,  FCFP's are expected to emerge at the 
bifurcations between RG trajectories leading to any one of the stable 
$N_a = 1$ fixed points.  While we are not able to provide an exact conformal boundary field theory description 
of the FCFP's, we can still access them   in the $\epsilon$-expansion framework. Indeed, they 
emerge as  nontrivial zeroes of the  $\beta$-functions at the right-hand side of Eqs. (\ref{1w.26}), 
with the corresponding boundary couplings satisfying  the equations 

\beq
\bar{t}_{a  }  \: \{ \epsilon - {\cal F} [ \nu ] \: \sum_{b \neq a} \bar{t}_{b }^2 \} = 0 
\:\:\:\: . 
\label{ntr.1}
\eneq
\noindent 
By inspection, we see that there is only a solution with all the $\bar{t}_a^* \neq 0$ (FCFP$_N$), corresponding to 

\beq
\bar{t}_1^* = \ldots = \bar{t}_N^* = t_* ( N )  = \sqrt{  \frac{\epsilon}{ ( N - 1 ) {\cal F} [ \nu ] } } 
\:\:\:\: . 
\label{ntr.2}
\eneq
\noindent 
Next (assuming $N \geq 3$), it is possible to have nontrivial solutions in which one 
$\bar{t}_a^*=0$, all the others being $\neq 0$ (FCFP$_{N-1}$). These are given by 

\begin{eqnarray}
&& \bar{t}_a^* = 0 \nonumber \\
&& \bar{t}_b^* = t_* (N-1) = \sqrt{ \frac{\epsilon}{ ( N - 2) {\cal F} [ \nu ] } } \:\: , \: (b \neq a )
\:\:\:\: . 
\label{ntr.3}
\end{eqnarray}
\noindent
Going ahead (assuming $N \geq 4$), we find $N (N-1) / 2 $ FCFP's in which two $\bar{t}_a^* = 0$, with all the 
others being $\neq 0$, etc. Remarkably, Eqs. (\ref{ntr.1}) do not exhibit solutions with just one 
$t_a^* \neq 0$ and all the others being $=0$, which gives us one more insight about the possible 
topology of the boundary phase diagram of the junction. To do so, we first of all note that, if all the $N$  
bare couplings are equal to each other, then the symmetry among them is not broken along the RG 
flow generated by Eqs. (\ref{1w.26}). In this case, we therefore expect the junction to flow 
towards the FCFP$_N$.  At variance,  a slight breaking of the symmetry between the couplings does,  in fact,
take the system out of the FCFP$_N$. To show this, let us assume that, in the 
vicinities of the  FCFP$_N$, the couplings 
are set so that $\bar{t}_{1, L} = t_* ( N ) - \rho$, while $\bar{t}_{a , L} = t_* ( N ) + \sigma$ for $a = 2 ,\ldots , N$, with 
$0 < \rho \ll 1$ and $0 < \sigma \ll 1$. On linearizing Eqs. (\ref{1w.26}), one obtains 

\begin{eqnarray}
 \frac{d \rho ( l) }{d l } &=& 2 \epsilon \sigma ( l ) \nonumber \\
 \frac{d \sigma ( l) }{d l } &=& \frac{2 \epsilon}{N-1} \: \rho ( l ) - 2 \epsilon \sigma ( l ) 
 \:\:\:\: . 
 \label{ntr.4}
\end{eqnarray}
\noindent
Once integrated, setting $\lambda_1 = - \epsilon + \epsilon \sqrt{1 + \frac{4}{N-1}}$, 
$\lambda_2 = - \epsilon - \epsilon \sqrt{1 + \frac{4}{N-1}}$, one finds

\begin{eqnarray}
 \rho ( l ) &=& e^{ \lambda_1 l } \: \left\{ \frac{1 + \sqrt{1 + \frac{4}{N-1}}}{2 \sqrt{1 + \frac{4}{N-1}}} \rho ( 0 ) + 
 \frac{1 }{  \sqrt{1 + \frac{4}{N-1}}} \sigma ( 0 ) \right\} + 
 e^{ \lambda_2 l} \: \left\{ \frac{- 1 + \sqrt{1 + \frac{4}{N-1}}}{2 \sqrt{1 + \frac{4}{N-1}}} \rho ( 0 ) - 
 \frac{1 }{  \sqrt{1 + \frac{4}{N-1}}} \sigma ( 0 ) \right\} 
 \label{ntr.5} \\
 \sigma ( l ) &=& \frac{ \lambda_1  e^{ \lambda_1 l } }{ 2 \epsilon  }  \: \left\{ \frac{1 + \sqrt{1 + \frac{4}{N-1}}}{2 \sqrt{1 + \frac{4}{N-1}}} \rho ( 0 ) + 
 \frac{1 }{  \sqrt{1 + \frac{4}{N-1}}} \sigma ( 0 ) \right\} + 
\frac{ \lambda_2 e^{ \lambda_2 l} }{ 2 \epsilon  } \: \left\{ \frac{- 1 + \sqrt{1 + \frac{4}{N-1}}}{2 \sqrt{1 + \frac{4}{N-1}}} \rho ( 0 ) - 
 \frac{1 }{  \sqrt{1 + \frac{4}{N-1}}} \sigma ( 0 ) \right\} \nonumber
 \:\:\:\: . 
\end{eqnarray}
\noindent
From Eqs. (\ref{ntr.5}) one infers that the RG trajectories flow towards the FCFP$_{N-1}$ fixed point, obtained  by decreasing $\bar{t}_1$ 
and symmetrically increasing all the other couplings. Conversely,   if 
  both $\rho ( 0 )$ and $\sigma ( 0 )$ are $<0$, then a direct flow to 
an $N_a=1$ fixed point is recovered. Now, the above analysis can be 
straightforwardly iterated, eventually generalizing to the $N$-wire junction the RG flow diagram derived in Ref. \cite{giuaf_1}. 
In Fig.\ref{flodia_1}, we draw a sketch of the minimal flow diagram for the junction. We see that,
for $1/2 < K < 1$, the RG trajectories flow away from the disconnected fixed point, either towards 
one of the $N_a=1$ fixed points, or towards some FCFP, depending on the symmetry between the 
initial values of the boundary couplings. Eventually, reducing the symmetry between the boundary couplings implies a 
flow between different FCFP's, till, when all the symmetries are removed, the system 
flows towards one of the maximally stable $N_a = 1$ fixed points. 

We now compute the $g$-function at the various fixed points of the $N$-wire junction, 
eventually arguing that   the corresponding results are consistent with the expected topology of the phase diagram only provided one 
properly accounts for the real fermionic modes, which is at the heart of our approach.

\noindent
 
\subsection{Impurity entropy at the fixed points of a junction between $N$ quantum wires and a topological 
superconductor} 
\label{impuN}

  To begin with, let us consider the disconnected fixed point, 
corresponding to type $N$ CIBC's at both boundaries in each channel. As a general 
remark we note that, in the junction we consider here, we have 2MM's (one at each boundary), as 
well as   $N$ KF's, for a total of $2+N$ real fermionic modes. Consistent with the discussion of 
section \ref{pha_1},   if $N$ is odd, we introduce the auxiliary wire,  with Luttinger parameters 
$\bar{u} , \bar{K}$, providing 
an extra Klein factor $\bar{\Gamma}$, so to make the total number of real fermionic zero mode operators  even.
From Eqs. (\ref{lutw.2},\ref{gd1.3},\ref{gd.4})  and taking into 
account the degeneracy factor associated with zero-mode real fermionic operators, for $N$ even, we eventually find 
for the partition function at the disconnected fixed point the result

\beq
 {\cal Z}_{\rm Disc} = 2^{1 + \frac{N}{2}}\: [ \eta ( q_c ) ]^{-1}  \: [ \eta ( q_r )]^{- ( N - 1 ) } 
\: {\cal Z}_{{\rm Disc} , e, 0} 
\;\;\;\; , 
\label{gd.5}
\eneq
with $q_{c , r} = e^{- \frac{  u_{c , r}   \beta \pi}{ \ell} }$, and

\beq
{\cal Z}_{{\rm Disc} , e, 0} =  \sum_{\{ m_{1 , \phi}  , \ldots , m_{N , \phi}  \} \in {\bf Z}} \: \exp \; \left\{ - \frac{\pi \beta}{2 \ell} \: 
\vec{m}_\phi^T {\bf B}_{N , e} \vec{m}_\phi  
\right\}
\:\:\:\: ,
\label{gd.10}
\eneq
\noindent
$\vec{m}_\phi^T = ( m_{1 , \phi } , \ldots , m_{N , \phi} )$,  and  the matrix ${\bf B}_{N,e}$ given by 

\beq
{\bf B}_{N,e} = [ {\bf M}_N ]^T  \:  \left[ \begin{array}{cccc} \frac{u_\rho}{K_\rho} & 0 & \ldots & 0 \\ 
0 & \frac{u_\sigma}{K_\sigma} & \ldots & 0 \\ \vdots & \vdots & \vdots & \vdots \\ 
0 & 0 & \ldots & \frac{u_\sigma}{K_\sigma} \end{array} \right] \: {\bf M}_N 
\:\:\:\: . 
\label{gd.11}
\eneq
\noindent
At variance, for $N$ odd, one obtains

\beq
{\cal Z}_{\rm Disc} =2^{1 + \frac{N+1}{2}}\: [\eta (\bar{q}) ]^{-1} \: [ \eta ( q_c ) ]^{-1}  \: [ \eta ( q_r )]^{- ( N - 1 ) } 
\: {\cal Z}_{{\rm Disc} ,o, 0} 
\;\;\;\; , 
\label{gd.12}
\eneq
with  $\bar{q} = e^{- \frac{ \bar{u}  \beta \pi}{ \ell} }$, and

\beq
{\cal Z}_{{\rm Disc} , o , 0} =  \sum_{\{ m_{1 , \phi}  , \ldots , m_{N , \phi} . m_{N+1 , \phi}  \} \in {\bf Z}} \: 
\exp \; \left\{ - \frac{\pi \beta}{2 \ell} \: \vec{m}_\phi^T  {\bf B}_{N , o} 
\vec{m}_\phi \right\}
\:\:\:\: ,
\label{gd.13}
\eneq
\noindent
and  the matrix ${\bf B}_{N,o}$ constructed from ${\bf B}_{N , e}$ by adding one row and one column with all the elements $=0$
except $[ {\bf B}_{N , o} ]_{N+1 , N+1 } =  \frac{\bar{u}}{\bar{K}}$.  To extract the $g$-function, we have to 
consider the partition function in   the $\ell \to \infty$ limit at fixed $\beta$, which can be 
readily done by employing the Poisson summation formula in the form 
presented in Eq. (\ref{math.16}). As a result, we obtain 

\begin{eqnarray}
 &&{\cal Z}_{{\rm Disc} } \longrightarrow_{\ell \to \infty} \:  2^{1 + \frac{N}{2}}\: e^{ \frac{\pi \ell}{6 \beta u_\rho} + \frac{\pi \ell (N-1) }{6 \beta u_\sigma } } \:
 [ K_\rho K_\sigma^{(N-1)} ]^\frac{1}{2} \;\; , \; ({\rm N \; even}) \nonumber \\
  && {\cal Z}_{{\rm Disc} } \longrightarrow_{\ell \to \infty} \: 2^{1 + \frac{N+1}{2}}\:
   e^{ \frac{\pi \ell}{6 \beta u_\rho} + \frac{\pi \ell (N-1) }{6 \beta u_\sigma } + \frac{\pi \ell}{6 \beta \bar{u} } } \:
 [ \bar{K} K_\rho K_\sigma^{(N-1)} ]^\frac{1}{2} \;\; , \; ({\rm N \; odd})
 \:\:\:\: , 
 \label{gd.14}
\end{eqnarray}
\noindent
from which we eventually obtain  for   $g_{\rm Disc}$ the result 

\begin{eqnarray}
 && g_{\rm Disc} = 2^{ \frac{N + 2}{4} } \:   [ K_\rho K_\sigma^{(N-1)} ]^\frac{1}{4} \;\; , \; ({\rm N \; even}) \nonumber \\
  && g_{\rm Disc} = 2^{ \frac{N + 3}{4} } \:  [ \bar{K}  K_\rho K_\sigma^{(N-1)} ]^\frac{1}{4} \;\; , \;({\rm N \; odd})
 \:\:\:\: . 
 \label{gd.15}
\end{eqnarray}
\noindent
To generalize Eqs. (\ref{gd.15}) to a fixed point with type $A$ CIBC's in the first $N_a$ channels and  
type $N$ in the remaining $N_n$ ones, we refer to Eq. (\ref{gd.5a}) of   \ref{bomo}
for the spectrum of the zero-mode operators. Accordingly, for $N$ even, we obtain that 
the zero-mode contribution to the total partition function is given by 

\beq
{\cal Z}_{(N_a , N_n) , e, 0} =  \sum_{\{ m_{a , \theta }  ,  m_{b , \phi}   \} \in {\bf Z}} \: 
\exp \; \left\{ - \frac{\pi \beta}{2 \ell} \:\left[  \vec{m}_\theta^T  \tilde{\bf B}_{N , e}
\vec{m}_\theta + \vec{m}_\phi^T  {\bf B}_{N , e}
\vec{m}_\phi  \right] 
\right\}
\:\:\:\: ,
\label{gd.a1}
\eneq
\noindent
with $\vec{m}_\theta^T = ( m_{1 , \theta} , \ldots , m_{N_a , \theta} , 0 , \ldots , 0 )$, 
$\vec{m}_\phi^T = ( 0 , \ldots , 0 , m_{N_a + 1 , \phi} , \ldots , m_{N , \phi} )$, and 
the matrix  $\tilde{\bf B}_{N , e}$ obtained from ${\bf B}_{N , e}$ in Eq. (\ref{gd.11}) by 
substituting $K_{c , r}^{-1}$ with $K_{c , r }$, respectively. Similarly, for $N$ odd, 
one obtains

\beq
{\cal Z}_{(N_a , N_n) , o, 0} =  \sum_{\{ m_{a , \theta }  ,  m_{b , \phi}   \} \in {\bf Z}} \: 
\exp \; \left\{ - \frac{\pi \beta}{2 \ell} \:\left[  \vec{m}_\theta^T  \tilde{\bf B}_{N , o}
\vec{m}_\theta + \vec{m}_\phi^T  {\bf B}_{N , o}
\vec{m}_\phi  \right] 
\right\}
\:\:\:\: ,
\label{gd.a2}
\eneq
\noindent
with $\tilde{\bf B}_{N , o}$ constructed from ${\bf B}_{N , o}$ by means of 
the same criterion used to build $\tilde{\bf B}_{N , e}$ from ${\bf B}_{N , e}$. 
Postponing, for the time being, the calculation of the degeneracy factors due to 
the zero-mode real fermion operators, $\delta_{e , o} [ N_a , N_n ]$, we now 
employ the approach of   \ref{duapoi} to compute the $g$-function 
from the results of Eqs. (\ref{gd.a1},\ref{gd.a2}). Taking into account that 
the contribution to the total partition function due to the oscillator modes 
does not depend on the specific CIBC's in the various channels, we 
obtain

\begin{eqnarray}
 &&{\cal Z}_{ N_a , N_n } \longrightarrow_{\ell \to \infty} \: \delta_{e} [ N_a , N_n] \: e^{ \frac{\pi \ell}{6 \beta u_\rho} + \frac{\pi \ell (N-1) }{6 \beta u_\sigma } } \:
 [  K_\sigma^{(N_n - N_a-1)}K_\rho^{-1} K^2 ]^\frac{1}{2} \: \left[ 1 + \frac{(N_n-1)K U}{\pi u} \right]^{-\frac{1}{2}} \:  ,  ({\rm N \; even}) 
 \label{gd.a3} \\
  && {\cal Z}_{ N_a , N_n } \longrightarrow_{\ell \to \infty}  \: \delta_{o} [ N_a , N_n] \:
   e^{ \frac{\pi \ell}{6 \beta u_\rho} + \frac{\pi \ell (N-1) }{6 \beta u_\sigma } + \frac{\pi \ell}{6 \beta \bar{u} } } \:
[ \bar{K}  K_\sigma^{(N_n - N_a-1)}K_\rho^{-1} K^2 ]^\frac{1}{2} \: \left[ 1 + \frac{(N_n-1)K U}{\pi u} \right]^{-\frac{1}{2}}  \; ,  ({\rm N \; odd})\nonumber
 \:\:\:\: . 
\end{eqnarray}
\noindent
In   \ref{stabl} we discuss in detail the calculation of   $\delta_{e , o } [ N_a , N_n ]$. Here, 
we just quote the final result for the $g$-function, which is

\begin{eqnarray}
  g_{ N_a , N_n } &=&  2^{\frac{3 N_a + N_n - 2}{4}   } \: 
 [  K_\sigma^{(N_n - N_a-1)}K_\rho^{-1} K^2 ]^\frac{1}{4} \: \left[ 1 + \frac{(N_n-1)K U}{\pi u} \right]^{-\frac{1}{4}} \;\; , \; ({\rm N \; even}) \nonumber \\
   g_{ N_a , N_n } &=& 2^{\frac{3 N_a + N_n -1}{2}  }\:
[ \bar{K}  K_\sigma^{(N_n - N_a-1)}K_\rho^{-1} K^2 ]^\frac{1}{4} \: \left[ 1 + \frac{(N_n-1)K U}{\pi u} \right]^{-\frac{1}{4}}  \;\; , \; ({\rm N \; odd})
 \:\:\:\: . 
 \label{gd.a4}
\end{eqnarray}
\noindent
On normalizing $g_{N_a , N_n}$ to $g_{\rm Disc}$, we obtain the ratio 

\beq
\rho_{N_a , N_n} = \frac{g_{N_a , N_n} }{g_{\rm Disc}} = 
\left[ \frac{2^{\frac{N_a}{2} - 1} }{ K_\rho^\frac{1}{2} K_\sigma^\frac{N_a}{2}} \right] 
\: \left\{ \frac{K^\frac{1}{2}}{\left[ 1 +  \frac{(N_n-1)K U}{\pi u} \right]^\frac{1}{4}} \right\}
\:\:\:\: , 
\label{gd.a5}
\eneq
\noindent
which gives back the result of Eq. (\ref{pd.3}) for $N_a = N_n = 1$ and the results of 
Eq. (\ref{pd.4}) for $N_a = 2, N_n = 0$. Besides the consistency check, a first important 
result is that one obtains 

\beq
\rho_{1,N-1} = \sqrt{ \frac{K}{2 K_\sigma K_\rho} } \left[ 1 +  \frac{(N-2)K U}{\pi u} \right]^{-\frac{1}{4}} 
= \frac{1}{\sqrt{2K}} \: \left\{ 1 - \frac{(N-1) \left( \frac{K U}{\pi u } \right)^2}{\left[ 1 +  \frac{(N-2)K U}{\pi u} \right] } \right\}^\frac{1}{4}
< 1
\:\:\:\: , 
\label{gd.a6}
\eneq
\noindent
as well as 

\beq 
\frac{\rho_{N_a+1,N_n - 1}}{\rho_{N_a , N_n}} = \sqrt{\frac{2}{K_\sigma}} \: \left[ \frac{1 + \frac{(N_n-1) K U}{\pi u}}{1 + \frac{(N_n-2) K U}{\pi u}} \right]^\frac{1}{4}
\;\;\;\; , 
\label{gd.a7}
\eneq
\noindent
which is $<1$ for $K_\sigma < 1$. Thus, the systematic calculation of the $g$-function at fixed points with given CIBC's  
ultimately confirms the phase diagram emerging from the perturbative RG approach combined with DEBC method. There
are $N$ equivalent stable fixed points, corresponding to $N_a = 1 , N_n = N-1$. These are separated by 
FCFP's that are expected to lie   along specific symmetry line in the boundary parameter space \cite{giuaf_1}. 
To compute the $g$-function at the FCFP's, we employ   the $\epsilon$-expansion method   
discussed in Ref. \cite{giuaf_1}. Specifically, we assume that 
$d_b = 1 - \epsilon$, with $0 < \epsilon \ll 1$ and eventually find that   the FCFP's are  all located at values of the boundary parameters $t_* \propto 
( {\cal F} ( 2 - K_\sigma^{-1} ) )^{-\frac{1}{2} }$, with the function ${\cal F}$ defined in Eq. (B.41) of 
Ref. \cite{giuaf_1} and reviewed here, in   \ref{epsilon_expansion}. In general, letting $M (\leq N)$ be the number 
of finite couplings $t_{* , M}$ at a FCFP,  we find $t_{* , M} = \sqrt{\frac{\epsilon}{(M - 1) {\cal F} ( \nu )}}$.
To proceed with the calculation of the corresponding value of the $g$-function, $g_{{\rm FCFP} , M}$,  we 
 go through exactly the same derivation of appendix G of Ref. \cite{giuaf_1}. As a 
 result, to leading order in the $t_{* ,M}$, we find    

\beq
g_{{\rm FCFP} , M} = g_{\rm Disc} \: \{ 1 - 2  \pi^2  M t_{* , M}^4 \} = g_{\rm Disc} \: \left\{ 1 - \frac{ 2  \pi^2 M \epsilon^2}{ (M-1) {\cal F} ( \nu ) }  \right\} 
\:\:\:\: . 
\label{gnt.3}
\eneq
\noindent
A remarkable consequence of Eq. (\ref{gnt.3}) is that, since, given two integers $M , M' $ both $\leq N$, we find 
$\frac{M}{M-1} \geq \frac{M'}{M' - 1}$, provided $M \leq M'$, $g_{{\rm FCFP} , M} / g_{{\rm FCFP} , M'}$ is $>1$ ($<1$) if $M > M'$ ($M< M'$), that 
is, if an RG trajectory takes place between two FCFP's, it must take the 
system towards the fixed point with the lower value of $M$, consistently with the result of 
Ref. \cite{giuaf_1} in the case $N=3, U=0$. 

\section{Conclusions}
\label{concl}
 
We discuss the method to  consistently compute the $g$-function at the boundary fixed points of the phase 
diagram of junctions between interacting quantum wires and/or topological superconductors, 
involving real fermionic modes in the corresponding boundary Hamiltonian (localized Majorana modes and/or Klein factors).
We show that, in doing the calculation,  one has to treat {\it all}  
of the real fermionic degrees of freedom  on the same footing, which is 
apparently a version of the Majorana-Klein hybridization phenomenon, which requires KF's to be 
considered as actual ``physical'' degrees of freedom in exactly the same way as MM's, when 
describing   junctions between interacting quantum wires 
and topological superconductors \cite{beritopo}. Incidentally, in our procedure for computing $g$, we also 
introduced a means to avoid ambiguities in counting the degrees of freedom associated with an odd total number 
of real fermions, by introducing an auxiliary wire, fully disconnected from the junction. The additional wire  has the 
 effect of providing an additional KF, which makes the total number of real fermionic modes 
always even. While  affecting the  value of $g$ at a specific fixed points, our procedure eventually 
gives back the right value of the {\it ratio} between $g$ computed at two different points.

By comparing the results  of perturbative RG approach and  DEBC-method 
with the  explicit calculation of the $g$-function and the implications of the 
$g$-theorem, we have mapped out a remarkable correspondence between the $N=2$ junction 
and the non-${\bf Z}_3$-symmetric Y3J, for suitably chosen values of 
the system parameters. In particular, we have employed the correspondence 
to recover informations about the phase diagram of the former system from known results about the 
phase diagram of the latter, and vice versa. Concerning the $N=2$ junction, we have shown  
  that the condition $K_\sigma > 1$ is enough to reverse its phase diagram  
 with respect to  the result of   Ref. \cite{giuaf_1} for $K_\sigma < 1$. Increasing $K_\sigma $ to values $>1$, we proved that the 
FCFP corresponds to the true stable phase of the system and is  identified with 
the  $M$-FCFP of the Y3J and, eventually, that, for large enough values of  $K_\sigma$, the 
$N=2$ junction undergoes a phase transition  to a phase with   perfect healing of the junction between wires-1 and -2,
with the MM decoupled from the two wires. Conversely, for the Y3J, 
we demonstrated the emergence of  a ``planar'' FCFP's (that is, with one of the 
boundary coupling strengths set to 0), which is a novel feature, so far not discussed for such systems.
 In addition, we were able to infer the phase diagram of the $N=2$-junction at zero inter-wire interaction 
and for $K>1$ in each wire, a regime which was not discussed in Ref. \cite{giuaf_1}. 

Despite being effective in deriving a number of results on the phase diagram of both 
systems, the correspondence between the $N=2$ junction 
and the non-${\bf Z}_3$-symmetric Y3J still presents a number of ``critical'' issues, 
which will have to be further analyzed, possibly with the help of a numerical approach 
to the problem, such as the one employed in Ref. \cite{rahmani}. In particular, 
issues related to our work that deserve to be further analyzed are:

- The failure of the $\epsilon$-expansion method to provide quantitative results 
about the FCFP, when applied to the Y3J in the ${\bf Z}_3$-symmetric  limit. As we show in  \ref{failure}, the 
perturbative $\beta$-function  for the boundary running coupling  in the Y3J contain terms that 
are all $\propto \epsilon$, which makes the perturbative RG approach not reliable for 
extracting informations about the FCFP, at variance to what happens in the $N=2$-junction 
\cite{giuaf_1}. (Yet, it must be stressed that, while not applicable in general, the 
$\epsilon$-expansion method works fine for the Y3J, as well, in some range of values of
the system's parameters, such as the one considered in section \ref{lul_0}, leading to 
Eqs. (\ref{puccy.y1bis},\ref{puccy.y2bis}) .)

 - The nature of the   FCFP that our correspondence predicts in the $N=2$ junction 
with no inter-wire interaction, $1<K<2$ in each wire, and for symmetric boundary couplings. 
In particular, it would be extremely interesting to figure out whether there is still just one  FCFP  and whether it  is continuously 
connected to the one we found in Ref. \cite{giuaf_1} for $1/2 < K < 1$, or if there is more 
than one FCFP's, possibly of some intrinsically different nature;
 
 - Whether the fact that the  correspondence extends to the  $g$-function, 
despite the fact that the physical nature 
of the real modes  in the Y3J and in the $N=2$ junction are fundamentally  different, 
is just an accident, or can apply, possibly in some different form, in similar systems.

Apart from the ones listed above, from our results, there are a number of issues that we left over and should be properly addressed, such as the 
relation between the validity of the $g$-theorem in the presence of real fermionic modes and the conservation 
of the total fermion parity, the explicit calculation of the $g$-functions at FCFP's where the 
$\epsilon$-expansion method fails, or the extension of our derivation to systems such as the ``Majorana-Kondo devices",
which at the same time encompass Majorana and (topological) Kondo physics \cite{erikson}. These topics are
 outside of the range of this work and we plan to address them in  forthcoming publications.

\vspace{0.5cm}

{\bf Acknowledgements} 

\vspace{0.5cm}

We thank  P. Sodano for valuable discussions. The research of I.A. is supported by NSERC Discovery Grant
04033-2016 and the Canadian Institute for Advanced Research.

 \appendix

\section{Lattice Hamiltonian, boundary conditions and mode expansion of the bosonic fields}
\label{bomo}

In this appendix,  we concisely review the bosonization procedure for the interacting lattice fermionic 
Hamiltonian for the junction between $N$ QW's and a TS. Following the bosonization procedure, 
we present  the mode expansion for the bosonic fields 
entering the corresponding TLL Hamiltonian  at given type $N$, or type $A$, 
boundary conditions at $x=0$ and at $x = \ell$. 

The lattice model  Hamiltonian for the $N$-wire junction is given by  
$H_{N , {\rm Fer}} = H_{0 , {\rm Fer} , N } + H_{I , {\rm Fer} , N} + H_{b , {\rm Fer} , N}$, with 

\begin{eqnarray}
 H_{0 , {\rm Fer}, N}  &=& \sum_{a = 1}^N \{ - J \sum_{ j = 1}^{\ell - 2 } \{ c_{j , a}^\dagger c_{j + 1 , a}   + c_{ j + 1 , a}^\dagger c_{j , a} \}
 - \mu \sum_{j = 1}^{\ell - 1} c_{ j , a}^\dagger c_{j , a} \}  \nonumber \\
  H_{I , {\rm Fer}, N}  &=& V \sum_{a = 1}^N \sum_{ j = 1}^{\ell - 2 } \left( c_{j , a}^\dagger c_{j , a } - \frac{1}{2} \right) 
   \left( c_{j +1  , a}^\dagger c_{j +1  , a } - \frac{1}{2} \right) + 
    U \sum_{a \neq b =  1}^N  \sum_{ j = 1}^{\ell - 1 } \left( c_{j , a}^\dagger c_{j , a } - \frac{1}{2} \right) 
   \left( c_{j   , b}^\dagger c_{j    , b } - \frac{1}{2} \right) \nonumber \\
  H_{b , {\rm Fer} , N}  &=&  -  \sum_{a = 1}^N t_{a } \gamma_L \: \{ c_{a , 1} - c_{a , 1}^\dagger \} 
  - i  \sum_{a = 1}^N t_{ a  }  \gamma_R  \: \{ c_{a , \ell - 1} +  c_{a , \ell - 1}^\dagger \} 
  \equiv H_{b , {\rm Fer} , N}^{(1)} + H_{b , {\rm Fer} , N}^{(\ell - 1)} 
   \:\:\:\: , 
   \label{bi.3}
\end{eqnarray}
\noindent
with $\gamma_L , \gamma_R$ being the localized MM's at the junctions between the QW's and the TS
and the $t_a$'s being all real and positive, as a possible phase can be always reabsorbed into an appropriate redefinition of 
the lattice fields. In the absence of interaction, retaining only low-energy, long-wavelength fermionic modes
allows for expanding the lattice fermion operators as 
$c_{j , a} \sim  \{ e^{ i k_f j} \psi_{R , a} ( x ) + e^{ - i k_f j } \psi_{L , a} ( x ) \}$,  
with the Fermi momentum $\pm k_f =  {\rm arccos} \left( - \frac{\mu}{2 J} \right)$. Bosonizing the 
chiral fermionic fields requires   introducing  $N$ pairs of canonically conjugate fields 
$\{ \phi_{R , a } ( x ) , \phi_{L , a } ( x ) \}$ ($a = 1 , \ldots , N$). In the noninteracting limit, they can   
be expressed in terms of $N$ pairs of chiral bosonic fields as 

\begin{eqnarray}
 \psi_{R , a } ( x ) &=& \Gamma_a \: e^{ i \sqrt{4 \pi} \phi_{R ,  a } ( x ) }    \nonumber \\
  \psi_{L  , a } ( x ) &=& \Gamma_a \: e^{ i \sqrt{4 \pi} \phi_{L ,  a } ( x ) } 
  \:\:\:\: . 
  \label{bi.9}
\end{eqnarray}
\noindent
with the chiral  bosonic fields  satisfying  the algebra 

\begin{eqnarray}
&& [ \phi_{R , a } ( x ) , \phi_{R , a} (x' ) ]  = -  [ \phi_{L , a } ( x ) , \phi_{L , a} (x' ) ] = \frac{i}{4} \epsilon ( x - x' ) \nonumber \\
&& [ \phi_{R , a } ( x ) , \phi_{L , a} ( x' ) ] =  - [ \phi_{L , a } ( x' ) , \phi_{R , a } ( x ) ]  = \frac{i}{4}  
\:\:\:\: , 
\label{bi.8}
\end{eqnarray}
\noindent
with all the other commutators equal to 0, and the 
$N$ KF's defined so that  $\{ \Gamma_a , \Gamma_{a'}  \} = 2 \delta_{a, a'}$, and,  if $H_{b , {\rm Fer} , N}$ contains localized MM's 
$\gamma_1 , \ldots , \gamma_M$,  requiring that  $\{ \gamma_j , \Gamma_a \} = 0 $, $\forall a, j$. In the interacting 
case, Eqs. (\ref{bi.9}) are replaced by

\begin{eqnarray}
 \psi_{R , a } ( x ) &=& \Gamma_a \: e^{ i \sqrt{  \pi} [ \phi_a ( x ) + \theta_a ( x ) ] }    \nonumber \\
  \psi_{L  , a } ( x ) &=& \Gamma_a \: e^{ i \sqrt{ \pi} [ \phi_a ( x ) - \theta_a ( x ) ]  } 
  \:\:\:\: . 
  \label{bi.9x}
\end{eqnarray}
\noindent
 with the (canonically conjugate) fields $ \{ \phi_a ( x ) , \theta_a ( x ) \}$ described by the 
 bulk Hamiltonian    $H_{N ,  {\rm B}} = H_{N , B ,  0 } + H_{N , B , {\rm Inter}}$ and 

\begin{eqnarray}
  H_{N , B ,  0 } &=&  \frac{u}{2}  \: \int_0^\ell  \: d x \: \sum_{a = 1}^N \: 
 [ K ( \partial_x \phi_a ( x ))^2 + K^{-1}  ( \partial_x \theta_a ( x ))^2 ] \nonumber \\
  H_{N , B , {\rm Inter}} &=&  \sum_{ a \neq b = 1}^N \frac{U}{ 2 \pi}  \: \int_0^\ell \: d x
  \: [ ( \partial_x \theta_a ( x )) ( \partial_x \theta_b ( x )) ]
  \:\:\:\: , 
  \label{lutw.1}
\end{eqnarray}
\noindent
with the Luttinger parameter $K$ and the plasmon velocity $u$ determined by the intra-wire interaction
$V$ and by the Fermi velocity in the wires, $v_f$.
By means of an appropriate orthogonal transformation,  $H_{N , {\rm B}}$ can be separated into  
independent terms by rotating to the basis of the center-of-mass fields $\Phi ( x ) , \Theta ( x )$ and 
the relative fields $\varphi_1 (x ) , \ldots , \varphi_{N - 1} ( x ) $ and $\vartheta_1 ( x ) , \ldots , 
\vartheta_{N-1} ( x )$, defined as

\beq
\left[ \begin{array}{c}
\Phi ( x ) \\ \varphi_1 ( x ) \\ \vdots \\ \varphi_{N-1} ( x )         
       \end{array} \right] = {\bf M}_N  \: \left[ \begin{array}{c}
\phi_1 ( x ) \\ \phi_2 ( x ) \\ \vdots \\ \phi_N ( x )                                                 
                                               \end{array} \right] 
\;\;\; , \;\;
\left[ \begin{array}{c}
\Theta ( x ) \\ \vartheta_1 ( x ) \\ \vdots \\ \vartheta_{N-1} ( x )         
       \end{array} \right] = {\bf M}_N \: \left[ \begin{array}{c}
\theta_1 ( x ) \\ \theta_2 ( x ) \\ \vdots \\ \theta_N ( x )                                                 
                                               \end{array} \right] 
\:\:\:\: , 
\label{1w.8bis}
\eneq
\noindent
with the matrix ${\bf M}_N$ only depending on $N$ and given by 

\beq
{\bf M}_N = \left[ \begin{array}{ccccc}
\frac{1}{\sqrt{N}} & \frac{1}{\sqrt{N}} & \frac{1}{\sqrt{N}} &  \ldots  & \frac{1}{\sqrt{N}} \\
\frac{1}{\sqrt{2}} & - \frac{1}{\sqrt{2}} & 0 & \ldots & 0 \\ 
\frac{1}{\sqrt{6}} & \frac{1}{\sqrt{6}} & - \frac{2}{\sqrt{6}} & \ldots & 0 \\
\ldots & \ldots & \ldots & \ldots & \ldots \\
\frac{1}{\sqrt{N ( N-1) }} & \frac{1}{\sqrt{N ( N-1)}} & \frac{1}{\sqrt{N ( N-1) }} & \ldots & - \frac{(N-1)}{\sqrt{N ( N-1)}}
                   \end{array} \right]
\:\:\:\: .
\label{1w.8ter}
\eneq
\noindent
In  terms of the rotated  fields, one obtains 

\beq
H_{N , {\rm B}} = \frac{u_\rho}{2}  \int_0^\ell \: d x \: [ K_\rho ( \partial_x \Phi ( x ))^2 +  K_\rho^{-1}  
( \partial_x \Theta ( x ))^2  ] + \frac{u_\sigma}{2} \: \int_0^\ell \: d x \: \sum_{a = 1}^{N - 1 } \: [ K_\sigma  ( \partial_x \varphi_a ( x ))^2 + 
 K_\sigma^{-1}   ( \partial_x \vartheta_a ( x ))^2 ]
 \:\:\:\: , 
 \label{lutw.2}
 \eneq
 \noindent
 with

 \begin{eqnarray}
 u_\rho K_\rho &=& u K \;\;\; , \;\; 
 \frac{u_\rho}{K_\rho} =   u  \left( K^{-1} + \frac{ (N-1) U}{\pi u } \right) \nonumber \\
 u_\sigma K_\sigma &=& u K\;\;\; , \;\; 
   \frac{u_\sigma}{K_\sigma} =    u   \left( K^{-1} - \frac{  U}{\pi u } \right) 
     \:\:\:\: . 
\label{1w.19}
\end{eqnarray}
\noindent 
Eqs. (\ref{1w.19}) yield  $( u_\rho , K_\rho)=  ( u_N , K_N)$ and 
$(u_\sigma , K_\sigma)=  ( u_0 , K_0)$, with $K_n = K / \sqrt{1 + \frac{(n-1)U K}{\pi u }}$ and 
$u_n = u \sqrt{1 + \frac{(n-1)U K}{\pi u }}$.
Note that, in particular, Eqs. (\ref{1w.19}) are consistently defined 
only as long as $\frac{K U}{\pi u} < 1$, which is our over-all assumption in this work.
Also, for a repulsive inter-wire interaction ($U>0$), by definition one  always has $K_\rho < 1$.  
As stated in section \ref{phadiaN2},   $H_{N , {\rm B}}$, for $N=2$,  corresponds to the bulk Hamiltonian of   Ref. \cite{giuaf_1},
with equal Luttinger parameters in the two wires, 
  $K_1 = K_2=K,u_1= u_2=u$.  Following the notation of Ref. \cite{giuaf_1}, 
we use  $N$ to denote open boundary conditions at both boundaries in a single bosonic channel corresponding  to a disconnected wire, 
which implies pure normal reflection at both boundaries.
This implies open boundary conditions for the lattice fermions \cite{affleck_eggert} or, in terms of 
the bosonic fields, pinning of $\theta_a ( x )$ at both boundaries, as  $\theta_a ( 0 ) =  \sqrt{\pi} n_{0 , a}$, 
$\theta_a ( \ell ) = \sqrt{\pi} n_{\ell , a}$, $n_{0 , a} , n_{\ell , a}  \in {\bf Z}$.  

As a simple,   paradigmatic, example we consider  
a single field $\phi ( x )$ which, together with its dual field $\theta ( x )$, is 
described by the TLL Hamiltonian 

\beq
H = \frac{u}{2} \: \int_0^\ell \: d x \: [ K ( \partial_x \phi ( x ))^2 + K^{-1} ( \partial_x \theta ( x ))^2 ] 
\:\:\:\: . 
\label{bomo.1}
\eneq
\noindent
Imposing Neumann boundary conditions on $\phi ( x )$ at both boundaries implies 
the mode expansions

\begin{eqnarray}
 \phi ( x ) &=& \phi_{   0}   + \sum_{ n = 1}^\infty \: \left\{ \frac{1}{\sqrt{K \pi n }} \cos \left[ \frac{\pi n x}{\ell} \right]
 [ \alpha_{ n } + \alpha_{ n }^\dagger ] \right\} \nonumber \\  
  \theta  ( x ) &=&   \theta_0 + \frac{  p_{  \phi} x}{\ell} +
  i \sum_{ n = 1}^\infty \: \left\{ \sqrt{\frac{K}{\pi n} } \sin \left[ \frac{\pi n x}{\ell} \right]
 [ \alpha_{  n  } - \alpha_{  n  }^\dagger ] \right\} 
 \:\:\:\: , 
 \label{bomo.2}
\end{eqnarray}
\noindent
with the oscillator modes  satisfying the algebra $ [ \alpha_{  n } , \alpha^\dagger_{  n'} ] = n \delta_{n,n'}$
  and the spectrum of the zero-mode operators given by   $p_\phi = \sqrt{\pi} m_\phi$, with $m_\phi$ relative integer. 
A mode expansion complementary to the one in Eqs. (\ref{bomo.2}) is recovered when imposing type $A$ boundary 
conditions on $\phi ( x )$ at both boundaries, and, accordingly, type $N$ boundary conditions on $\theta ( x )$. In
this case, one obtains

\begin{eqnarray}
 \phi ( x ) &=& \phi_{   0}   +  \frac{  p_{  \theta} x}{\ell} + i \sum_{ n = 1}^\infty \: \left\{ \frac{1}{\sqrt{K \pi n }} \sin \left[ \frac{\pi n x}{\ell} \right]
 [ \alpha_{ n } - \alpha_{ n }^\dagger ] \right\} \nonumber \\  
  \theta  ( x ) &=&   \theta_0 +  
   \sum_{ n = 1}^\infty \: \left\{ \sqrt{\frac{K}{\pi n} } \cos \left[ \frac{\pi n x}{\ell} \right]
 [ \alpha_{  n  } + \alpha_{  n  }^\dagger ] \right\} 
 \:\:\:\: , 
 \label{bomo.3}
\end{eqnarray}
\noindent
with the eigenvalues of $p_\theta$ equal to $\sqrt{\pi} m_\theta$, and $m_\theta$ relative integer. 

For the $N$-wire junction, when all the wires satisfy type $N$ boundary conditions at both boundaries, 
one obtains the mode expansion  
  
\begin{eqnarray}
 \Phi ( x ) &=& \Phi_{   0}   + \sum_{ n = 1}^\infty \: \left\{ \frac{1}{\sqrt{K_\rho \pi n }} \cos \left[ \frac{\pi n x}{\ell} \right]
 [ \alpha_{ c ,   n } + \alpha_{c ,   n }^\dagger ] \right\} \nonumber \\ 
 \varphi_a ( x ) &=& \varphi_{ a ,  0}   + \sum_{ n = 1}^\infty \: \left\{ \frac{1}{\sqrt{K_\sigma \pi n }} \cos \left[ \frac{\pi n x}{\ell} \right]
 [ \alpha_{ a , n } + \alpha_{ a , n }^\dagger ] \right\} \nonumber \\
  \Theta  ( x ) &=&   \Theta_0 + \frac{  p_{  \Phi} x}{\ell} +
  i \sum_{ n = 1}^\infty \: \left\{ \sqrt{\frac{K_\rho}{\pi n} } \sin \left[ \frac{\pi n x}{\ell} \right]
 [ \alpha_{ c , n  } - \alpha_{c ,  n  }^\dagger ] \right\} \nonumber \\
 \vartheta_a  ( x ) &=&   \vartheta_{a , 0} + \frac{  p_{ a, \varphi} x}{\ell} +
  i \sum_{ n = 1}^\infty \: \left\{ \sqrt{\frac{K_\sigma}{\pi n} } \sin \left[ \frac{\pi n x}{\ell} \right]
 [ \alpha_{a , n  } - \alpha_{a , n  }^\dagger ] \right\}
 \:\:\:\: , 
 \label{gd1.3}
\end{eqnarray}
\noindent
with the oscillator modes  satisfying the algebra $ [ \alpha_{b , n } , \alpha_{b' , n'}^\dagger  ] = n \delta_{b , b'}
\delta_{n, n'}$, and the spectrum of the zero-mode operators given by   

\beq
\left[ \begin{array}{c} p_\Phi \\ p_{1 , \varphi} \\ \vdots \\ p_{N-1 , \varphi} \end{array} \right] 
= {\bf M}_N \: \left[ \begin{array}{c} \sqrt{\pi} m_{1 , \phi} \\ \sqrt{\pi} m_{2 ,\phi} \\ \vdots \\ 
\sqrt{\pi} m_{N , \phi} \end{array} \right] 
\:\:\:\: , 
\label{gd.4}
\eneq
\noindent
with  $m_{1 , \phi} , \ldots , m_{N , \phi}$ relative integers. In the general case in
which one has type $A$ CIBC's at both boundaries in the first $N_a$ channels and 
type $N$ CIBC's in the remaining $N_n$ ones, Eqs. (\ref{gd.4}) generalize to

\beq
\left[ \begin{array}{c} p_\Theta \\ p_{1 , \vartheta} \\ \vdots \\ p_{N-1 , \vartheta} \end{array} \right] 
= {\bf M}_N \: \left[ \begin{array}{c} \sqrt{\pi} m_{1 , \theta} \\ \vdots \\ \sqrt{\pi} m_{N_a ,\theta} \\ \vdots \\ 0 \end{array} \right]  \;\;\; , \;\;
\left[ \begin{array}{c} p_\Phi \\ p_{1 , \varphi} \\ \vdots \\ p_{N-1 , \varphi} \end{array} \right] 
= {\bf M}_N \: \left[ \begin{array}{c} 0 \\ \vdots  \\  \sqrt{\pi} m_{N_a+1 , \phi} \\ \vdots  \\ \sqrt{\pi} m_{N ,\phi}  \end{array} \right]
\:\:\:\: , 
\label{gd.5a}
\eneq
\noindent
for the eigenvalues of the zero-mode operators of the $ ( \Phi , \varphi_1 , \ldots , \varphi_{N-1} )$ and of 
the $ ( \Theta , \vartheta_1 , \ldots , \vartheta_{N-1} )$-fields respectively, with 
$m_{1 , \theta} , \ldots , m_{N_a , \theta}$ and $m_{N_a + 1 , \phi} , \ldots , m_{N , \phi}$ 
relative integers. 
 
\section{Stabilization of fixed points with $N_a > 1$}
\label{stabl}

In order to rigorousely define the algorithm we use in the main text to 
count the degrees of freedom associated with real fermionic zero-mode operators 
at a fixed points with type $A$ CIBC's in the first $N_a$ channels, type $N$ CIBC's in the remaining 
$N_n$ channels, we now concisely review the approach employed in Ref. \cite{giuaf_1} to 
artificially stabilize the $N_a = 2 , N_n = 0$ fixed point in the 
$N=2$ junction and eventually  extend it to the $N$-wire junction, with 
a generic $N$. As a reference model Hamiltonian, we consider $H_{N , {\rm Fer}}$
in Eq. (\ref{bi.3}) taken in the noninteracting limit. To recover type $N$ CIBC's in the last 
$N_n = N - N_a$ channels,   we   set $t_{N_a + 1} = \ldots = t_N = 0$. Next, we rewrite  the 
complex fermion lattice operators in the first $N_a$ channels in terms of 
real fermion lattice operators $ \{ \xi_{j , a } , \eta_{j , a } \}$ as 
 
 \begin{eqnarray}
 c_{j , a } &=& \frac{1}{2} \{ \xi_{j , a } + i \eta_{ j , a } \} \nonumber \\
 c_{j , a}^\dagger &=& \frac{1}{2} \{ \xi_{j , a } - i \eta_{ j , a } \}
\:\:\:\: , 
\label{2a.1}
\end{eqnarray}
\noindent
with $a = 1 , \ldots , N_a$. In terms of the real fermion lattice operators, the 
boundary Hamiltonian $H_{b , {\rm Fer} , N}  $ can 
be rewritten as

\beq
H_{b , {\rm Fer} , N}  = - 2 i \sum_{a = 1}^{N_a} t_{a  } \{ \gamma_L \eta_{1 , a }  + \gamma_R \xi_{\ell - 1 , a } \} 
\:\:\:\: . 
\label{2a.2}
\eneq
\noindent
Now, we note that, besides $H_{b , {\rm Fer} , N}$, the  operators $\{ \eta_{1 , a }  , \xi_{\ell - 1 , a } \}$  enter  
$H_{0 , {\rm Fer} , N}$ in the term $H^{'}$, given by

 \beq
 H^{'} = \sum_{a = 1}^{N_a} \left\{ - \frac{i J}{2} \left[  \xi_{2 , a } \eta_{ 1 , a } + \xi_{\ell - 1 , a } \eta_{\ell - 2 , a} \right] - 
 \frac{i \mu}{2} [ \xi_{1,a} \eta_{ 1 , a } + \xi_{\ell - 1 , a } \eta_{\ell - 1 , a } ]  - \frac{ i J }{2}
 [ \xi_{1,a} \eta_{2,a} + \xi_{\ell - 2 , a } \eta_{\ell - 1 , a } ]   \right\} 
 \:\:\:\: . 
 \label{2a.3}
 \eneq
 \noindent
The key point, now, is to regard Eq. (\ref{2a.3}) as a special case of the generic 
Hamiltonian  $\tilde{H}'$, defined as 
 
 \beq
 \tilde{H}' =  \sum_{a = 1}^{N_a}  \left\{ - \frac{i J_{A ,a} }{2} \left[  \xi_{2 , a } \eta_{ 1 , a } + \xi_{\ell - 1 , a } \eta_{\ell - 2 , a} \right]
 - \frac{i \mu_{a} }{2}  [ \xi_{1,a} \eta_{ 1 , a } + \xi_{\ell - 1 , a } \eta_{\ell - 1 , a } ]  - \frac{ i J_{B , a }  }{2}
[ \xi_{1,a} \eta_{2,a} + \xi_{\ell - 2 , a } \eta_{\ell - 1 , a } ] \right\} 
 \:\:\:\: , 
 \label{2a.4}
 \eneq
 \noindent 
with $J_{A ,a }= J_{B ,a } = J$ and $\mu_{a} = \mu$. To stabilize 
a fixed point with type $A$ CIBC's in the first $N_a$ channels, 
we therefore fine-tune the parameters of $\tilde{H}'$ as 

\begin{eqnarray} 
 J_{A , 1} &=& \ldots = J_{A , N_a } = 0 \nonumber \\ 
 \mu_1 &=& \ldots = \mu_{N_a}  = 0 \nonumber \\ 
 J_{B , 1 } &=&  \ldots = J_{B , N_a } = J 
 \:\:\:\: . 
 \label{2a.5}
\end{eqnarray}
\noindent
As a result, QW's from 1 to $N_a$  are separately coupled to a Majorana mode at their endpoints, 
respectively given by $\xi_{1,a}$ and by $\eta_{\ell - 1,a}$, which makes them all renormalize to $A$ 
boundary conditions at both boundaries. As highlighted in section \ref{pha_1}, this leaves one 
unpaired real fermion zero-mode operator for each one of the first $N_a$ QW's. In addition, 
there are $N_n$ KF's from the remaining decoupled $N_n$ QW's plus, if $N$ is odd, 
the auxiliary KF $\bar{\Gamma}$. To fully account for all the degrees of freedom one has eventually 
to consider the two MM's at the endpoints of the TS's and the $2N_a$ real fermion operators 
$\{ \eta_{1 , a } , \xi_{\ell - 1 , a } \}$, with $a = 1 , \ldots , N_a$. They are coupled to 
each other via $H_{b , {\rm Fer} , N}$ in Eq. (\ref{2a.2}), which implies additional 
$2N_a-2$ real fermion zero-mode operators. Taking all this into account, we eventually 
provide the results for $\delta_e [ N_a , N_n]$ and $\delta_o [ N_a , N_n] $ as 

\begin{eqnarray}
 \delta_e [ N_a , N_n ] &=& 2^{\frac{3 N_a}{2} + \frac{N_n-2}{2} } \nonumber \\
  \delta_e [ N_a , N_n ] &=& 2^{\frac{3 N_a}{2} + \frac{N_n -1}{2} }
  \;\;\;\; , 
  \label{last}
\end{eqnarray}
\noindent
which are the formulas we use when computing the $g$-function in section \ref{impuN}.

In employing the method we develop here for the   derivation of section \ref{j3} for the Y3J, we 
note that, when considering the two boundary Hamiltonian for this specific system, due to the 
fact that only three real fermion modes actually enter the boundary interaction 
(the three KF's from the three QW's), one recovers a sort of 
``special case'', corresponding to $N=2$ and, formally,  $\gamma_L = \gamma_R = \hat{\gamma}$. In this case,
$H_{b , {\rm Fer} , N}$ in Eq.(\ref{2a.2}) reduces to

\beq
\hat{H}_{b , {\rm Fer} , 2}  = - 2 i \sum_{a = 1}^{2} t_{a  } \{ \hat{\gamma} [  \eta_{1 , a }  +  \xi_{\ell - 1 , a } ] \} 
\:\:\:\: , 
\label{2a.2addi}
\eneq
\noindent
which has three zero-energy  real fermion eigenmodes, differently from the case of two distinct MM's 
 at the two boundaries 

Before concluding this appendix, we concisely review how to use the construction detailed above 
to derive the set of allowed boundary operators at the $A \otimes A$ fixed point of the 
$N=2$ junction. To do so, we resort to the one-boundary version of the system, with the 
corresponding boundary Hamiltonian, $H_{b,F,2}^{(1)}$, simply given by $H_{b , {\rm Fer} , N=2}$ in 
Eq. (\ref{2a.2}), with the coupling to $\gamma_R$ dropped off. Accordingly, we 
simplify Eq. (\ref{2a.4}), by also assuming that the couplings are the same in each channel, to 

 \beq
 \tilde{H}^{'}  \to   - \frac{i J_{A  } }{2} \left[  \xi_{2 ,1 } \eta_{ 1 , 1 } + \xi_{2, 2} \eta_{1 , 2} \right]
 - \frac{i \mu  }{2}  [ \xi_{1,1} \eta_{ 1 , 1 } + \xi_{1 , 2 } \eta_{ 1 , 2 } ]  - \frac{ i J_{B  }  }{2}
[ \xi_{1,1} \eta_{2,1} + \xi_{1, 2 } \eta_{  1 , 2 } ]  
 \:\:\:\: , 
 \label{2a.4bis}
 \eneq
 \noindent 
and, to   stabilizes the  $A \otimes A$ fixed point,  we set $J_A = \mu = 0$. This makes 
 $H_{b,F,2}^{(1)}$ fully decouple from the bulk of the system. In particular, on rewriting it as 
 
 \beq
 H_{b,F,2}^{(1)} = - 2 i t \gamma_R \gamma 
\;\;\;\; , 
\label{2a.xx1}
\eneq
\noindent
with $t= \sqrt{t_1^2 + t_2^2}$ and $\gamma = \frac{t_1}{t} \eta_{1,1} + \frac{t_2}{t} \eta_{1 ,2 }$, 
we recover a real-fermionic zero-mode operator, $\tilde{\gamma} = - \frac{t_2}{t} \eta_{1,1} + 
\frac{t_1}{t} \eta_{1 , 2}$, which is fully decoupled from the system, as long as $J_A = \mu  =0$. 
Turning on $J_A$ and $\mu$, the corresponding contribution to $\tilde{H}'$, $\delta \tilde{H}'$, can 
be written as 

\beq
\delta \tilde{H}' = - \frac{i }{2 t} \: \{[ t_1 ( J_A \xi_{2,1} + \mu \xi_{1,1} )  + t_2 ( J_A \xi_{2,2}
+ \mu \xi_{1,2} )] \gamma + [ - t_2 ( J_A \xi_{2,1} + \mu \xi_{1,1} ) + 
t_1 ( J_A \xi_{2,2} + \mu \xi_{1,2} ) ] \tilde{\gamma} \} 
\:\:\:\: . 
\label{2a.xx2}
\eneq
\noindent
At large values of $t$, $\gamma$ is locked together with $\gamma_L$ so, in analogy to what happens 
with the residual coupling to the MM at the $A \otimes N$ fixed point, the term $\propto \gamma$ in 
$\delta \tilde{H}'$ only contributes to second-order in the corresponding boundary couplings (see 
Ref. \cite{giuaf_1} for a detailed discussion about this point). To this order, it gives 
rise to the boundary operators arising from fermion bilinears at the $A \otimes A$ fixed point:
the inter-channel normal boundary backscattering and the inter-channel boundary pairing 
operator, the intra-channel normal backscattering operator in both channels. 
At variance, $\tilde{\gamma}$  is decoupled from other real fermionic modes. Accordingly, the 
term in $\delta \tilde{H}'$ that is $\propto \tilde{\gamma}$ does act as an effective 
residual coupling to the MM, eventually leading to the boundary operators 
$\tilde{V}_{1 , {\rm Res}  }, \tilde{V}_{2 , {\rm Res}}$ of section \ref{phadiaN2}.

\section{Duality and Poisson summation formula}
\label{duapoi}

As illustrated in the main text, at a given fixed point, the  $g$-function corresponding to 
type $A$ boundary conditions can be extracted from the partition function  ${\cal Z}_{AA}=
\sum_n\exp \left[ - \frac{ x_{AA}^n \beta u}{\ell} \right]$, by sending  $\ell \to  \infty$ 
at fixed $\beta$. To recast ${\cal Z}_{AA}$  
in a form suitable for taking such a limit, one has to make a combined use of 
the duality properties of the Dedekind function, as well as of Poisson's summation formula (PSF), 
which we review in this appendix. 

Given a  complex number $\tau = \tau_x + i \tau_y$, with $\tau_y > 0$, and setting $q = e^{   \pi i \tau}$,
the Dedekind function $\eta ( q )$ is defined as  
$\eta ( q ) = e^{ \frac{\pi i \tau}{24}} \: \prod_{n = 1}^\infty \: [1 - q^n ]$. $\eta ( q )$ is known 
to exhibit the duality property (which is relevant to our 
derivation) $\eta ( q ) = \frac{1}{\sqrt{-i \tau}} \: \eta ( \tilde{q} )$, 
with $\tilde{q} = e^{- \frac{2 \pi i }{\tau}}$. In fact, once one sets $\tau = \frac{i \beta \pi}{u \ell}$, 
one obtains the needed change of variable  in the $\eta$-function. 

Moving to PSF, for  a single-variable 
function $f ( x )$, it is defined starting from the quantity $F$, given by 

\beq
F = \sum_{ n \in {\bf Z}} \: f ( n ) 
\;\:\:\: . 
\label{math.3}
\eneq
\noindent
Defining the Fourier transform of $f ( x )$, $\hat{f} ( p )$, as   

\beq
\hat{f} ( p ) =   \int_{- \infty}^\infty \: d x \: e^{ - i p x } \: f ( x ) 
\;\;\;\; , 
\label{math.6}
\eneq
\noindent
PSF states the identity  

\beq
F =  \sum_{ n \in {\bf Z}} \: f ( n )  =   \sum_{m \in {\bf Z}} \hat{f} ( m ) 
 \;\;\;\; .
 \label{math.7}
 \eneq
 \noindent
  PSF can be readily extended to a sum over 
 a generic $d$-dimensional Bravais lattice $\Lambda$. Indeed, given a function of $d$ variables,
 $f ( x_1 , \ldots , x_d )$, we set  
 
 \beq
 F = \sum_{ {\bf R} \in \Lambda } f ( {\bf R} )
 \:\:\:\: . 
 \label{math.8}
 \eneq
 \noindent
 Assuming that, $\forall \: {\bf R} \in \Lambda$, $\exists \: ( n_1 , \ldots , n_d )$ such that 
 ${\bf R} = \sum_{ i = 1}^d n_i {\bf R}_i$, we therefore obtain 
 
 \beq
 F = \sum_{ n_1 , \ldots , n_d \in {\bf Z}} \: f ( \sum_{i = 1}^d n_i {\bf R}_i ) 
 \:\:\:\: . 
 \label{math.9}
 \eneq
 \noindent
 Let us define a $d \times d$-matrix ${\bf A}$ such that, $\forall x_1 , \ldots , x_d \in {\bf R}^d$, one 
 gets 
 
 \beq
 \sum_{ i = 1}^d x_i {\bf R}_i = {\bf A} \cdot \left[ \begin{array}{c} x_1 \\ \vdots \\ x_d \end{array} \right]
 \:\:\:\: . 
 \label{math.10}
 \eneq
 \noindent
 We note that we obtain 
 
 \beq
  \frac{1}{ ( 2 \pi)^\frac{d}{2}} \: \int_{ - \infty}^\infty \: d x_1 \ldots dx_d \:
 e^{ - i [ p_1 x_1 + \ldots + p_d x_d ] }    f \left(  
 {\bf A} \cdot \left[ \begin{array}{c} x_1 \\ \vdots \\ x_d \end{array} \right]  \right) = [ {\rm det} {\bf A} ]^{-1}
  \hat{f} \left(  
 {\bf A}^{-1}  \cdot \left[ \begin{array}{c} p_1 \\ \vdots \\ p_d \end{array} \right]  \right)   
 \:\:\:\: , 
 \label{math.13}
 \eneq 
 \noindent
 with $\hat{f} ( p_1 , \ldots , p_d )$ being the multidimensional Fourier transform of 
 $f ( x_1 , \ldots , x_d )$.  On explicitly performing the integral, we eventually get 
 
 \beq
 F  
 = \frac{  1}{ {\rm det} {\bf A}  } 
  \: \sum_{  {\bf K} \in \Lambda^* } \:     \hat{f} ( {\bf K} ) 
 \:\:\:\: , 
 \label{math.14}
 \eneq
 \noindent
 with $\Lambda^*$ being the dual lattice of $\Lambda$.    Now, in the specific problem we consider, we typically 
 obtain 
 
 \beq
 f ( x_1 , \ldots , x_d ) = \exp \left[ - \frac{ \pi \beta u}{ 2 K \ell } ( x_1^1 + \ldots + x_d^2 ) \right]
 \Rightarrow \hat{f} ( p_1 , \ldots , p_d ) =   
 \left( \frac{ 2 \ell K}{ \beta u} \right)^\frac{d}{2} \: 
 \exp \left[ - \frac{ \ell K}{2 \pi \beta u}  ( p_1^2 + \ldots + p_d^2 ) \right]
 \:\:\:\: . 
 \label{math.15}
 \eneq
 \noindent
 Eq. (\ref{math.15}) eventually implies 
 
 \beq
 \sum_{ {\bf R} \in \Lambda} \exp \left[ - \frac{ \pi \beta u}{ 2 K \ell }  | {\bf R} |^2 \right]
=  \frac{1}{{\rm det} {\bf A}  } \: 
 \left( \frac{ 2 \ell K}{ \beta u} \right)^\frac{d}{2}\:  \sum_{ {\bf K} \in \Lambda^*} 
  \exp \left[ - \frac{ \ell K}{2 \pi \beta u}  | {\bf K} |^2 \right]  
  \:\:\:\: . 
  \label{math.16}
  \eneq
  \noindent
  Finally, using the duality of the Dedekind function, we get 
  
  \beq
  \eta ( e^{ - \frac{ \pi \beta u}{ \ell   }  } ) = \sqrt{\left( \frac{2 \ell  }{\beta u } \right) } \: \eta ( e^{ - \frac{4 \pi \ell  }{ \beta u}}) 
  \;\:\:\: , 
  \label{math.17}
  \eneq
  \noindent
which eventually implies 

\beq
\eta^{-d} (  e^{ - \frac{ \pi \beta u}{ \ell   }  } )    \sum_{ {\bf R}\in \Lambda} \exp \left[ - \frac{ \pi \beta u}{ 2 K \ell }  | {\bf R}|^2 \right]
= \frac{K^\frac{d}{2}}{ {\rm det} {\bf A} }\: \eta^{-d} (  e^{ - \frac{4 \pi \ell  }{ \beta u}} )   
\:  \sum_{ {\bf K}\in \Lambda^*} 
  \exp \left[ - \frac{ \ell K}{2 \pi \beta u}  | {\bf K}|^2 \right]  
  \:\:\:\: . 
  \label{math.18}
  \eneq
  \noindent
Eq. (\ref{math.18}) is the key equation we use to compute the $g$-function at the various fixed points 
of the systems we study in our paper.

\section{Review of the $\epsilon$-expansion approach to junction of quantum wires}
\label{epsilon_expansion}

In this appendix, we review the $\epsilon$-expansion approach to  the RG equations and to the calculation of 
the $g$-function at the FCFP's in   a junction between $N$ interacting quantum wires and 
a topological superconductor and in the Y3J. In the former case, 
we revisit and generalize the derivation discussed in Ref. \cite{giuaf_1}. In the latter 
case, we highlight the peculiarites of the $\epsilon$-expansion applied to the 
Y3J, estimate the $g$-function at the FCFP of such a system \cite{oca} and eventually 
discuss the analogies with the $N$-wire junction.

\subsection{Renormalization group equations for  a junction between $N$ quantum wires and 
a topological superconductor}
\label{renge}

To encompass FCFP's, the $\beta$-functions for the boundary couplings of a junction between 
$N$ quantum wires and a topological superconductor   must include nonlinear terms in the boundary 
couplings themselves. This requires    employing an adapted version of Cardy's method to derive nonlinear
contributions to  the $\beta$-function in perturbed conformal field theories from two-point OPE's \cite{cardy_1}, 
to terms  involving up to three-point OPE's in the boundary interaction operators \cite{aflud_2,giuaf_1}.
The starting point is   the full partition function ${\cal Z}$ written   as 
a power series of the boundary action   $S_b$ as 

\beq
{\cal Z} = {\cal Z}_0 \: \sum_{n = 0}^\infty \: \frac{ (-1)^n }{ n ! } \: \langle {\bf T}_\tau S_b^n \rangle_0
\:\:\:\: , 
\label{renge.1}
\eneq
\noindent
with $S_b = \int_{- \frac{\beta}{2}}^\frac{\beta}{2} \: d \tau \: H_b ( \tau )$, $H_b ( \tau )$ being the boundary Hamiltonian 
in imaginary time $\tau$, ${\bf T}_\tau$ being the imaginary time-ordering operator, ${\cal Z}_0$ being 
the partition function computed at the reference point corresponding to the absence of boundary interactions (that is, 
the disconnected fixed point), and $\langle \ldots \rangle_0$ denoting averages computed at the reference fixed point. In addition, 
to regularize  diverging contributions arising at short imaginary-time distances, one introduces a hard-core short imaginary time cutoff $\tau_0$
by requiring $|\tau_i-\tau_j|\geq \tau_0$ for all $i\neq j$.  To derive the RG  equations, we  increase 
$\tau_0$ to $\tau_0 + \delta \tau$ (which corresponds to   reducing the cut-off $D_0$ in energy domain) and  derive
 how the $\bar{t}_a$'s correspondingly change. Using
 as boundary Hamiltonian $H_{b , {\rm B},N}^{(1)}$ in Eq. (\ref{1w.20}), we see that, to second order in the  $ \bar{t}_a$'s, 
 only terms not involving $\gamma_L$ can be generated which, clearly, do not contribute any renormalization to the boundary 
 couplings. Therefore, to find the leading nonlinear correction to $\beta$-functions, one has to go  to 
 third order in the  $ \bar{t}_a$'s.  In doing so, one has to systematically subtract terms equal to 
$H_b ( \tau )$ times a free-energy correction arising to second order in the boundary couplings.
Considering that  a boundary operator 
product factorizes into a product of real fermion operators (MM's and/or KF's) and 
a product of bosonic vertex operators, we therefore begin by reviewing 
the  following rules for the fermionic OPE's \cite{giuaf_1}:

\begin{eqnarray}
 {\bf T}_\tau [ \gamma_L ( \tau_1 ) \gamma_L ( \tau_2 ) \gamma_L ( \tau_3 ) ] &=& 
 \epsilon ( \tau_1 , \tau_2 ,\tau_3 ) \gamma_L \nonumber \\
 {\bf T}_\tau [ \Gamma_a ( \tau_1 ) \Gamma_a ( \tau_2 ) \Gamma_a ( \tau_3 ) ] &=& 
  \epsilon ( \tau_1 , \tau_2 ,\tau_3 ) \Gamma_a \nonumber \\
  {\bf T}_\tau  [ \Gamma_a ( \tau_1 ) \Gamma_a ( \tau_2 ) \Gamma_b ( \tau_3 ) ] &=& 
  \epsilon ( \tau_1 , \tau_2 )  \Gamma_b 
  \;\;\;\; , 
  \label{renge.3}
\end{eqnarray}
\noindent
with $a \neq b$ and $\epsilon ( \tau_1 , \tau_2 , \tau_3 ) $, $\epsilon ( \tau_1 , \tau_2 )$ being fully
antisymmetric functions of their arguments. Moreover, we also need the rules for the OPE's between
bosonic vertex operators, which we here review in the large-system size, zero-temperature limit, and 
which are given by 

\begin{eqnarray}
 {\bf T}_\tau [ e^{ i \sqrt{\pi} \phi_a ( \tau_1 )} e^{ i \sqrt{\pi} \phi_a ( \tau_2 ) } e^{ - i \sqrt{\pi} \phi_a ( \tau_3 ) } ]
 &\to&_{\tau_1 \sim \tau_2 \sim \tau_3 } \: \left| \frac{\tau_1 - \tau_2}{ ( \tau_1 - \tau_3 ) ( \tau_2 - \tau_3 )} \right|^{ 2 d_b } 
  e^{ i \sqrt{\pi} \phi_a (  \tau_3 ) } \nonumber \\
   {\bf T}_\tau [ e^{ i \sqrt{\pi} \phi_a ( \tau_1 )} e^{ - i \sqrt{\pi} \phi_a ( \tau_2 ) } e^{  i \sqrt{\pi} \phi_b ( \tau_3 ) } ] 
   &\to&_{\tau_1 \sim \tau_2 \sim \tau_3 } \: \left| \frac{ \tau_1 - \tau_3 }{\tau_2 - \tau_3 } \right|^{ \frac{1}{  N K_\rho} - \frac{1}{  N K_\sigma} } 
   \: \left| \frac{1}{ \tau_1 - \tau_2 }\right|^{ 2 d_b  }  \: e^{  i \sqrt{\pi} \phi_b ( \tau_3 ) }
   \:\:\:\: , 
   \label{renge.4}
\end{eqnarray}
\noindent
with $ a \neq b$, et cetera. Following the derivation of Ref. \cite{giuaf_1}, we also make the assumption that 
the coupling to the Majorana mode is a slightly relevant operator, that is, that we have $1 - ( 2 d_b )^{-1} = \epsilon$, 
with $0 < \epsilon \ll 1$. Now, to leading order in $\epsilon$, we set $d_b^{-1} = 2$ in the integrals involving 
the OPE's at the right hand side of Eq. (\ref{renge.4}). Accordingly, due to the remarkable identity 

\begin{eqnarray} 
 && \left| \frac{\tau_1 - \tau_2}{( \tau_1 - \tau_3 ) (\tau_2 - \tau_3 )} \right|^2 + 
  \left| \frac{\tau_1 - \tau_3}{( \tau_1 - \tau_2 ) (\tau_2 - \tau_3 )} \right|^2  + 
  \left| \frac{\tau_2 - \tau_3}{( \tau_1 - \tau_3 ) (\tau_1 - \tau_2 )} \right|^2 = \nonumber \\
  && \frac{1}{ | \tau_1 - \tau_2 |^2} +  \frac{1}{ | \tau_1 - \tau_3 |^2} +  \frac{1}{ | \tau_2 - \tau_3 |^2} 
  \:\:\:\: , 
  \label{renge.5}
\end{eqnarray}
\noindent
after subtracting the free energy correction times $H_b$, as discussed above, 
we see that, to leading order in $\epsilon$,  no renormalization of the $ t_a$'s arises that is $\propto t_a^3$. 
Instead, a nonzero renormalization of the boundary coupling arises from corrections to 
the boundary action that can be derived by a straightforward generalization of the 
analysis done in Ref. \cite{giuaf_1} for $N=2$. In particular, the relevant correction 
turns out to be given by 

\begin{eqnarray}
&& \delta S_b^{(3)} = 2 i \sum_{a = 1}^N \int \: d \tau_1 \: \frac{ t_a }{\tau_0^\epsilon}  \gamma_L ( \tau_1 ) \Gamma_a ( \tau_1 ) 
\cos [ \sqrt{\pi} \phi_a ( \tau_1 ) ] \: \frac{1}{2} \: \sum_{ b \neq a = 1}^N \: \frac{ t_b^2 }{\tau_0^{2 \epsilon}} \: \int \: d \tau_2 d \tau_3 \times \nonumber \\ 
&& \left\{ \left| \frac{\tau_1 - \tau_2}{\tau_1 - \tau_3} \right|^\nu + \left| \frac{\tau_1 - \tau_3}{\tau_1 - \tau_2} \right|^\nu \epsilon ( \tau_1 - \tau_2 ) 
\epsilon ( \tau_1 - \tau_3 ) - 1 
\right\} \frac{1}{| \tau_1 - \tau_2 |^2} \: \prod_{i < j = 1}^N \theta ( | \tau_i - \tau_j | - \tau_0 ) 
\:\:\:\: , 
\label{renge.6}
\end{eqnarray}
\noindent
with $\nu = \frac{1}{ N K_\rho} - \frac{1}{ N K_\sigma}$ and the cutoff function explicitly denoted.  
At this point, by analogy with Ref. \cite{giuaf_1}, one differentiates Eq. (\ref{renge.6}),
 obtaining the nonlinear corrections to the $\beta$-functions for the running couplings. 
As a result, one gets 

\beq
\frac{d \bar{t}_a}{d \ln ( \tau / \tau_0 ) } = \frac{d \bar{t}_a}{d \ln \left( \frac{ D_0 }{ D } \right) } 
= \epsilon \bar{t}_a - {\cal F} ( \nu ) \bar{t}_a \sum_{b \neq a = 1}^N \bar{t}_b^2
\:\:\:\: , 
\label{renge.7}
\eneq
\noindent
with 

\begin{eqnarray}
{\cal F}(\nu )&=& 6-2\int_1^\infty dx\biggl[{(x+1)^\nu +(x+1)^{-\nu}\over x^2}-{x^\nu +x^{-\nu}\over (x+1)^2}
\nonumber \\ &+& \left({x+1\over x}\right)^\nu +\left({x+1\over x}\right)^{-\nu}-2 \biggr] 
\:\:\:\: , 
\label{renge.8}
\end{eqnarray}
\noindent
that is, Eq. (\ref{1w.26}) of the main text.

\subsection{Renormalization group equations for the running coupling strengths at a   Y-junction of 
three spinless interacting quantum wires }
\label{failure}

We now extend the $\epsilon$-approach to derive the RG equations for the   Y3J with 
$1 < K < 3$ and, in general, boundary couplings all different from each other.
To do so,   we start from the   boundary Euclidean action $S_b$, which is now given by

\begin{eqnarray}
S_b &=& -   \frac{ \bar{t}_{2,1} }{\tau_0^\epsilon} \: \int \: d \tau \: 
\Gamma_1 ( \tau)  \Gamma_2( \tau) [ e^{ - i \sqrt{\pi} \phi_{1 , 2} ( \tau )  } -   e^{  i \sqrt{\pi} \phi_{1 , 2} ( \tau )  } ]  
 -   \frac{ \bar{t}_{3,2} }{\tau_0^\epsilon} \: \int \: d \tau \: \Gamma_2( \tau) \Gamma_3 ( \tau)[ e^{ - i \sqrt{\pi} \phi_{2 , 3} ( \tau )  } - 
 e^{  i \sqrt{\pi} \phi_{2 , 3} ( \tau )  } ]  \nonumber \\
 &-& \frac{ \bar{t}_{1,3} }{\tau_0^\epsilon} \: \int \: d \tau \:  \Gamma_3( \tau) \Gamma_1 ( \tau) [ e^{ - i \sqrt{\pi} \phi_{3 , 1} ( \tau )  } -  
 e^{  i \sqrt{\pi} \phi_{3 , 1} ( \tau )  } ] 
 \;\;\;\; , 
 \label{perty.2}
\end{eqnarray}
\noindent
with $\phi_{a , b} ( \tau ) = \phi_a ( 0 , \tau) - \phi_b ( 0 , \tau)$, $\epsilon =1 - \frac{1}{K}$ and $\bar{t}_{a+1,a}$ denoting the 
dimensionless coupling strengths. To implement the $\epsilon$-expansion, we assume $K>1$ and 
$0 < 1 - K^{-1} \ll 1$. The key ingredients of our derivation are  the OPE's between the operators entering $S_b$, which are given by

\begin{eqnarray}
 && \langle {\bf T}_\tau \Gamma_a ( \tau_1 ) \Gamma_b ( \tau_2 ) \rangle = \delta_{a ,b } \epsilon ( \tau_1 - \tau_2  ) \nonumber \\
 && \langle {\bf T}_\tau e^{\pm i \sqrt{\pi} \phi_a ( \tau_1 ) } e^{ \mp i \sqrt{\pi} \phi_b ( \tau_2 ) } \rangle = \frac{ \delta_{a , b } }{ 
| \tau_1 - \tau_2 |^\frac{1}{K}  }
  \:\:\:\: .
 \label{perty.6}
\end{eqnarray}
\noindent
In principle, nonzero contributions to the $\beta$-function may arise to ${\cal O} ( \bar{t}_a \bar{t}_b  )$. To check 
whether this is the case, we  consider the OPE 

\begin{eqnarray}
&& \sum_{ a , b = 1,2,3} \:\bar{t}_{a+1 , a} \bar{t}_{b + 1 , b}  {\bf T}_\tau \{ \Gamma_a ( \tau_1 ) \Gamma_{a+1} ( \tau_1 ) 
[ e^{ - i \sqrt{\pi} \phi_{a , a+1} ( \tau_1 ) } - e^{   i \sqrt{\pi} \phi_{a , a+1} ( \tau_1 ) }  ]
\Gamma_b ( \tau_2 ) \Gamma_{ b + 1} ( \tau_2 ) 
[ e^{ - i \sqrt{\pi} \phi_{b , b+1} ( \tau_2 ) } - e^{   i \sqrt{\pi} \phi_{b , b+1} ( \tau_2 ) }  ] \}
\nonumber \\
&&=_{\tau_1 \sim \tau_2}   - 2 \{ \bar{t}_{2,1}^2 + \bar{t}_{3,2}^2 + \bar{t}_{1,3}^2\}  \:   |  \tau_1 - \tau_2 |^{ - \frac{2}{K}}   \theta ( | \tau_1 - \tau_2 | - \tau_0 )  - \epsilon ( \tau_1 - \tau_2 ) 
\times \nonumber \\
&&\sum_{a \neq b =1,2,3} \bar{t}_{a+1 , a} \bar{t}_{b + 1 , b} 
{\bf T}_\tau \: \{ \Gamma_a ( \tau_1 ) \Gamma_{b} ( \tau_2 ) [ e^{ - i \sqrt{\pi} [ \phi_a ( \tau_1 ) - \phi_b ( \tau_2 ) ] } + 
e^{  i \sqrt{\pi} [ \phi_a ( \tau_1 ) - \phi_b ( \tau_2 ) ] } \} \: | \tau_1 - \tau_2 |^{ - \frac{1}{K}}  
\theta ( | \tau_1 - \tau_2 | - \tau_0 ) ,  
\label{perty.x2}
\end{eqnarray}
\noindent
with $3 + 1 \equiv 1$.  The   term on the right-hand side of Eq. (\ref{perty.x2}) merely 
 renormalizes the total free energy. Therefore, to find nonlinear contributions to 
 the $\beta$-functions for the boundary couplings, we need to go to third order in the $\bar{t}_{a+1 , a}$'s. This requires considering the 
 three-point OPE's between operators entering $S_b$.   In  doing this, we find 
three different contributions, which we separately discuss in the following 
  
\begin{itemize}
 \item {\bf Term number 1:}
 
Taking into account the symmetries effective under integrating over the imaginary times
and taking the  $\beta \to \infty$-limit, one obtains a correction to the boundary action, 
$\delta S_b^{(1)}$, given by 
 
 \begin{eqnarray}
&& \delta S_b^{(1)} = 
 \frac{\bar{t}_{2,1} ( \bar{t}_{3,2}^2 + \bar{t}_{1,3}^2 ) }{2} \: \int\: d \tau_1 \: d \tau_2 \: d \tau_3 \: 
\epsilon ( \tau_1 -\tau_3 ) \epsilon ( \tau_2 - \tau_3 )  \: \prod_{i<j=1}^3 \theta ( | \tau_i - \tau_j | - \tau_0 )\: \Gamma_1 ( \tau_3 ) \Gamma_2 ( \tau_3) 
 \times \nonumber \\
 && \Biggl\{ \frac{ |  \tau_1 - \tau_3 |^\frac{1}{K} }{  | \tau_1 - \tau_2 |^\frac{2}{K} 
 | \tau_2 - \tau_3 |^\frac{1}{K} } +  \frac{ | \tau_2 - \tau_3 |^\frac{1}{K}  }{  | \tau_1 - \tau_2 |^\frac{2}{K} 
  | \tau_1 - \tau_3 |^\frac{1}{K} } \Biggr\} 
 \: [ e^{ - i \sqrt{\pi} \phi_{1,2} ( \tau_3 )} - e^{ i \sqrt{\pi} \phi_{1,2} ( \tau_3 ) } ] 
  +   \nonumber \\
 &&  \frac{\bar{t}_{3,2} ( \bar{t}_{1,3}^2 + \bar{t}_{2,1}^2 ) }{2} \: \int\: d \tau_1 \: d \tau_2 \: d \tau_3 \: 
\epsilon ( \tau_1 -\tau_3 ) \epsilon ( \tau_2 - \tau_3 )\: \prod_{i<j=1}^3 \theta ( | \tau_i - \tau_j | - \tau_0 ) \: \Gamma_2 ( \tau_3 ) \Gamma_3 ( \tau_3) 
 \times \nonumber \\
 && \Biggl\{ \frac{  | \tau_1 - \tau_3 |^\frac{1}{K} }{  | \tau_1 - \tau_2 |^\frac{2}{K} 
  | \tau_2 - \tau_3 |^\frac{1}{K} } +  \frac{ | \tau_2 - \tau_3 |^\frac{1}{K} }{   | \tau_1 - \tau_2 |^\frac{2}{K} 
   |  \tau_1 - \tau_3 |^\frac{1}{K} } \Biggr\} 
 \: [ e^{ - i \sqrt{\pi} \phi_{2,3} ( \tau_3 ) }- e^{ i \sqrt{\pi} \phi_{2,3} ( \tau_3 ) } ] 
 +   \nonumber \\
 &&  \frac{\bar{t}_{1,3} ( \bar{t}_{2,1}^2 + \bar{t}_{3,2}^2 ) }{2} \: \int\: d \tau_1 \: d \tau_2 \: d \tau_3 \: 
\epsilon ( \tau_1 -\tau_3 ) \epsilon ( \tau_2 - \tau_3 ) \: \prod_{i<j=1}^3 \theta ( | \tau_i - \tau_j | - \tau_0 )   \: \Gamma_3 ( \tau_3 ) \Gamma_1 ( \tau_3) 
 \times \nonumber \\
 && \Biggl\{ \frac{ | \tau_1 - \tau_3 |^\frac{1}{K}  }{   | \tau_1 - \tau_2 |^\frac{2}{K} 
  | \tau_2 - \tau_3 |^\frac{1}{K} } +  \frac{  | \tau_2 - \tau_3 |^\frac{1}{K}  }{  | \tau_1 - \tau_2 |^\frac{2}{K} 
  | \tau_1 - \tau_3 |^\frac{1}{K} } \Biggr\} 
 \: [ e^{ - i \sqrt{\pi} \phi_{3,1} ( \tau_3 ) }- e^{ i \sqrt{\pi} \phi_{3,1} ( \tau_3 ) } ] 
\:\:\:\:,
\label{pp.5}
\end{eqnarray}
\noindent
minus a term  given by the  product of $S_b$ times the free energy correction to ${\cal O} (\bar{t}_a^2 )$
  Performing the subtraction and 
using shifted integration variables, we eventually find   

\begin{eqnarray}
 && \delta S_b^{(1)} = \int \: d w_1 \: d w_2 \: \theta ( | w_1 | - \tau_0 ) \: \theta ( | w_2 | - \tau_0 ) \: \theta ( | w_1 - w_2 | - \tau_0 ) \times \nonumber \\
 && \Biggl\{  \: \frac{\epsilon ( w_1 ) \epsilon ( w_2 )}{ | w_1 - w_2 |^\frac{2}{K} }
 \: \left[ \left| \frac{ w_1 }{w_2} \right|^\frac{1}{K} + \left| \frac{w_2}{w_1} \right|^\frac{1}{K} \right] - \frac{2}{| w_1 - w_2 |^\frac{2}{K}} \Biggr\} 
 \times \nonumber \\
 && \sum_{a = 1}^3 \: \frac{\bar{t}_{a+1 , a} ( \bar{t}_{a+2,a+1}^2 + \bar{t}_{a , a + 2}^2 ) }{2} \: \int \: d \tau_3 \: \Gamma_a ( \tau_3 ) \Gamma_{a+1} ( \tau_3 ) 
 \: [ e^{ - i \sqrt{\pi} \phi_{a , a+1} ( \tau_3 ) } - e^{ i \sqrt{\pi} \phi_{a , a+1} ( \tau_3 ) } ] 
 , 
 \label{puccy.1}
\end{eqnarray}
\noindent
with $w_{1,2} = \tau_{1,2} - \tau_3$.  
On differentiating $\delta S_b^{(1)}$ with respect to $\tau_0$, we obtain 

\beq
 \frac{\partial \delta S_b^{(1)} }{ \partial \tau_0} = \frac{1}{\tau_0^{\frac{2}{K} - 1}} \: {\cal A} \left[ \frac{1}{K}  \right] 
 \:  \sum_{a = 1}^3 \:  \bar{t}_{a+1,a} ( \bar{t}_{a+2 , a+1}^2 + \bar{t}_{a , a + 2}^2 )   \: \int \: d \tau_3 \: \Gamma_a ( \tau_3 ) \Gamma_{a+1} ( \tau_3 ) 
 \: [ e^{ - i \sqrt{\pi} \phi_{a , a+1} ( \tau_3 ) } - e^{ i \sqrt{\pi} \phi_{a , a+1} ( \tau_3 ) } ] 
 \:\:\:\: ,
 \label{puccy.2}
\eneq
\noindent
with

\begin{eqnarray}
 {\cal A} [ \nu ]  &=&- 2 \int_{\frac{3}{2}}^\infty \: d z \: \Biggl\{ \frac{1}{ \left( z - \frac{1}{2} \right)^{2 \nu} }
 \: \left[ \frac{1}{ \left( z + \frac{1}{2} \right)^\nu } +  \left( z + \frac{1}{2} \right)^\nu   + 2 \right] 
 - \frac{1}{ \left( z+ \frac{1}{2} \right)^{2 \nu} }
 \: \left[ \frac{1}{ \left( z - \frac{1}{2} \right)^\nu } +  \left( z - \frac{1}{2} \right)^\nu - 2  \right]  
  \nonumber \\
  &+& \left( \frac{z + \frac{1}{2}}{z - \frac{1}{2} } \right)^\nu + 
   \left( \frac{z - \frac{1}{2}}{z +  \frac{1}{2} } \right)^\nu - 2 \Biggr\} 
 \:\:\:\: . 
 \label{puccy.3}
\end{eqnarray}
\noindent
Apparently,  ${\cal A} [ \nu]$ in  Eq. (\ref{puccy.3})  generalizes  the ${\cal F}$-function of Ref. \cite{giuaf_1} to 
the Y3J. Assuming    $K^{-1} = 1 - \epsilon$ and expanding  to leading order in 
$\epsilon$, one finds ${\cal A} [ \nu = 1   -   \epsilon  ] \approx c \epsilon + {\cal O} ( \epsilon^2 )$, 
with the coefficient $c \approx 16.45$.

  \item {\bf Term number 2:}
  
 Already accounting for the symmetries that become effective under integrating over the imaginary times, 
this corresponds to 
 
 \begin{eqnarray}
&&  -  \bar{t}_{2,1} \bar{t}_{3,2} \bar{t}_{1,3}   \:  {\bf T}_\tau \:  \{ \Gamma_1 ( \tau_1 ) \Gamma_{2} ( \tau_1 ) \Gamma_2 ( \tau_2 ) 
\Gamma_3 ( \tau_2 ) \Gamma_3 ( \tau_3 ) \Gamma_1 ( \tau_3 ) \} \times \nonumber \\
&& {\bf T}_\tau \: \{ [ e^{ - i \sqrt{\pi} \phi_{1 , 2} ( \tau_1 ) } - e^{ i \sqrt{\pi} \phi_{1 ,2} ( \tau_1 ) } ] 
[ e^{ - i \sqrt{\pi} \phi_{2 ,3} ( \tau_2 ) } - e^{ i \sqrt{\pi} \phi_{2 , 3} ( \tau_2 ) } ] 
[ e^{ - i \sqrt{\pi} \phi_{3 , 1} ( \tau_3 ) } - e^{ i \sqrt{\pi} \phi_{3 , 1} ( \tau_3 ) } ] \}
 , 
\label{pp.6}
\end{eqnarray}
\noindent
and, clearly, it  cannot contribute any further renormalization to the $\bar{t}_a$'s.

  \item {\bf Term number 3:}

This corresponds to  a correction to $S_b$ given by

\begin{eqnarray}
&&    \delta  S_b^{(3)} =  \frac{1}{6} \: \sum_{a = 1}^3\:\bar{t}_{a+1,a}^3 \: \int_0^\beta \: d \tau_1 d \tau_2 \: d \tau_3 \: {\bf T}_\tau \{ \Gamma_a ( \tau_1 ) \Gamma_{a+1} ( \tau_1 )\Gamma_a ( \tau_2 ) \Gamma_{a+1} ( \tau_2 )   \Gamma_a ( \tau_3 ) \Gamma_{a+1} ( \tau_3 )  \}  \times \nonumber \\
&& {\bf T}_\tau \{ [ e^{ - i \sqrt{\pi} \phi_{a , a+1} ( \tau_1 ) } - e^{ i \sqrt{\pi} \phi_{a , a+1} ( \tau_1 ) } ] 
 [ e^{ - i \sqrt{\pi} \phi_{a , a+1} ( \tau_2 ) } - e^{ i \sqrt{\pi} \phi_{a , a+1} ( \tau_2 ) } ] 
  [ e^{ - i \sqrt{\pi} \phi_{a , a+1} ( \tau_3 ) } - e^{ i \sqrt{\pi} \phi_{a , a+1} ( \tau_3 ) } ] \} 
  \times \nonumber \\
  && \theta ( | \tau_1 - \tau_2 | - \tau_0 ) \: \theta ( | \tau_2 - \tau_3 | - \tau_0 ) \: \theta ( | \tau_2 - \tau_3 | - \tau_0 )  = \nonumber \\
   &&  - \frac{1}{6} \: \sum_{a = 1}^3\:\bar{t}_{a+1,a}^3 \: \int_0^\beta \: d \tau_1 d \tau_2 \: d \tau_3 \:     \Gamma_a ( \tau_3 ) \Gamma_{a+1} ( \tau_3 )    \times \nonumber \\
&& {\bf T}_\tau \{ [ e^{ - i \sqrt{\pi} \phi_{a , a+1} ( \tau_1 ) } - e^{ i \sqrt{\pi} \phi_{a , a+1} ( \tau_1 ) } ] 
 [ e^{ - i \sqrt{\pi} \phi_{a , a+1} ( \tau_2 ) } - e^{ i \sqrt{\pi} \phi_{a , a+1} ( \tau_2 ) } ] 
  [ e^{ - i \sqrt{\pi} \phi_{a , a+1} ( \tau_3 ) } - e^{ i \sqrt{\pi} \phi_{a , a+1} ( \tau_3 ) } ] \} 
  \times \nonumber \\
  && \theta ( | \tau_1 - \tau_2 | - \tau_0 ) \: \theta ( | \tau_2 - \tau_3 | - \tau_0 ) \: \theta ( | \tau_2 - \tau_3 | - \tau_0 )     
  \:\:\:\: ,
  \label{pp.4}
\end{eqnarray}
\noindent
again minus a term given by the product of $S_b$ times the free energy correction to ${\cal O} (\bar{t}_a^2 )$.  As a 
result, taking into account the OPE

\beq
e^{ i a \sqrt{\pi} \phi_{a , a+1} ( \tau_1 ) } e^{ i b \sqrt{\pi} \phi_{a , a+1} ( \tau_2 ) } e^{ i c \sqrt{\pi}  \phi_{a , a+1} ( \tau_3 ) } 
\approx_{ \tau_1 \sim \tau_2 \sim \tau_3 } 
\: | \tau_1 - \tau_2 |^\frac{2 ab}{K} \: | \tau_1 - \tau_3 |^{ \frac{2 a c }{K}} \: | \tau_2 - \tau_3 |^{\frac{2 b c }{K}}
\:\:\:\: , 
\label{pp4one}
\eneq
\noindent
we obtain  

\begin{eqnarray}
 &&  \delta  S_b^{(3)} =\sum_{a = 1}^3 \:  \frac{\bar{t}_{a+1,a}^3}{6} \: \int \:d w_1 d w_2 \: \times 
 \label{pp.4bis} \\
 && \Biggl\{ 
 \frac{ | w_1 - w_2 |^\frac{2}{K} }{ | w_1 |^\frac{2}{K} \:  | w_2 |^\frac{2}{K} } 
 + \frac{ | w_1 |^\frac{2}{K} }{ | w_1 - w_2 |^\frac{2}{K} \:  | w_2 |^\frac{2}{K} } 
 +   \frac{ | w_2 |^\frac{2}{K} }{ |  w_1 - w_2 |^\frac{2}{K} \: | w_1 |^\frac{2}{K} }
 \theta ( | w_1 | - \tau_0 ) \theta ( | w_2 | - \tau_0 ) \theta ( | w_1 - w_2 | - \tau_0 ) 
 \nonumber \\
 && - \frac{6}{| w_1 - w_2 |^\frac{2}{K}} \theta ( | w_1 - w_2 | - \tau_0 ) \Biggr\} \: \int_0^\beta \: 
 d \tau_3 \: \sum_{a = 1}^3 \:  \Gamma_a ( \tau_3 ) \Gamma_{a+1} ( \tau_3 )  \: 
 [ e^{ - i \sqrt{\pi} \phi_{a , a+1} ( \tau_3 ) } - e^{ i \sqrt{\pi} \phi_{a , a+1} ( \tau_3 ) } ] \nonumber
 \:\:\:\: .
\end{eqnarray}
\noindent 
\end{itemize}
On separately considering  the contributions from positive- and negative- values of the integration variables, 
we may rewrite Eq. (\ref{pp.4bis}) as

\begin{eqnarray}
 &&  \delta  S_b^{(3)} =\sum_{a = 1}^3 \:  \frac{\bar{t}_{a+1,a}^3}{3} \: \int 
 d \tau_3 \:   \Gamma_a ( \tau_3 ) \Gamma_{a+1} ( \tau_3 )  \: 
 [ e^{ - i \sqrt{\pi} \phi_{a , a+1} ( \tau_3 ) } - e^{ i \sqrt{\pi} \phi_{a , a+1} ( \tau_3 ) } ]  \: \int_{\tau_0}^\infty  \:d w_1 d w_2 \: \times \nonumber \\
 && \Biggl\{ 
 \frac{ | w_1 - w_2 |^\frac{2}{K} }{   w_1^\frac{2}{K} \:   w_2^\frac{2}{K} } 
 + \frac{   w_1^\frac{2}{K} }{ | w_1 - w_2 |^\frac{2}{K} \:    w_2^\frac{2}{K} } 
 +   \frac{ w_2^\frac{2}{K} }{ |  w_1 - w_2 |^\frac{2}{K} \:  w_1^\frac{2}{K} }
 - \frac{6}{| w_1 - w_2 |^\frac{2}{K}} \Biggr\} \theta ( | w_1 - w_2 | - \tau_0 )  \nonumber \\
 &&\sum_{a = 1}^3 \:  \frac{\bar{t}_{a+1,a}^3}{3}  \: \int 
 d \tau_3 \:   \Gamma_a ( \tau_3 ) \Gamma_{a+1} ( \tau_3 )  \: 
 [ e^{ - i \sqrt{\pi} \phi_{a , a+1} ( \tau_3 ) } - e^{ i \sqrt{\pi} \phi_{a , a+1} ( \tau_3 ) } ]\: \int_{\tau_0}^\infty  \:d w_1 d w_2 \: \times \nonumber \\
 && \Biggl\{ 
 \frac{ | w_1+ w_2 |^\frac{2}{K} }{   w_1^\frac{2}{K} \:   w_2^\frac{2}{K} } 
 + \frac{   w_1^\frac{2}{K} }{ | w_1 + w_2 |^\frac{2}{K} \:    w_2^\frac{2}{K} } 
 +   \frac{ w_2^\frac{2}{K} }{ |  w_1 + w_2 |^\frac{2}{K} \:  w_1^\frac{2}{K} }
 - \frac{6}{| w_1 + w_2 |^\frac{2}{K}}  \Biggr\}
 \:\:\:\: ,
 \label{pp.4bisca}
\end{eqnarray}
\noindent
On differentiating $\delta S_b^{(3)}$ with respect 
to $\tau_0$, we therefore get 

\beq
 \frac{\partial \delta S_b^{(3)} }{ \partial \tau_0} = \frac{1}{\tau_0^{\frac{2}{K} - 1}} \: {\cal B} \left[ \frac{1}{K}  \right] 
 \:  \sum_{a = 1}^3 \:  \bar{t}_{a+1,a}^3   \: \int \: d \tau_3 \: \Gamma_a ( \tau_3 ) \Gamma_{a+1} ( \tau_3 ) 
 \: [ e^{ - i \sqrt{\pi} \phi_{a , a+1} ( \tau_3 ) } - e^{ i \sqrt{\pi} \phi_{a , a+1} ( \tau_3 ) } ] 
 \:\:\:\: ,
 \label{pp.4ter}
 \eneq
 \noindent
with 

\begin{eqnarray} 
{\cal B} [ \nu ] &=& - \frac{2}{3} \: \int_\frac{3}{2}^\infty \: d z \: 
 \Biggl\{ \left( \frac{z-1/2}{z+1/2} \right)^{2 \nu} + \left( \frac{z + 1/2}{z-1/2} \right)^{2 \nu} 
 + \frac{1}{  ( z^2 - 1/4 )^{2 \nu} } - \frac{6}{(z-1/2)^{2 \nu} } \Biggr\} \nonumber \\
 &-& \frac{2}{3} \: \int_\frac{3}{2}^\infty \: d z \: 
 \Biggl\{ \left( \frac{z+1/2}{z-1/2} \right)^{2 \nu} + \left( \frac{z - 1/2}{z + 1/2} \right)^{2 \nu} 
 + \frac{1}{  ( z^2 - 1/4 )^{2 \nu} } - \frac{6}{(z+1/2)^{2 \nu} } \Biggr\} \nonumber \\
 &-&  \frac{2}{3} \: \int_\frac{3}{2}^\infty \: d z \: 
 \Biggl\{ \left( \frac{z+1/2}{z-1/2} \right)^{2 \nu} + \left( \frac{z - 1/2}{z + 1/2} \right)^{2 \nu} 
 + \frac{1}{  ( z^2 - 1/4 )^{2 \nu} } - 6   \Biggr\}
 \;\:\:\: . 
 \label{pp.4quater}
 \end{eqnarray}
 \noindent
 Setting $K^{-1} = 1 - \epsilon$ and expanding to leading order in $\epsilon$ one obtains ${\cal B} [ 1 - \epsilon ] \approx b \epsilon + {\cal O} ( \epsilon^2 )$,  with  
  $b \sim 26.32$. Thus, eventually putting together the contributions from Term 1 and Term 3, we 
find that the  perturbative RG equations for the running couplings are given by 

\begin{eqnarray}
 \frac{d \bar{t}_{2,1}}{d \ln ( \tau / \tau_0 ) } &=& \epsilon \{ \bar{t}_{2,1} - \bar{t}_{2,1} [ b ( \bar{t}_{2,1} )^2 + c ( ( \bar{t}_{3,2} )^2 + 
 ( \bar{t}_{1,3} )^2 ) ] \} \nonumber \\
  \frac{d \bar{t}_{3,2}}{d \ln ( \tau / \tau_0 ) } &=& \epsilon \{ \bar{t}_{3,2} - \bar{t}_{3,2} [ b ( \bar{t}_{3,2} )^2 + c ( ( \bar{t}_{1,3} )^2 +
  ( \bar{t}_{2,1} )^2 ) ] \} \nonumber \\
  \frac{d \bar{t}_{1,3}}{d \ln ( \tau / \tau_0 ) } &=& \epsilon \{ \bar{t}_{1,3} - \bar{t}_{1,3} [ b ( \bar{t}_{1,3} )^2 + c ( ( \bar{t}_{2,1} )^2 +
  (  \bar{t}_{3,2} )^2 ) ] \} \nonumber \\
 \:\:\:\: . 
 \label{puccy.x1}
\end{eqnarray}
\noindent
From Eqs. (\ref{puccy.x1}), we find a FCFP at 
$\bar{t}_{2,1} = \bar{t}_{3,2} = \bar{t}_{1,3} = t_* = 1 /\sqrt{  b + 2 c }$, that is independent of $\epsilon$.  
Additional fixed points are found at $\bar{t}_{2,1} = 0 , \bar{t}_{3,2} = \bar{t}_{1,3} = 1 / \sqrt{b + c}$, plus 
permutations of the three indices. In the main text, however, we discuss the reliability of this 
result and use alternative methods, such as the DEBC-approach of the explicit calculation of the 
$g$-function in combination with the $g$-theorem, to make more accurate predictions about 
FCFP's in the phase diagram of the 3YJ. 

For the purpose of discussing the relation between the $N=2$ junction and the Y3J, it is important to 
generalize Eqs. (\ref{puccy.x1}) to the case of unequal Luttinger parameters in the three wires. Specifically, 
we now assume that wire-1 and -2 are characterized by a Luttinger parameter $K$, while wire-3 is characterized by 
a Luttinger parameter $K_3$ with, in general, $K_3 \neq K$. In this case, following the same 
procedure we followed above, yields the generalization of Eqs. (\ref{puccy.x1}) to 

\begin{eqnarray}
  \frac{d \bar{t}_{2,1}}{d \ln ( \tau / \tau_0 ) } &=&    \left( 1 - \frac{1}{K} \right)  \bar{t}_{2,1} - 
  {\cal B} \left[ \frac{1}{K} \right] ( \bar{t}_{2,1} )^3 - {\cal C} \left[ \frac{1}{K} , \frac{1}{K_3} \right]   \bar{t}_{2,1} ( ( \bar{t}_{3,2} )^2 + ( \bar{t}_{1,3})^2 ) \equiv 
  \hat{\beta}_1 [ \bar{t}_{2,1} , \bar{t}_{3,2} , 
  \bar{t}_{1,3} ] \nonumber \\
    \frac{d \bar{t}_{3,2} }{d \ln ( \tau / \tau_0 ) } &=& \left( 1 - \frac{1}{2K} - \frac{1}{2 K_3} \right) \bar{t}_{3,2} -{\cal B} \left[ \frac{1}{2K} + \frac{1}{2 K_3} \right] ( \bar{t}_{3,2} )^3 
    -  {\cal C} \left[ \frac{1}{K} , \frac{1}{K_3} \right]   \bar{t}_{3,2} ( \bar{t}_{1,3} )^2  -  {\cal C} \left[ \frac{1}{K} , \frac{1}{K} \right]   \bar{t}_{3,2} ( 
    \bar{t}_{2,1} )^2 \nonumber \\
    &\equiv& \hat{\beta}_2  [ \bar{t}_{2,1} , \bar{t}_{3,2} , 
  \bar{t}_{1,3} ]   \nonumber \\
    \frac{d \bar{t}_{1,3} }{d \ln ( \tau / \tau_0 ) } &=& \left( 1 - \frac{1}{2K} - \frac{1}{2 K_3} \right) \bar{t}_{1,3} -{\cal B} \left[ \frac{1}{2K} + \frac{1}{2 K_3} \right] ( \bar{t}_{1,3} )^3 
    -  {\cal C} \left[ \frac{1}{K} , \frac{1}{K_3} \right]   \bar{t}_{1,3} ( \bar{t}_{3,2} )^2  -  {\cal C} \left[ \frac{1}{K} , \frac{1}{K} \right]   \bar{t}_{1,3}  (
    \bar{t}_{2,1} )^2   \nonumber \\
    &\equiv& \hat{\beta}_3  [ \bar{t}_{2,1} , \bar{t}_{3,2} , 
  \bar{t}_{1,3} ] 
    \: , 
    \label{puccy.y1}
\end{eqnarray}
\noindent
with the function ${\cal B} [  \nu ]$ defined in Eq. (\ref{pp.4quater}) and ${\cal C} [ \nu_1 , \nu_2 ]$ generalizing 
${\cal A} [ \nu ]$ in Eq. (\ref{puccy.3}) to

\begin{eqnarray}
 {\cal C} [ \nu_1 , \nu_2 ]  &=&- 2 \int_{\frac{3}{2}}^\infty \: d z \: \Biggl\{ \frac{1}{ \left( z - \frac{1}{2} \right)^{ \nu_1 + \nu_2 } }
 \: \left[ \frac{1}{ \left( z + \frac{1}{2} \right)^{\nu_2} } +  \left( z + \frac{1}{2} \right)^{\nu_2}    + 2 \right] 
 - \frac{1}{ \left( z+ \frac{1}{2} \right)^{ \nu_1 + \nu_2 } }
 \: \left[ \frac{1}{ \left( z - \frac{1}{2} \right)^{\nu_2}  } +  \left( z - \frac{1}{2} \right)^{\nu_2} - 2  \right]  
  \nonumber \\
  &+& \left( \frac{z + \frac{1}{2}}{z - \frac{1}{2} } \right)^{\nu_2}  + 
   \left( \frac{z - \frac{1}{2}}{z +  \frac{1}{2} } \right)^{\nu_2} - 2 \Biggr\} 
 \:\:\:\: , 
 \label{puccy.y2}
\end{eqnarray}
\noindent
which  implies  ${\cal A} [ \nu ] = {\cal C} [ \nu , \nu ]$, as well as ${\cal F} [ \nu ] = {\cal C} [ 2 - \nu , \nu ]$.

 \end{document}